\documentclass[11pt,reqno]{amsproc}
\allowdisplaybreaks
\usepackage[only,llbracket,rrbracket,llparenthesis,rrparenthesis]{stmaryrd} 
\usepackage{fullpage}
\usepackage[semicolon,square,authoryear]{natbib}
\numberwithin{equation}{section}
\usepackage{cite}
\usepackage{enumerate}
\usepackage{caption}
\usepackage{mathrsfs}
\usepackage{cancel}

\usepackage{psfrag}
\usepackage{graphicx}
\DeclareGraphicsExtensions{.eps}
\usepackage{epstopdf}
\usepackage{subfigure}
\usepackage{float}
\usepackage{soul}
\usepackage[utf8]{inputenc}
\usepackage[table]{xcolor}
\usepackage[most]{tcolorbox}
\usepackage[debug=false, colorlinks=true, pdfstartview=FitV, linkcolor=blue, citecolor=blue, urlcolor=blue]{hyperref}

\newtheorem{theorem}{\bf Theorem}[section]

\usepackage{morefloats}

\newlength{\drop}
\definecolor{amethyst}{rgb}{0.6, 0.4, 0.8}
\definecolor{burgundy}{rgb}{0.5, 0.0, 0.13}

\newcommand\ddfrac[2]{\frac{\displaystyle #1}{\displaystyle #2}}
\title{On interface conditions for flows in coupled free-porous media}

\author{\textbf{K.~B.~Nakshatrala} and 
\textbf{M.~S.~Joshaghani}\\
{\small 
Department of Civil and Environmental 
Engineering, University of Houston, 
Houston, Texas 77204.\\
\textbf{Correspondence to:}~knakshatrala@uh.edu}}

\begin{document}

%===========================;
%  Title page of the paper  ;
%===========================;
\begin{titlepage}
  \drop=0.1\textheight
  \centering
  \vspace*{\baselineskip}
  \rule{\textwidth}{1.6pt}\vspace*{-\baselineskip}\vspace*{2pt}
  \rule{\textwidth}{0.2pt}\\[\baselineskip]
  {\LARGE \textbf{\color{burgundy}
    On interface conditions for flows in 
    coupled free-porous \\[0.2\baselineskip] 
    media}}\\[0.2\baselineskip]
    \rule{\textwidth}{0.4pt}\vspace*{-\baselineskip}\vspace{3.2pt}
    \rule{\textwidth}{1.6pt}\\[0.5\baselineskip]
    \scshape
    An e-print of this paper is available on arXiv:~1902.02510. \par
    \vspace*{0.3\baselineskip}
    Authored by \\[0.1\baselineskip]
    {\Large Kalyana~B.~Nakshatrala\par}
    {\itshape Department of Civil \& Environmental Engineering \\
    University of Houston, Houston, Texas 77204--4003. \\ 
    \textbf{phone:} +1-713-743-4418, \textbf{e-mail:} knakshatrala@uh.edu \\
    \textbf{website:} \url{http://www.cive.uh.edu/faculty/nakshatrala}\par}
    \vspace*{0.1\baselineskip}
    {\Large Mohammad~S.~Joshaghani \par}
    {\itshape Graduate Student, University of Houston. \par}
    \vspace*{0.1\baselineskip}
    %%
	%-----------;
	%  Figures  ;
	%-----------;
    \begin{figure}[H]
    \vspace{-0.12in}
      \centering
      \includegraphics[scale=0.7]{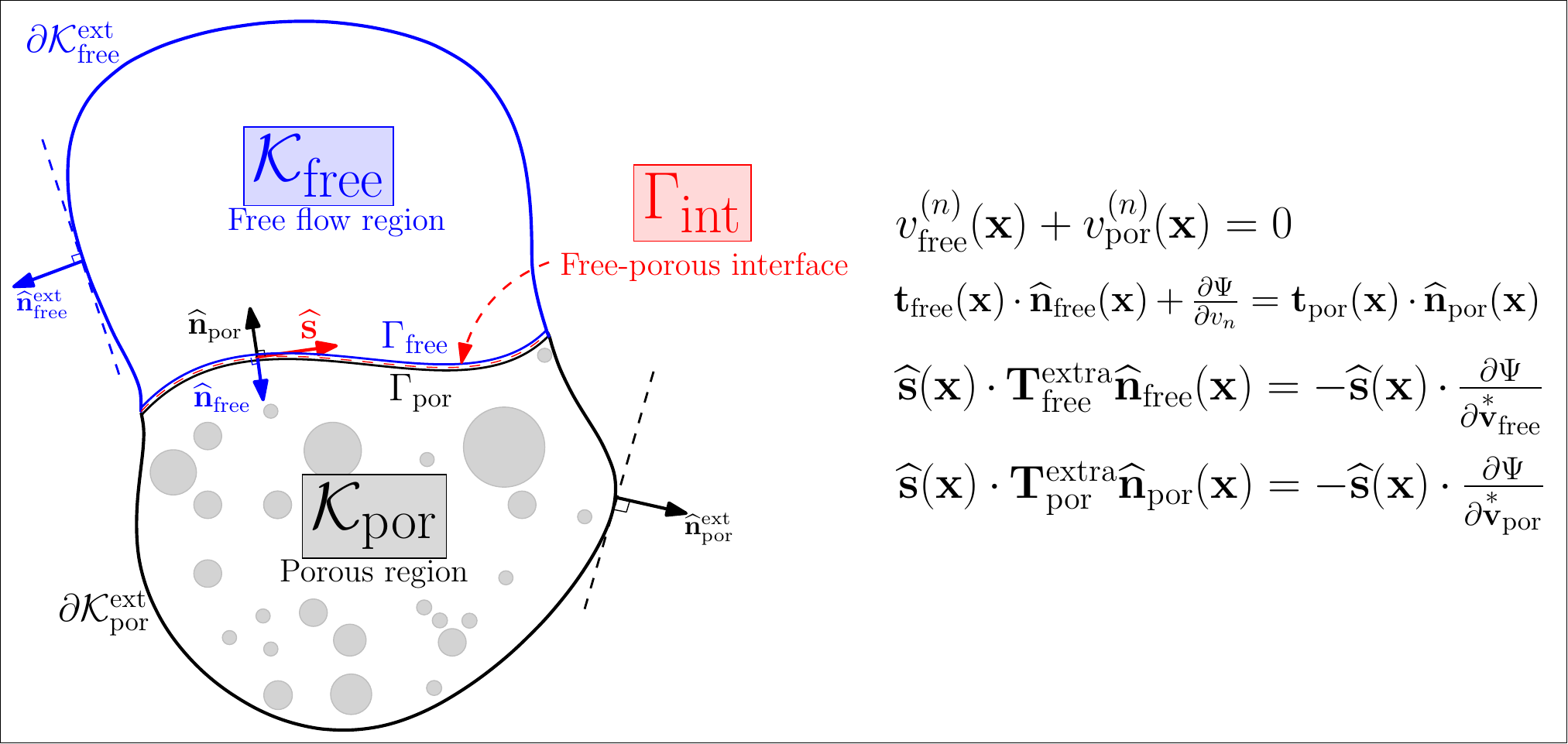}

      \emph{The derived interface conditions are summarized 
        along with a pictorial description of
        the problem, which pertains to the flow of an incompressible 
        fluid in coupled free-porous media. 
        $\Psi$ is the power expended density along the 
        interface. $\mathbf{v}_{\mathrm{free}}$
        and $\mathbf{v}_{\mathrm{por}}$ are the velocities
        in the free and porous regions, respectively.
        A superposed asterisk on a (vectorial) quantity denotes 
        its tangential component along the interface. 
        $v_n$ is the normal component of the velocity at the 
        interface from the free region into the porous region. 
        $\mathbf{T}_{\mathrm{free}}^{\mathrm{extra}}$ and 
        $\mathbf{T}_{\mathrm{por}}^{\mathrm{extra}}$,
        respectively, denote the extra Cauchy stresses in the
        free and porous regions.
        $\mathbf{t}_{\mathrm{free}}$ and $\mathbf{t}_{\mathrm{por}}$,
        respectively, denote the tractions on the free and porous 
        sides of the interface with outward normals 
        $\widehat{\mathbf{n}}_{\mathrm{free}}$ and 
        $\widehat{\mathbf{n}}_{\mathrm{por}}$. A unit 
        tangential vector along the interface is denoted 
        by $\widehat{\mathbf{s}}$.}
    \end{figure}
    \vfill
        {\scshape 2019} \\
        {\small Computational \& Applied Mechanics Laboratory} \par
\end{titlepage}

%=========================;
%  Abstract of the paper  ;
%=========================;
\begin{abstract}
\vspace{-0.1in}
Many processes in nature (e.g., physical and biogeochemical processes in hyporheic zones, and arterial mass transport) occur near the interface of free-porous media. A firm understanding of these processes needs an accurate prescription of flow dynamics near the interface which (in turn) hinges on an appropriate description of interface conditions along the interface of free-porous media. Although the conditions for the flow dynamics at the interface of free-porous media have received considerable attention, many of these studies were empirical 
and lacked a firm theoretical underpinning. In this paper, we derive a complete and self-consistent set of conditions for flow dynamics at the interface of free-porous media. We first propose a principle of virtual power by incorporating the virtual power expended at the interface of free-porous media. Then by appealing to the calculus of variations, we obtain a complete set of interface conditions for flows in coupled free-porous media. A noteworthy feature of our approach is that the derived interface conditions apply to a wide variety of porous media models. We also show that the two most popular interface conditions -- the Beavers-Joseph condition and the Beavers-Joseph-Saffman condition -- are special cases of the approach presented in this paper. The proposed principle of virtual power also provides a minimum power theorem for a class of flows in coupled free-porous media, which has a similar mathematical structure as the ones enjoyed by flows in uncoupled free and porous media.  
\end{abstract}

%=========================;
%  Keywords of the paper  ;
%=========================;
\keywords{coupled free-porous media; principle of virtual power; interface conditions; internal constraints; calculus of variations; minimum power principle}

\maketitle

%==================================;
%  Include all the sections below  ;
%==================================;

%*************************************************;
%                                                 ;
%  NAME                                           ;
%    S0_ICs_Problem_statement.tex                 ;
%                                                 ;
%  WRITTEN BY                                     ;
%    Kalyana Babu Nakshatrala                     ;
%                                                 ;
%*************************************************;
\section*{PROBLEM STATEMENT}
\begin{figure}[h]
  \centering
  \includegraphics[scale=0.8]{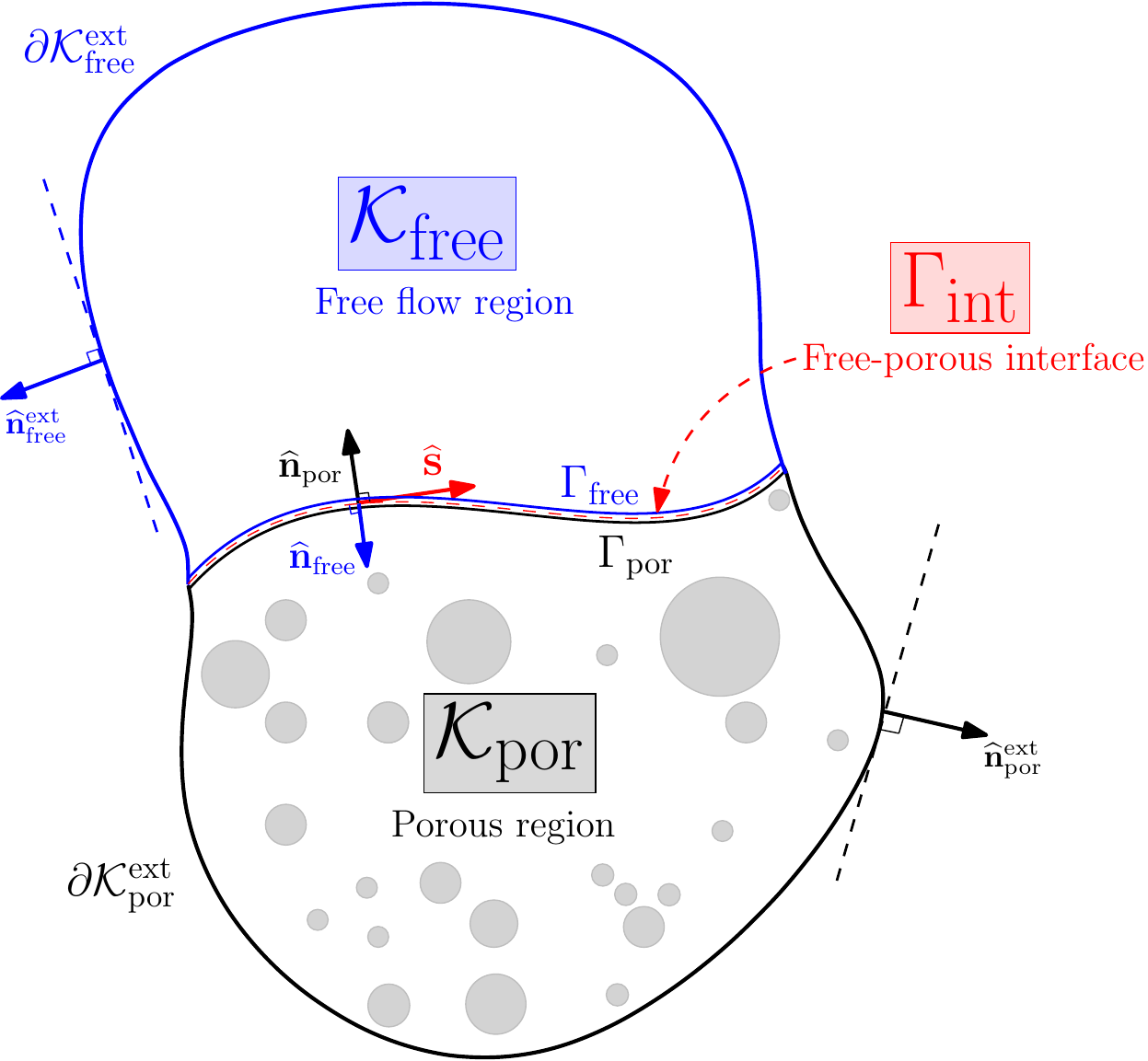}
  \caption{A pictorial description of coupled 
  free-porous media. The free flow region 
  $\mathcal{K}_{\mathrm{free}}$ and the 
  porous region $\mathcal{K}_{\mathrm{por}}$ 
  share a common interface $\Gamma_{\mathrm{int}}$.
    The outward unit normal vector to
    $\mathcal{K}_{\mathrm{free}}$ at the
    interface is denoted by
    $\widehat{\mathbf{n}}_{\mathrm{free}}
    (\mathbf{x})$. A similar notation
    holds for
    $\widehat{\mathbf{n}}_{\mathrm{por}}(\mathbf{x})$,
    which is equal to
    $-\widehat{\mathbf{n}}_{\mathrm{free}}(\mathbf{x})$.
    The side of $\Gamma_{\mathrm{int}}$ that shares with
    $\mathcal{K}_{\mathrm{free}}$ is noted by
    $\Gamma_{\mathrm{free}}$, and a similar
    notation holds for $\Gamma_{\mathrm{por}}$.
    The external boundaries of the free and porous
    regions are, respectively, denoted by $\partial
    \mathcal{K}_{\mathrm{free}}^{\mathrm{ext}}$ and
    $\partial \mathcal{K}_{\mathrm{por}}^{\mathrm{ext}}$. 
    The corresponding unit outward normals to
    these external boundaries are denoted by
    $\widehat{\mathbf{n}}_{\mathrm{free}}^{\mathrm{ext}}$
    and $\widehat{\mathbf{n}}_{\mathrm{por}}^{\mathrm{ext}}$.  
    A unit tangent vector on the interface is denoted by
    $\widehat{\mathbf{s}}$. \label{Fig:Schematic_domain}}
\end{figure}

Let us consider a domain which consists of two 
non-overlapping regions: a porous region and a 
free flow region. \emph{The interface} is the surface 
that demarcates these two regions.
\textbf{Fig.~\ref{Fig:Schematic_domain}} provides 
a pictorial description. Now consider the situation 
in which an incompressible fluid flows in this domain 
with the porous solid to be rigid. 
%---------------------;
%  Problem statement  ;
%---------------------;
The central question pertaining flows in coupled 
free-porous media is:
\begin{quote}
  \emph{Given the domain, free flow and
  porous regions, boundary conditions on the
  external boundaries, properties of the
  incompressible fluid (e.g., the coefficient
  of dynamic viscosity, true density),
  and properties of the rigid porous medium
  (e.g., porosity, permeability), what
  is the set of conditions appropriate 
  at the interface?}
\end{quote}

%----------------------------------------------------------;
%  Figure 1: A pictorial description of free-porous media  ;
%----------------------------------------------------------;

%*****************************************************;
%                                                     ;
%  NAME                                               ;
%    S1_ICs_Intro.tex                                 ;
%                                                     ;
%  WRITTEN BY                                         ;
%    Kalyana Babu Nakshatrala                         ;
%    Mohammad S. Joshaghani                           ;
%                                                     ;
%*****************************************************;
\section{INTRODUCTION AND MOTIVATION}
\label{Sec:S1_ICs_Intro}
%==========================;
%  Subsection: Motivation  ;
%--------------------------;
\subsection{Motivation} Many important science 
and engineering problems involve flows in a domain 
which comprises free flow and porous regions. In 
these problems, a plethora of vital processes takes 
place near the interface of free flow and porous 
regions. One has to capture these processes 
accurately to discern the overall dynamics and 
all the interactions in the entire domain. We now 
discuss two such problems, which have motivated 
us to undertake the research presented in this 
paper\footnote{Professor Lallit Anand has informed 
us that an appropriate set of conditions at the interface 
of porous and free regions is also important in the studies 
on Lithium-ion batteries.}.  
  
\begin{figure}[h]
    \vspace{-0.5in}
  \centering
  \subfigure{
    \includegraphics[scale=0.5]{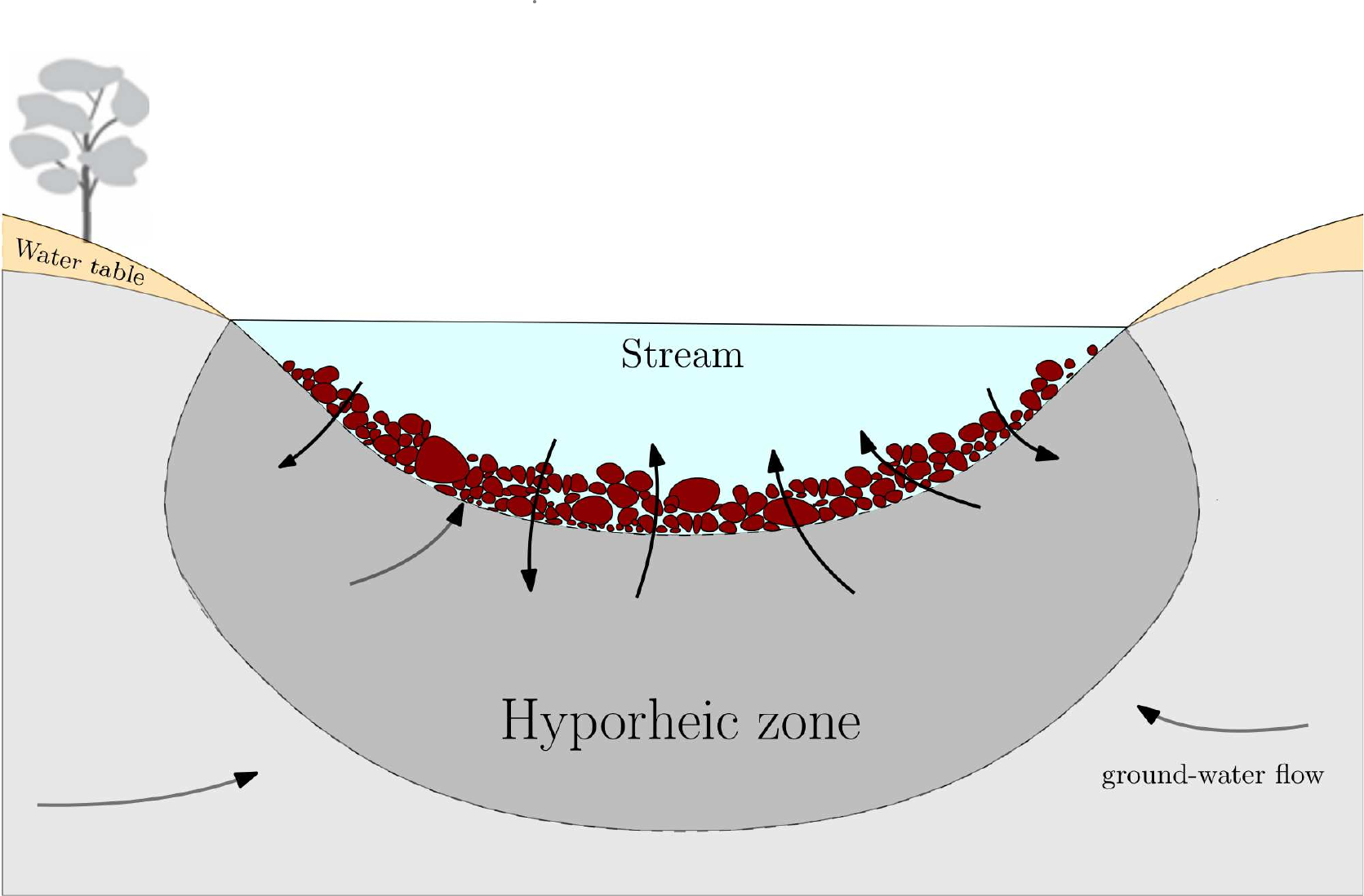}}
  \subfigure{
    \includegraphics[scale=0.3]{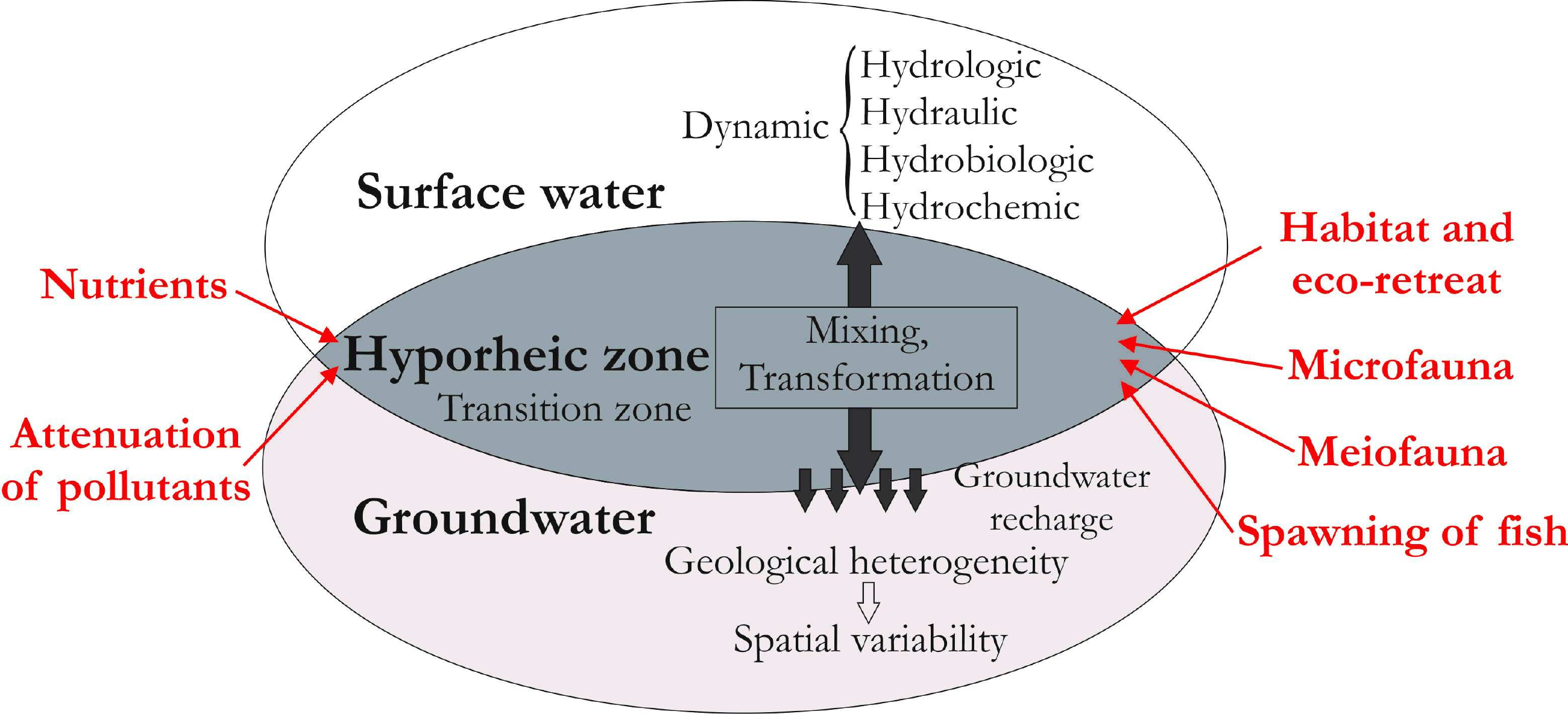}}
  \caption{The top figure (adapted 
    from the US Geological Survey 
    \citep{thomas1998ground}) shows 
    a typical hyporheic zone. 
    The size of a hyporheic zone can
    vary tens of meters vertically
    to hundreds of meters laterally. 
    The bottom figure (adapted from 
    the British Environment Agency 
    \citep{buss2009hyporheic}) depicts 
    a myriad of important processes 
    that take place in a hyporheic 
    zone which affect the processes 
    at the watershed scale (varying 
    from tens to hundreds 
    of kilometers) and hence affect 
    the overall ecosystem.
    \label{Fig:Hyporheic_zone}}
\end{figure}

The \emph{first problem} pertains to surface-subsurface
interactions of large water systems. Groundwater
and surface water interactions between rivers
and streams are vital to flora and fauna, water
distribution, and environmental factors which 
all affect the whole food chain \citep{jones1996surface,
  sophocleous2002interactions}. 
For example, mixing at the interface of 
groundwater and surface water is critical 
for nutrient transport and the carbon \& 
nitrogen (C\&N) cycles; both are vital to 
an ecosystem \citep{dwivedi2017impact}.
%
%  Importance of hyporheic zones  
The interactions between groundwater and surface
water greatly depend on the flow dynamics in
the hyporheic zone (see
\textbf{Fig.~\ref{Fig:Hyporheic_zone}}). Several
physical and biogeochemical processes take place
in the hyporheic zone, and these processes are in
turn coupled with the processes that take place
in the free and subsurface zones. 
Therefore, the success of a predictive modeling
of surface-subsurface interactions will rest on the 
accurate modeling of the flow dynamics at the 
interface of free-porous regions.
    
The \emph{second problem} pertains to the arterial mass 
transfer---the transport of atherogenic macromolecules, 
such as low-density lipoproteins (LDL), from bulk blood 
flow into artery walls and vice versa \citep{2006_sun,wada1999theoretical}. 
Accumulation of LDL at the interface of bulk blood flow and the endothelial 
layer---the part of lumen next to the blood flow---is a primary cause of 
various cardiovascular diseases; for example, atherosclerotic lesions 
within the intima of arteries \citep{caro1971atheroma,hoff1975localization}. 
A firm understanding of this complex process will enable physicians to 
administer better therapeutic procedures. Mechanics can play an important 
role to gain a good understanding of this problem. However, any such an 
effort has to address accurately the complex flow dynamics at the interface 
of bulk blood flow and (porous) arterial walls.

%------------------------------------------;
%  Figure 2: Hyporheic zone and processes  ;
%------------------------------------------;

Obtaining a self-consistent,
independent and a complete set of conditions
at the interface---which we will refer
to as \emph{the interface conditions}\footnote{We
  believe that the usage of interface
conditions is more appropriate than the
two alternatives: \emph{jump conditions}
and \emph{boundary conditions}.
As discussed in Appendix \ref{Sec:App_ICs_JCs}, the 
jump conditions (which are the balance laws across
a singular surface) do not furnish a workable set of 
conditions for flows in coupled free-porous
media; especially when the porous solid is rigid,
which is the case in this paper. Moreover, the set of conditions derived
in this paper (given by equation \eqref{Eqn:Compact_form_of_interface})
does not entirely stem from the jump conditions.
Since the interface $\Gamma_{\mathrm{int}}$
is not an external boundary to the domain
of interest (which consists of both
free and porous regions, i.e., $\Omega$),
it is not appropriate to refer to these
conditions as boundary conditions 
for coupled flows. 
}---for flows in coupled free-porous media
is far from settled.
Before we elaborate on some prior works and 
present our approach, we now outline what 
should be the nature of the interface conditions. 
We portray the \emph{character of interface conditions}\footnote{Influenced by the lecture
  ``\emph{The Character of Physical Law}'' given
  by \citet{feynman1967character}, we mimic the
  terminology and employ the phrase: the character
  of interface conditions, in our discussion
  on the general physical and mathematical nature of
  interface conditions.} as follows:  
\begin{enumerate}[(i)]
\item Interface conditions may directly
  stem from the balance laws and the associated
  jump conditions. For example, the no-penetration
  boundary condition at a stationary impervious
  wall, commonly employed in fluid mechanics, 
  stems from the jump condition
  associated with the balance of mass. 
\item Alternatively, they may be constitutive 
  specifications. If this is the case, they
  should be compatible with the balance laws
  and satisfy the essential invariance properties 
  (e.g., the principle of material frame-indifference
  or the Galilean invariance).
\item It is needless to say that they
  should agree with the experiments.
\item They should apply to a wide variety
  of problems.
\item They should give rise to mathematical
  models (i.e., boundary value problems and
  initial boundary value problems) that are
  mathematically well-posed.
\end{enumerate}

This paper fills the gap in our understanding of interface conditions for flows in coupled free-porous media. Our treatment of the problem will be at the continuum (or the so-called Darcy) scale.
The specific aims of this paper are twofold.
\emph{First}, to develop a framework for obtaining
appropriate conditions for coupled flow dynamics
at the interface of free-porous media. \emph{Second},
to recover some popular conditions available in the 
literature for coupled flows as special cases of the 
proposed framework.
Our approach will utilize the principle of virtual
power and the theory of interacting continua, invoke
a geometric argument to enforce the internal constraints,
impose the principle of material frame-indifference
on all the constitutive relations and use the standard
results from the calculus of variations. 

Over the last three decades, the principle of virtual
power has been extended with respect to its domain of
applicability, which was re-ignited by
\citet{germain1973method} and was further developed 
by \citet{maugin1980method}. Currently, the principle of 
virtual power has been employed for a wide variety of problems in 
mechanics, ranging from viscoplasticity \citep{anand2005theory}, 
gradient theories \citep{gurtin2005theory} to coupled problems 
\citep{fried2007thermomechanics}. 
A significant extension of this principle is to pose on
an arbitrary subset of the domain and obtain the Cauchy's
fundamental theorem for the stress (which relates the Cauchy
stress with the traction on a surface) as a consequence
\citep{podio2009virtual,
  fosdick2011principle}. Although such an extension 
  (defining the principle on an arbitrary subset) is not 
  essential to derive the interface conditions, we 
  will still show how to extend the proposed framework 
  to recover the Cauchy's fundamental theorem but will
  relegate such a discussion to one of the appendices.

  The theory of interacting continua, TIC,
  (also known as the mixture theory) is a
  mathematical framework to develop continuum
  models for a homogenized response of a mixture
  of (interacting) constituents \citep{Bowen}.
  The overall idea of TIC is to model a mixture
  of constituents as a collection of superposed
  continua.  
  Two inherent assumptions of TIC are the
  treatment of each constituent as a continuum 
  and the coexistence of all constituents in the
  space occupied by the mixture \citep{truesdell2012rational}.
  The second assumption
  can be thought as follows: at every point in the space
  occupied by the mixture, there is a particle from
  each of its constituents.\footnote{The
    coexistence of all constituents at a
    point in space may seem like a violation
    of the reality. However, it is no different from the fact that
  a spatial point in a continuum description
  is (in reality) made of several atoms, electrons,
  and elementary particles. It is thus essential 
  to be aware of the scale at which the modeling
  is done and at the same time recognize that TIC
  is a form of homogenized theory.}
  Each constituent has balance laws similar
  to that of a single continuum. However,
  the balance laws will contain terms
  which account the interactions due
  to the presence of other constituents.
  We will appeal to the TIC framework
  to model the porous media. 
    
The last piece in our proposed framework
is to systematically enforce internal
constraints, which in our case arise
due to the incompressibility of the
fluid in both regions, under the
principle of virtual power.
There are several approaches proposed in the
literature to enforce internal constraints.
The most popular approach, which is commonly 
referred to as the Truesdell-Noll approach 
\citep[\S30]{truesdell2013non}, is built upon 
two \emph{a priori} constitutive
assumptions: (i) the stress is decomposed into 
active and reactive components, and
(ii) the reactive component performs \emph{no}
work under a motion consistent with the
constraint. 
Alternatively, we employ the approach put forth by
\citet{carlson2004geometrically} to enforce the
internal constraint.
An attractive feature of this approach is that
the two assumptions made under the Truesdell-Noll
approach can be obtained as mathematical
consequences rather than \emph{a priori}
constitutive assumptions.
This approach hinges on the direct sum provided
by the projection theorem; however, the approach
can be easily explained by a simple geometrical
argument: \emph{If a vector $\boldsymbol{a}$ is
  perpendicular to every vector $\boldsymbol{b}$
  that is (in turn) perpendicular to a vector
  $\boldsymbol{c}$ then $\boldsymbol{a}$ and
  $\boldsymbol{c}$ are collinear.}
\citet{carlson2004geometrically} have employed the geometric argument
in the context of hyperelasticity (which is a non-dissipative
model) by utilizing the underlying energy balance formalism.
Herein, we show the principle of virtual power nicely
blends with the geometric argument for flows in
coupled free-porous media.

%===============================================;
%  Subsection: Scope and outline of this paper  ;
%-----------------------------------------------;
\subsection{Scope and an outline of this paper}
The plan for the rest of this paper is as follows. 
We will first outline some of the experimental observations 
and discuss some important prior works 
(\textbf{\S\ref{Sec:S2_ICs_Prior}}). 
We propose a principle of virtual power for
coupled flows by taking into account the virtual
power expended at the interface of free-porous
media (\textbf{\S\ref{Sec:S3_ICs_Proposed_framework}}). 
Using this principle, we obtain a complete
set of interface conditions which capture the
prior experimental observations 
(\textbf{\S\ref{Sec:S4_ICs_Derivation}}). 
We then show the popular conditions -- 
Beavers-Joseph and Beavers-Joseph-Saffman conditions -- 
to be special cases of the proposed framework. This
discussion will particularly reveal the assumptions and the
validity of these popular conditions for flows in coupled 
free-porous media
(\textbf{\S\ref{Sec:S5_ICs_Special_cases}}).
We also show that a class of flows in coupled 
free-porous media enjoys a minimum power 
theorem (\textbf{\S\ref{Sec:S6_ICs_MPT}}). 
We then employ the minimum power theorem 
to establish the uniqueness of solutions
under certain assumptions on the internal
dissipation and the power expended density
along the interface
(\textbf{\S\ref{Sec:S7_ICs_Uniqueness}}). 
We end the paper with a discussion on the main
findings (\textbf{\S\ref{Sec:S8_ICs_CR}}).

%*****************************************************;
%                                                     ;
%  NAME                                               ;
%    S2_ICs_Prior.tex                                 ;
%                                                     ;
%  WRITTEN BY                                         ;
%    Kalyana Babu Nakshatrala                         ;
%    Mohammad S. Joshaghani                           ;
%                                                     ;
%*****************************************************;
\section{EXPERIMENTAL OBSERVATIONS AND PRIOR WORKS}
\label{Sec:S2_ICs_Prior}
\begin{figure}[h]
  \includegraphics[scale=0.8]{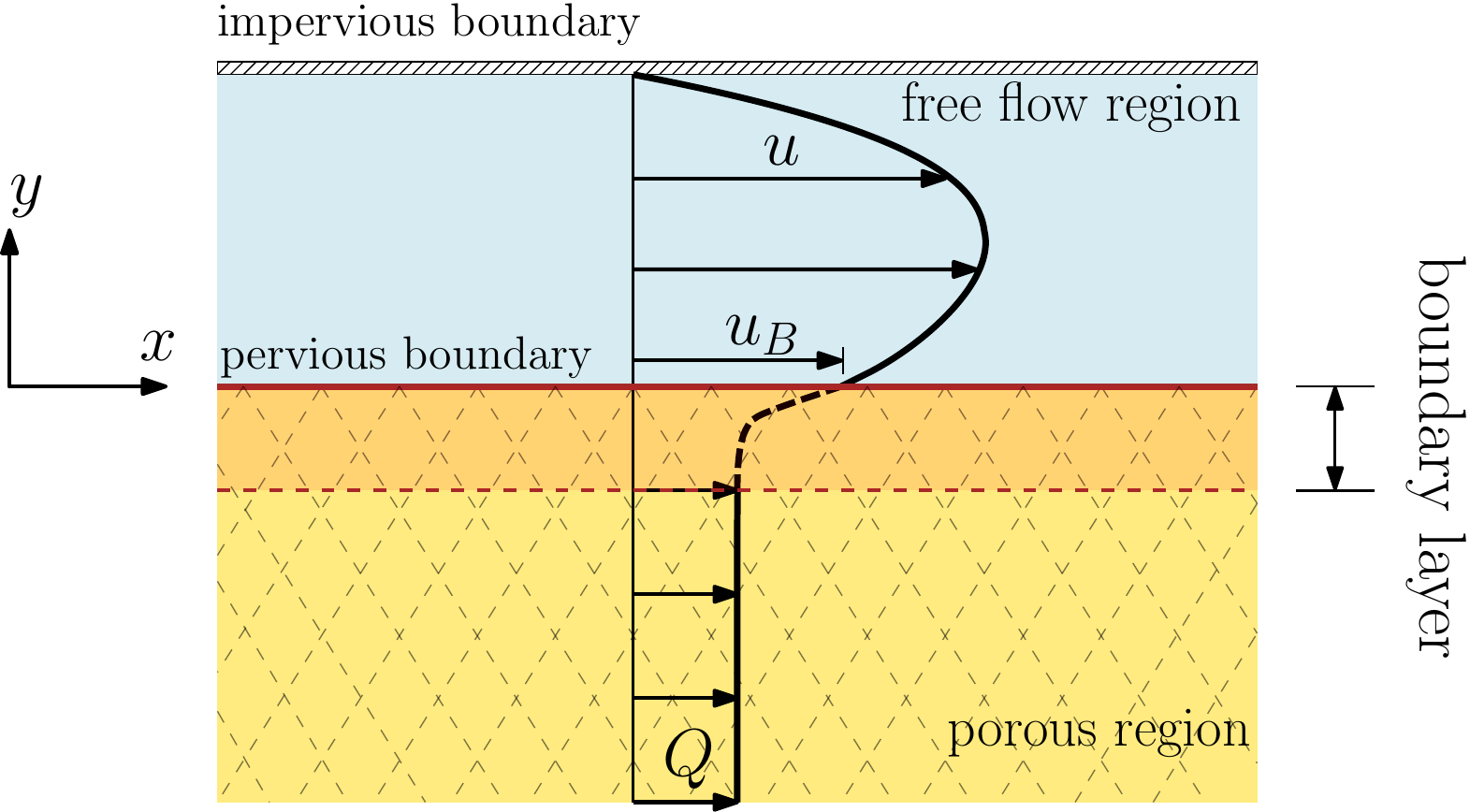}
  \caption{A pictorial description of the rectilinear
    flow in a horizontal channel between an impervious
    upper wall and a pervious lower wall (at $y = 0$).
    The corresponding velocity profile that was conjectured
    and used by \citet{1967_Beavers_JFM} 
    in their mathematical model for such flows
    is also shown. $Q$ is the discharge velocity 
    (and not the true or seepage velocity) in the
    porous medium. \label{Fig:ICs_BJ_velocity_profile}}
\end{figure}

The two most popular approaches are
the Beavers-Joseph (BJ) condition
\citep{1967_Beavers_JFM} and the
Beavers-Joseph-Saffman (BJS) condition
\citep{1971_Saffman_SAM}.
The experiments conducted by \citet{1967_Beavers_JFM}
provided the following two pieces of information
regarding flows near the interface of coupled 
free-porous media:
\begin{enumerate}[(i)]
\item The no-slip condition, commonly
  used for free flows at a boundary, is no
  longer satisfied at the interface.
\item There is a jump in the tangential
  components of velocity on either side of
  the interface. 
\end{enumerate}
\citet{1967_Beavers_JFM} also proposed an
empirical relation, which advocates that
the shear stress on the free flow side
of the interface is linearly proportional
to the jump in the tangential velocities
across the interface.
Based on the velocity profile and
the notation introduced in
\textbf{Fig.~\ref{Fig:ICs_BJ_velocity_profile}},
the BJ condition takes the following form: 
%--------------------------;
%  Equation: BJ condition  ;
%--------------------------;
\begin{align}
\label{Eqn:ICs_BJ_original_equation}
  u_B - Q = \left(\frac{k^{1/2}}{\alpha}\right)
  \left.\frac{\partial u}{\partial y}\right\vert_{y=0^{+}}
\end{align}
where $y = 0^{+}$ is the boundary limit point
from the free flow region, $k$ denotes the
isotropic permeability of the porous medium,
and $\alpha$ is a constant that depends only
on the properties of the fluid and the porous
material.

%------------------------;
%  Figure: BJ condition  ;
%------------------------;

Later, \citet{1971_Saffman_SAM} performed
a statistical analysis and suggested a
modification to the BJ condition, and
this new condition is popularly referred
to as the BJS condition.
Specifically, using a one-dimensional
geometrical setting and assuming uniform
pressure gradient in the porous medium,
\citet{1971_Saffman_SAM} argued that
the velocity on the porous medium side
is a higher-order term compared to the
velocity on the free flow side of the
interface, and hence one can neglect
the higher-order term.
The BJS condition takes the following form:
%---------------------------;
%  Equation: BJS condition  ;
%---------------------------;
\begin{align}
  u_B = \left(\frac{k^{1/2}}{\alpha}\right) 
  \left.\frac{\partial u}{\partial y}
  \right\vert_{y = 0^{+}} + O(k) 
\end{align}
where $O(\cdot)$ is the standard
``big O notation,'' which describes
the limiting behavior of a function
when the argument tends towards a
particular value.

Although these two approaches have laid the
foundation for much of the works in this
field, they suffer from some drawbacks,
which became clear because of new
experimental and numerical studies.
\emph{First},  the slip coefficients under
the BJ and BJS conditions are independent
of the velocities in the free flow and
porous regions.
However, \citet{2011_Liu_Prosperetti}
have shown the linear dependence of
the slip coefficient on the Reynolds
number, so the slip coefficient can
depend on the velocities. 
\emph{Second}, their primary interest is
free flows in a region with a part of
its boundary to be pervious due to a
juxtaposed porous medium. Their approaches  
were aimed at replacing the slip condition
with an alternate boundary condition
which is appropriate for free flows
due to a pervious boundary.
Their intended aim is also clear from the titles
of these works\footnote{The title of the paper
by Beavers and Joseph is ``\emph{Boundary
  conditions at a naturally permeable wall},''
and the title of the paper by Saffman is
``\emph{On the boundary condition at the
  surface of a porous medium}.''}.
Thus their
treatments do not provide sufficient
information to study flows in coupled
free-porous media, as there was no
discussion on appropriate boundary
conditions for the flows in the porous 
region. 
\emph{Third}, their treatment of
the boundary conditions is rather
ad hoc\footnote{To quote from
  \citep[p.~199]{1967_Beavers_JFM}:
  ``\emph{...we relate the slip velocity
    to the exterior flow by the
    \emph{ad hoc} boundary condition
    \[
    \left.\frac{du}{dy}\right|_{y=0^{+}}
    = \beta (u_B - Q) \qquad \qquad \qquad
    \qquad (2)
    \]
    where $0^{+}$ is a boundary limit point
    from the exterior fluid.}''} and are
not amenable to generalization to other
porous media models.

One can find in the literature great efforts
towards extending these two empirical conditions;
for example, see \citep{1987_Larson_JFM,1992_Sahraoui_IJHMT}.
However, to the authors' best knowledge, the literature does not
address all the issues laid out earlier under
the character of interface conditions. For
example, do the BJ/BJS conditions stem from
the jump conditions, are they constitutive
specifications, or do they combine jump
conditions and constitutive specifications?
If they are constitutive specifications, what
is the rationale behind them? Are they compatible
with the balance laws? Are these conditions valid
for other porous media models? In the subsequent
sections, we will answer all these questions and
present a framework for getting a complete set of
interface conditions (not just boundary conditions
for free flows due to the presence of a pervious
boundary) suitable for modeling flows in coupled
free-porous media.

%*****************************************************;
%                                                     ;
%  NAME                                               ;
%    S3_ICs_Framework.tex                             ;
%                                                     ;
%  WRITTEN BY                                         ;
%      Kalyana Babu Nakshatrala  	              ;
%      Mohammad S. Joshaghani                         ;
%                                                     ;
%*****************************************************;
\section{THE PROPOSED FRAMEWORK}
\label{Sec:S3_ICs_Proposed_framework}
%========================================;
%  Subsection: Notation and definitions  ;
%----------------------------------------;
\subsection{Notation and definitions}
Consider a domain $\Omega \subset \mathbb{R}^{nd}$
in which an incompressible fluid flows, where
``$nd$'' denotes the number of spatial dimensions 
and $\mathbb{R}$ denotes the set of real numbers.
A spatial point in the domain is denoted by
$\mathbf{x}$. The gradient and divergence
operators with respect to $\mathbf{x}$ are,
respectively, denoted by $\mathrm{grad}[\cdot]$
and $\mathrm{div}[\cdot]$.
The domain consists of two non-overlapping but 
adjoining regions: a porous region and a free flow 
region. See \textbf{Fig.~\ref{Fig:Schematic_domain}}
for a pictorial description.

%================================;
%  Subsubsection: The interface  ;
%--------------------------------;
\subsubsection{The interface}
The interface---the surface that demarcates the two 
regions---is denoted by $\Gamma_{\mathrm{int}}$.
The face of $\Gamma_{\mathrm{int}}$ that
is adjacent to the free flow region
is denoted by $\Gamma_{\mathrm{free}}$,
and the face of $\Gamma_{\mathrm{int}}$
that is adjacent to the porous region
is denoted by $\Gamma_{\mathrm{por}}$.
Note that $\Gamma_{\mathrm{int}}$, for our purposes,
has a zero thickness\footnote{In some
  applications involving multiphase
  fluids and heterogeneous mixtures,
  it will be necessary to treat the
  thickness of an interface to be of
  finite-size (albeit small) across
  which material and thermodynamic
  properties change sharply. For
  example, see \citep{berg2010introduction}.},
and the faces $\Gamma_{\mathrm{free}}$ and
$\Gamma_{\mathrm{por}}$ have been introduced
for mathematical convenience.
The unit outward normal on $\Gamma_{\mathrm{free}}$
emanating away from the free flow region is
denoted by $\widehat{\mathbf{n}}_{\mathrm{free}}$.
Similarly, the unit outward normal on
$\Gamma_{\mathrm{por}}$ emanating away
from the porous region is denoted by
$\widehat{\mathbf{n}}_{\mathrm{por}}$.
Clearly, these normals on the
interface satisfy:
%-------------------------------;
%  Equation: Interface normals  ;
%-------------------------------;
\begin{align}
  \widehat{\mathbf{n}}_{\mathrm{free}}(\mathbf{x})
  + \widehat{\mathbf{n}}_{\mathrm{por}}(\mathbf{x}) 
  = \mathbf{0}
  \quad \forall \mathbf{x} \in
  \Gamma_{\mathrm{int}}
\end{align}
A unit tangent vector on $\Gamma_{\mathrm{int}}$
is denoted by $\widehat{\mathbf{s}}$. 

%================================;
%  Subsection: Free flow region  ;
%--------------------------------;
\subsubsection{Free flow region} 
We denote the region in which free flow
occurs by $\mathcal{K}_{\mathrm{free}}$,
and its whole boundary and external
boundary are, respectively, denoted by
$\partial \mathcal{K}_{\mathrm{free}}$ and
$\partial \mathcal{K}_{\mathrm{free}}^{\mathrm{ext}}$.
We thus have:
%------------------------------------------------;
%  Equation: Whole boundary of free flow region  ;
%------------------------------------------------;
\begin{align}
  \partial \mathcal{K}_{\mathrm{free}} =
  \partial \mathcal{K}_{\mathrm{free}}^{\mathrm{ext}}
  \cup \Gamma_{\mathrm{free}}
  \quad \mathrm{and} \quad
  \partial \mathcal{K}_{\mathrm{free}}^{\mathrm{ext}}
  \cap \Gamma_{\mathrm{free}} = \emptyset
\end{align}
The unit outward normal to the external
boundary $\mathcal{K}_{\mathrm{free}}^{\mathrm{ext}}$
is denoted by
$\widehat{\mathbf{n}}_{\mathrm{free}}^{\mathrm{ext}}$.
We denote the velocity vector field in the
free flow region by $\mathbf{v}_{\mathrm{free}}
(\mathbf{x})$, and the corresponding pressure
field by $p_{\mathrm{free}}(\mathbf{x})$.
Mathematically, $\mathbf{v}_{\mathrm{free}}:\mathcal{K}_{\mathrm{free}} 
\cup \partial \mathcal{K}_{\mathrm{free}}
\rightarrow \mathbb{R}^{nd}$ and $p_{\mathrm{free}}:\mathcal{K}_{\mathrm{free}}
\cup \partial \mathcal{K}_{\mathrm{free}}
\rightarrow \mathbb{R}$.
We denote the specific body force and the
stress tensor in the free flow region by
$\mathbf{b}_{\mathrm{free}}(\mathbf{x})$ and
$\mathbf{T}_{\mathrm{free}}$, respectively.
The external boundary
$\partial \mathcal{K}_{\mathrm{free}}^{\mathrm{ext}}$ 
is divided into two parts: $\Gamma_{\mathrm{free}}^{v}$
and $\Gamma_{\mathrm{free}}^{t}$, such that
%----------------------------;
%  Equation: Well-posedness  ;
%----------------------------;
\begin{align}
  \Gamma^{v}_{\mathrm{free}} \cup \Gamma^{t}_{\mathrm{free}}
  = \partial \mathcal{K}_{\mathrm{free}}^{\mathrm{ext}}
  \quad \mathrm{and} \quad
  \Gamma^{v}_{\mathrm{free}} \cap \Gamma^{t}_{\mathrm{free}} = \emptyset 
\end{align}
$\Gamma^{v}_{\mathrm{free}}$ is the part of the
external boundary of the free flow region
on which velocity boundary condition is
prescribed, and $\Gamma^{t}_{\mathrm{free}}$
is that part of the external boundary of
the free flow region on which traction boundary
condition is prescribed. We thus have:
%------------------------------------------------;
%  Equation: Whole boundary of free flow region  ;
%------------------------------------------------;
\begin{align}
\label{Eqn:ICs_decomposition_of_whole_boundary_Kfree}
  \partial \mathcal{K}_{\mathrm{free}}
  = \partial \mathcal{K}_{\mathrm{free}}^{\mathrm{ext}}
  \cup \Gamma_{\mathrm{free}}
  = \Gamma_{\mathrm{free}}^{v}
  \cup \Gamma_{\mathrm{free}}^{t}
  \cup \Gamma_{\mathrm{free}}
\end{align}
We denote the prescribed velocity vector on
$\Gamma_{\mathrm{free}}^{v}$ by
$\mathbf{v}_{\mathrm{free}}^{\mathrm{p}}(\mathbf{x})$,
and the prescribed traction on $\Gamma_{\mathrm{free}}^{t}$
by $\mathbf{t}_{\mathrm{free}}^{\mathrm{p}}(\mathbf{x})$.

%=============================;
%  Subsection: Porous region  ;
%-----------------------------;
\subsubsection{Porous region}
We denote the porous region by
$\mathcal{K}_{\mathrm{por}}$, and its whole boundary
and external boundary are, respectively, denoted
by $\partial \mathcal{K}_{\mathrm{por}}$ and $\partial
\mathcal{K}_{\mathrm{por}}^{\mathrm{ext}}$. 
Similar to the free flow region, we have
%---------------------------------------------;
%  Equation: Whole boundary of porous region  ;
%---------------------------------------------;
\begin{align}
\label{Eqn:ICs_decomposition_of_whole_boundary_Kpor}
    \partial \mathcal{K}_{\mathrm{por}} =
  \partial \mathcal{K}_{\mathrm{por}}^{\mathrm{ext}}
  \cup \Gamma_{\mathrm{por}}
  \quad \mathrm{and} \quad
  \partial \mathcal{K}_{\mathrm{por}}^{\mathrm{ext}}
  \cap \Gamma_{\mathrm{por}} = \emptyset
\end{align}
The unit outward normal to the external boundary
$\mathcal{K}_{\mathrm{por}}^{\mathrm{ext}}$ is denoted
by $\widehat{\mathbf{n}}_{\mathrm{por}}^{\mathrm{ext}}$.
\emph{The porous solid is assumed to be rigid,
  and its motion can be ignored.\footnote{Some
    terms pertaining to the porous region will
    tacitly involve the velocity of the porous
    solid.
    Three such cases will
    be the virtual velocities
    in the virtual power expended due to the
    interactions
    \eqref{Eqn:ICs_internal_virtual_power_por_interactions}, the interaction term
    $\mathbf{i}_{\mathrm{por}}$ itself and the power
    expended density along the interface $\Psi$.
    A quantity that appears in these cases
    will be the velocity of the fluid in the porous
    region with respect to the velocity of the
    porous solid.
    If the motion of the
    porous solid is taken to be zero, which
    can be done by choosing a specific frame of
    reference, its explicit dependence will
    not be apparent. For example, the interaction
    force under the Darcy model (which assumed
    the porous solid to be rigid) is commonly
    written as 
    \[
    \mathbf{i}_{\mathrm{por}} = \mu \mathbf{K}^{-1}
    \mathbf{v}_{\mathrm{por}}
    \]
    but in fact it needs to be interpreted as
    \[
    \mu \mathbf{K}^{-1}(\mathbf{v}_{\mathrm{por}}
    - \mathbf{v}_{\mathrm{por}}^{\mathrm{(solid)}})
    \]
    (In the above equations, $\mathbf{v}_{\mathrm{por}}^{\mathrm{(solid)}}$
    is the velocity of the porous solid, $\mathbf{K}$ is
    the permeability of the porous region, and $\mu$ is
    the coefficient of viscosity.)
    Noting the dependence on the velocity of the
    porous solid will be particularly important
    when we invoke a change of observer to obtain
    constitutive restrictions, or when we require
    the internal virtual power expended
    to vanish under a superimposed rigid body
    motion on the virtual velocities. In such
    cases, the actual velocity of the porous
    solid and its virtual counterpart will not
    be zero.}}
We denote the porosity by
$\phi_{\mathrm{por}}(\mathbf{x})$.
We denote the discharge velocity and the
pressure of the fluid in the porous region
by $\mathbf{v}_{\mathrm{por}}(\mathbf{x})$ and
$p_{\mathrm{por}}(\mathbf{x})$, respectively.
It is important to note that the discharge
velocity is equal to the true (or seepage)
velocity times the porosity. We denote the
specific body force and the stress of the
fluid in the porous region by
$\mathbf{b}_{\mathrm{por}}(\mathbf{x})$ and
$\mathbf{T}_{\mathrm{por}}$, respectively. 
We denote the interaction term for the fluid
in the porous region by $\mathbf{i}_{\mathrm{por}}$,
which accounts for the momentum supply due to
the coexistence of the other constituent --
the porous solid. As mentioned earlier, the
interaction term should be interpreted in the
context of TIC.
The external boundary
$\partial \mathcal{K}_{\mathrm{por}}^{\mathrm{ext}}$ 
is divided into two parts: $\Gamma_{\mathrm{por}}^{v}$
and $\Gamma_{\mathrm{por}}^{t}$, such that
%----------------------------;
%  Equation: Well-posedness  ;
%----------------------------;
\begin{align}
  \Gamma^{v}_{\mathrm{por}} \cup \Gamma^{t}_{\mathrm{por}}
  = \partial \mathcal{K}_{\mathrm{por}}^{\mathrm{ext}}
  \quad \mathrm{and} \quad
  \Gamma^{v}_{\mathrm{por}} \cap
  \Gamma^{t}_{\mathrm{por}} = \emptyset 
\end{align}
$\Gamma^{v}_{\mathrm{por}}$ is the part of the
external boundary of the porous region
on which velocity boundary condition is
prescribed, and $\Gamma^{t}_{\mathrm{por}}$
is that part of the external boundary of
the porous region on which traction boundary
condition is prescribed. We thus have:
%---------------------------------------------;
%  Equation: Whole boundary of porous region  ;
%---------------------------------------------;
\begin{align}
  \partial \mathcal{K}_{\mathrm{por}}
  = \partial \mathcal{K}_{\mathrm{por}}^{\mathrm{ext}}
  \cup \Gamma_{\mathrm{por}}
  = \Gamma_{\mathrm{por}}^{v}
  \cup \Gamma_{\mathrm{por}}^{t}
  \cup \Gamma_{\mathrm{por}}
\end{align}
We denote the prescribed velocity on
$\Gamma_{\mathrm{por}}^{v}$ by
$\mathbf{v}_{\mathrm{por}}^{\mathrm{p}}(\mathbf{x})$\footnote{Under
  Darcy equations, only the normal component of the velocity
  vector field can be prescribed on the boundary. In such cases,
  the velocity boundary condition will be of the form: 
  $\mathbf{v}_{\mathrm{por}}(\mathbf{x}) \cdot
  \widehat{\mathbf{n}}_{\mathrm{por}}(\mathbf{x}) =
  v_{\mathrm{por}}^{\mathrm{p}}(\mathbf{x})$ on 
  $\Gamma_{\mathrm{por}}^{v}$. On the other
  hand, under mathematical models like the Darcy-Brinkman model, the whole
  velocity vector field can be prescribed on the boundary.
  The mathematical reason is that the Darcy equations
  contain at most first-order spatial derivative of the
  velocity field. On the
  other hand, the Darcy-Brinkman model gives rise to
  governing equations which contain a second-order
  spatial derivative of the velocity field.
  \label{ICs:Different_vel_BCs}} and the prescribed
traction on $\Gamma_{\mathrm{por}}^{t}$
by $\mathbf{t}_{\mathrm{por}}^{\mathrm{p}}(\mathbf{x})$.

%===================================;
%  Subsubsection: Fluid properties  ;
%-----------------------------------;
\subsubsection{Fluid properties}
The dynamic coefficient of viscosity of the
fluid is denoted by $\mu$. The true density
of the fluid in the free flow region is denoted
by $\gamma_{\mathrm{free}}$, and the corresponding
quantity of the fluid in the porous region
is denoted by $\gamma_{\mathrm{por}}$. Note that
the bulk density of the fluid in porous media
is equal to the true density of the fluid times
the porosity of the porous medium. The interface
conditions are derived under the realistic case
of $\gamma_{\mathrm{free}} = \gamma_{\mathrm{por}}
= \gamma$. 

%==============================================================;
%  Subsubsection: Kinematically admissible and virtual fields  ;
%--------------------------------------------------------------;
\subsubsection{Kinematically admissible and virtual fields}
\label{Subsec:ICs_kinematic_virtual}
We introduce the following space for the pairs 
of vector fields defined on free flow and porous 
regions:
%--------------------------------------------;
%  Equation: General space of vector fields  ;
%--------------------------------------------;
\begin{align}
    \label{Eqn:ICs_general_space_of_vector_fields}
    \mathcal{W} := 
    \left\{(\mathbf{w}_{\mathrm{free}}(\mathbf{x}),
    \mathbf{w}_{\mathrm{por}}(\mathbf{x})) 
    \; \vert \; 
    \mathbf{w}_{\mathrm{free}} : 
    \mathcal{K}_{\mathrm{free}} \cup \partial \mathcal{K}_{\mathrm{free}} 
    \rightarrow \mathbb{R}^{nd}, 
    \mathbf{w}_{\mathrm{por}} : 
    \mathcal{K}_{\mathrm{por}} \cup \partial \mathcal{K}_{\mathrm{por}} 
    \rightarrow \mathbb{R}^{nd}
    \right\}
\end{align}
For a given pair of vector fields 
$(\mathbf{w}_{\mathrm{free}},\mathbf{w}_{\mathrm{por}}) 
\in \mathcal{W}$, we introduce the following normal components: 
%--------------------------------------------;
%  Equation: Notation for normal components  ;
%--------------------------------------------;
\begin{subequations}
  \begin{align}
    w_{\mathrm{free}}^{(n)}(\mathbf{x}) 
    &:= \mathbf{w}_{\mathrm{free}}(\mathbf{x}) 
    \cdot \widehat{\mathbf{n}}_{\mathrm{free}}(\mathbf{x}) \\
    w_{\mathrm{por}}^{(n)}(\mathbf{x}) 
    &:= \mathbf{w}_{\mathrm{por}}(\mathbf{x}) 
    \cdot \widehat{\mathbf{n}}_{\mathrm{por}}(\mathbf{x}) 
  \end{align}
\end{subequations}
and the following decomposition: 
%---------------------------------------;
%  Equation: Decomposition of velocity  ;
%---------------------------------------;
\begin{subequations}
    \label{Eqn:ICs_tangential_velocities}
  \begin{align}
  \label{Eqn:ICs_tangential_velocities_vfree}
    \mathbf{w}_{\mathrm{free}}(\mathbf{x}) 
    &= w_{\mathrm{free}}^{(n)}(\mathbf{x}) 
    \widehat{\mathbf{n}}_{\mathrm{free}}(\mathbf{x}) 
    + \overset{*}{\mathbf{w}}_{\mathrm{free}}(\mathbf{x}) \\
    \label{Eqn:ICs_tangential_velocities_vpor}
    \mathbf{w}_{\mathrm{por}}(\mathbf{x}) 
    &= w_{\mathrm{por}}^{(n)} 
    (\mathbf{x}) \widehat{\mathbf{n}}_{\mathrm{por}}(\mathbf{x}) 
    + \overset{*}{\mathbf{w}}_{\mathrm{por}}(\mathbf{x})
  \end{align}
\end{subequations}
where $\overset{*}{\mathbf{w}}_{\mathrm{free}}(\mathbf{x})$ 
and $\overset{*}{\mathbf{w}}_{\mathrm{por}}(\mathbf{x})$ 
denote the corresponding tangential components of the 
vector fields. 

We refer to a pair of vector fields
$(\mathbf{w}_{\mathrm{free}}(\mathbf{x}),
\mathbf{w}_{\mathrm{por}}(\mathbf{x})) 
\in \mathcal{W}$ to be \emph{kinematically 
  admissible} if the following properties are 
satisfied:
%----------------------------------------------;
%  Enumerate: Kinematically admissible fields  ;
%----------------------------------------------;
\begin{enumerate}[(i)]
\item $\mathrm{div}[\mathbf{w}_{\mathrm{free}}]=0 \;
  \mathrm{in} \; \mathcal{K}_{\mathrm{free}}$ and
  $\mathrm{div}[\mathbf{w}_{\mathrm{por}}]
  = 0 \; \mathrm{in} \; \mathcal{K}_{\mathrm{por}}$, 
\item $w_{\mathrm{free}}^{(n)}(\mathbf{x}) + 
  w_{\mathrm{por}}^{(n)}(\mathbf{x}) = 0$ on
  the interface $\Gamma_{\mathrm{int}}$, and 
\item $\mathbf{w}_{\mathrm{free}}(\mathbf{x})$
  and $\mathbf{w}_{\mathrm{por}}(\mathbf{x})$
  satisfy the velocity boundary conditions on the
  external boundary (i.e., on $\Gamma_{\mathrm{free}}^{v}$ and
  $\Gamma_{\mathrm{por}}^{v}$, respectively).
\end{enumerate}
We denote the set of all kinematically
admissible pairs of vector fields by
$\mathcal{V}$. Certainly, the exact
velocity fields are kinematically
admissible; that is $(\mathbf{v}_{\mathrm{free}}
(\mathbf{x}),\mathbf{v}_{\mathrm{por}}(\mathbf{x}))
\in \mathcal{V}$. 

We refer to a pair of vector fields 
$(\mathbf{w}_{\mathrm{free}}(\mathbf{x}),
\mathbf{w}_{\mathrm{por}}(\mathbf{x})) 
\in \mathcal{W}$ to be a pair of 
\emph{virtual vector fields} if the 
first two properties under kinematical
admissibility are met, and
$\mathbf{w}_{\mathrm{free}}(\mathbf{x})$ and
$\mathbf{w}_{\mathrm{por}}(\mathbf{x})$ vanish
on $\Gamma_{\mathrm{free}}^{v}$ and
$\Gamma_{\mathrm{por}}^{v}$, respectively. 
We denote the set of all pairs of virtual 
vector fields by $\widetilde{\mathcal{V}}$.

%===================================================;
%  Subsubsection: Fields under a rigid body motion  ;
%---------------------------------------------------;
\subsubsection{Fields under a rigid body motion} 
Consider a \emph{superimposed} rigid
body motion of the entire domain\footnote{\label{FN:ICs_RBM}One
  should not confuse the expression
  \eqref{Eqn:ICs_RBM_expression} with
  that of a Euclidean transformation
  between two frames of reference (i.e.,
  two observers). We will deal the latter
  aspect in a subsequent section 
  when we discuss the principle of material
  frame-indifference for constitutive
  relations.
  For the current discussion, it is
  important to note that a single
  observer looks at two motions
  ($\mathbf{x}^{'}$ and $\mathbf{x}$)
  which differ by a rigid body motion.}:
%-------------------------------;
%  Equation: Rigid body motion  ;
%-------------------------------;
\begin{align}
  \label{Eqn:ICs_RBM_expression}
  \mathbf{x}^{'}(t) \leftarrow \mathbf{Q}(t)
  \mathbf{x} + \mathbf{c}(t) 
  \quad \forall \mathbf{x} \in \Omega
\end{align}
where $t$ denotes the time,
$\mathbf{c}(t)$ is a translational
vector, and $\mathbf{Q}(t) \in \mathrm{SO(3)}$
is a rotation at each instance of time\footnote{$\mathrm{SO(3)}$ is a
    group of all rotations about the origin
    of $\mathbb{R}^{3}$ -- the three-dimensional
    Euclidean space -- under the operation of
    composition; e.g., see \citep{marsden2013introduction}.}. 
The subspace
$\mathcal{W}_{\mathrm{rigid}} \subseteq
\mathcal{W}$ that is spanned by the
vector fields generated by such
a rigid body motion at a given instance
of time $t$ takes the following
form: 
\begin{align}
  \mathcal{W}_{\mathrm{rigid}} &:=
  \left\{(\mathbf{w}_{\mathrm{free}}(\mathrm{x}),
  \mathbf{w}_{\mathrm{por}}(\mathrm{x})) \in \mathcal{W} 
  \; \vert \;
  \mathbf{w}_{\mathrm{free}}(\mathbf{x}) =
  \mathbf{v}(\mathbf{x},t)\big\vert_{\mathbf{x}
    \in \mathcal{K}_{\mathrm{free}} \cup \partial
    \mathcal{K}_{\mathrm{free}}}, \right. \notag \\
  &\left. \qquad \qquad \qquad 
  \mathbf{w}_{\mathrm{por}}(\mathbf{x}) =
  \mathbf{v}(\mathbf{x},t)\big\vert_{\mathbf{x}
    \in \mathcal{K}_{\mathrm{por}} \cup \partial
    \mathcal{K}_{\mathrm{por}}} 
  \; \mbox{where} \;
  \mathbf{v}(\mathbf{x},t) = \dot{\mathbf{Q}}(t)
  (\mathbf{x} - \mathbf{x}_0) + \mathbf{v}_0
  \right\}
\end{align}
where $\dot{\mathbf{Q}}(t)$ denotes the
time derivative of $\mathbf{Q}(t)$.
It is important to note that, at each instance
of time, the vector fields
$(\mathbf{w}_{\mathrm{free}}(\mathbf{x}),
\mathbf{w}_{\mathrm{por}}(\mathbf{x})) \in
\mathcal{W}_{\mathrm{rigid}}$ satisfy: 
\begin{align}
  \mathrm{grad}[\mathbf{w}_{\mathrm{free}}]
  = \dot{\mathbf{Q}}(t) \mathbf{Q}^{\mathrm{T}}(t) 
  \in \mathrm{skew}[\mathcal{K}_{\mathrm{free}}] 
  \quad \mathrm{and} \quad 
  \mathrm{grad}[\mathbf{w}_{\mathrm{por}}]
  = \dot{\mathbf{Q}}(t) \mathbf{Q}^{\mathrm{T}}(t) 
  \in \mathrm{skew}[\mathcal{K}_{\mathrm{por}}]
\end{align}
where $\mathrm{skew}[\cdot]$ denotes the
space of skew-symmetric tensor fields on
the indicated spatial region. 

%=================================================;
%  Subsubsection: Other notation for convenience  ;
%-------------------------------------------------;
\subsubsection{Other notation for convenience}
We occasionally use the following notation:
\begin{align}
  \mathbf{L}_{\mathrm{free}} = \mathrm{grad}[\mathbf{v}_{\mathrm{free}}],
  \mathbf{L}_{\mathrm{por}} = \mathrm{grad}[\mathbf{v}_{\mathrm{por}}],
  \mathbf{D}_{\mathrm{free}} = \frac{1}{2}\left(\mathbf{L}_{\mathrm{free}}
  + \mathbf{L}_{\mathrm{free}}^{\mathrm{T}} \right)
  \; \mathrm{and} \;
  \mathbf{D}_{\mathrm{por}} = \frac{1}{2}\left(\mathbf{L}_{\mathrm{por}}
  + \mathbf{L}_{\mathrm{por}}^{\mathrm{T}} \right)
\end{align}

%===================================================;
%  Subsection: Proposed principle of virtual power  ;
%===================================================;
\subsection{Proposed principle of virtual power}
The mathematical statement of the proposed 
principle of virtual power for flows in coupled 
free-porous media, which will be in the form 
of balance of virtual power, can be written as 
follows:

%-------------------------------------;
%  Tcolorbox: Proposed virtual power  ;
%-------------------------------------;
\begin{tcolorbox}[breakable]
  Find $(\mathbf{v}_{\mathrm{free}}(\mathbf{x}),
  \mathbf{v}_{\mathrm{por}}(\mathbf{x}))
  \in \mathcal{V}$
  such that the following two properties are met: 
  \begin{alignat}{2}
    \label{Eqn:ICs_proposed_PVP_P1}
    \mathrm{(P1)} & \qquad \qquad \qquad \qquad
    \mathscr{P}^{\mathrm{(internal)}} =
    \mathscr{P}^{\mathrm{(external)}}
    &&\quad \forall (\mathbf{w}_{\mathrm{free}}(\mathbf{x}),
    \mathbf{w}_{\mathrm{por}}(\mathbf{x}))
    \in \widetilde{\mathcal{V}}
    \\
    \label{Eqn:ICs_proposed_PVP_P2}
    \mathrm{(P2)} & \qquad \qquad \qquad \qquad 
    \mathscr{P}^{\mathrm{(internal)}} 
    = 0
    &&\quad \forall(\mathbf{w}_{\mathrm{free}}(\mathbf{x}),
    \mathbf{w}_{\mathrm{por}}(\mathbf{x})) \in
    \mathcal{W}_{\mathrm{rigid}}
  \end{alignat}
  \noindent where the internal virtual power expended
  (i.e., virtual stress power) in the free
  flow region is given by
  %----------------------------------------------------;
  %  Equation: Internal virtual power under free flow  ;
  %----------------------------------------------------;
  \begin{align}
    \label{Eqn:ICs_internal_virtual_power_free}
    \mathscr{P}^{\mathrm{(internal)}}_{\mathrm{free}} &:=
    \int_{\mathcal{K}_{\mathrm{free}}} \mathbf{T}_{\mathrm{free}}
    \cdot \mathrm{grad}[\mathbf{w}_{\mathrm{free}}] \; \mathrm{d}
    \Omega
  \end{align}
  The internal virtual power expended in
  the porous region is written as follows: 
  %---------------------------------------------------------;
  %  Equation: Internal virtual power in the porous region  ;
  %---------------------------------------------------------;
  \begin{align}
    \label{Eqn:ICs_internal_virtual_power_por}
    \mathscr{P}^{\mathrm{(internal)}}_{\mathrm{por}} &:= 
    \mathscr{P}^{\mathrm{(internal)}}_{\mathrm{por,\;stress}} +
    \mathscr{P}^{\mathrm{(internal)}}_{\mathrm{por,\;interactions}} 
  \end{align}
  where virtual stress power in the porous
  region is defined as follows:
  %------------------------------------------------------;
  %  Equation: Virtual power due to porous stress power  ;
  %------------------------------------------------------;
  \begin{align}
    \label{Eqn:ICs_internal_virtual_power_por_stress}
    \mathscr{P}^{\mathrm{(internal)}}_{\mathrm{por,\;stress}} &:=
    \int_{\mathcal{K}_{\mathrm{por}}} \mathbf{T}_{\mathrm{por}}
    \cdot \mathrm{grad}[\mathbf{w}_{\mathrm{por}}] \; \mathrm{d}
    \Omega
  \end{align}
  and the internal virtual power expended due
  to interactions between the constituents in
  the porous region is written as follows:
  %------------------------------------------------------;
  %  Equation: Virtual power due to porous interactions  ;
  %------------------------------------------------------;
  \begin{align}
    \label{Eqn:ICs_internal_virtual_power_por_interactions}
    \mathscr{P}^{\mathrm{(internal)}}_{\mathrm{por,\;interactions}}
    &:= \int_{\mathcal{K}_{\mathrm{por}}} \mathbf{i}_{\mathrm{por}}
    \cdot \left(\mathbf{w}_{\mathrm{por}} -
    \cancelto{0}{\mathbf{w}_{\mathrm{por}}^{\mathrm{(solid)}}}\right)
        \; \mathrm{d} \Omega
    = \int_{\mathcal{K}_{\mathrm{por}}} \mathbf{i}_{\mathrm{por}}
    \cdot \mathbf{w}_{\mathrm{por}} \; \mathrm{d} \Omega
  \end{align}
  In the above equation, $\mathbf{w}_{\mathrm{por}}^{\mathrm{(solid)}}$
  denotes the vector field associated with the porous solid. Since
  we assumed the porous solid to be rigid and neglected its motion,
  this term becomes zero. However, when we invoke vanishing of the
  internal virtual power under a rigid body motion of the entire
  domain, it becomes important to acknowledge the presence of this
  term, as it will not be zero in that situation. 
  The internal virtual power expended at the
  interface is written as follows:
  %-------------------------------------------------;
  %  Equation: Internal virtual power at interface  ;
  %-------------------------------------------------;
  \begin{align}
    \label{Eqn:ICs_internal_virtual_power_interface}
    \mathscr{P}^{\mathrm{(internal)}}_{\mathrm{int}}
    := \int_{\Gamma_{\mathrm{int}}} \delta \Psi
    \; \mathrm{d} \Gamma
  \end{align}
  where $\delta \Psi$ denotes the virtual
  power expended density at the interface
  and depends on both the true velocity
  fields and their virtual counterparts.
  The total internal virtual power expended
  takes the following form: 
  %------------------------------------------;
  %  Equation: Total internal virtual power  ; 
  %------------------------------------------;
  \begin{align}
    \label{Eqn:ICs_total_internal_virtual_power}
    \mathscr{P}^{\mathrm{(internal)}} :=
    \mathscr{P}^{\mathrm{(internal)}}_{\mathrm{free}} +
    \mathscr{P}^{\mathrm{(internal)}}_{\mathrm{por}} +
    \mathscr{P}^{\mathrm{(internal)}}_{\mathrm{int}}
  \end{align}
  The total external virtual power expended
  takes the following form: 
  %------------------------------------------;
  %  Equation: Total external virtual power  ; 
  %------------------------------------------;
  \begin{align}
    \label{Eqn:ICs_external_virtual_power}
    \mathscr{P}^{\mathrm{(external)}}
    &:= \underbrace{\int_{\mathcal{K}_{\mathrm{free}}}
      \gamma \mathbf{b}_{\mathrm{free}} \cdot 
      \mathbf{w}_{\mathrm{free}}
      \; \mathrm{d} \Omega
      + \int_{\Gamma_{\mathrm{free}}^{t}}
      \mathbf{t}_{\mathrm{free}}^{\mathrm{p}}
      \cdot \mathbf{w}_{\mathrm{free}}
      \; \mathrm{d} \Gamma}_{\substack{
        \mbox{external virtual power expended} \\
        \mbox{on the free flow region}}}
    \nonumber \\
    &+ \underbrace{\int_{\mathcal{K}_{\mathrm{por}}}
      \gamma \phi_{\mathrm{por}} \mathbf{b}_{\mathrm{por}}
      \cdot \mathbf{w}_{\mathrm{por}} \; \mathrm{d} \Omega
      + \int_{\Gamma_{\mathrm{por}}^{t}} \mathbf{t}_{\mathrm{por}}^{\mathrm{p}}
      \cdot \mathbf{w}_{\mathrm{por}}
      \; \mathrm{d} \Gamma}_{\substack{
        \mbox{external virtual power expended} \\
        \mbox{on the porous region}}} 
  \end{align}
\end{tcolorbox}

We will show that an appropriate set
of interface conditions can be derived
by prescribing a functional form for
$\delta \Psi$, and this prescription
will be a constitutive specification.
We place the following restrictions on
$\delta \Psi$, and these restrictions
are based on either invariance
requirements, physical properties
or convenience.

\begin{enumerate}[(i)]
  %============================;
  %  Item: Exact differential  ;
  %----------------------------;
\item \textit{Exact differential.}~
  In equation \eqref{Eqn:ICs_internal_virtual_power_interface},
  $\delta \Psi$ need not be an exact
  differential. However, for \emph{convenience}
  we assume $\delta \Psi$ to be an exact
  differential.
  This implies that there exists a functional
  $\Psi$, which will be referred to as the
  power expended at the interface, such that
  $\delta \Psi$ is a G\^ateaux variation of
  $\Psi$. Mathematically, if $\delta \Psi$
  depends on a set of variables, which is
  collectively denoted by $\boldsymbol{\chi}$,
  and a set of the corresponding virtual
  variables, $\delta \boldsymbol{\chi}$, then
  %-------------------------;
  %  Equation: \delta \Psi  ;
  %-------------------------;
  \begin{align}
    \delta \Psi[\boldsymbol{\chi};\delta
    \boldsymbol{\chi}] =
    \left[\frac{d}{d \epsilon} \Psi[\boldsymbol{\chi}
      + \epsilon \delta \boldsymbol{\chi}] \right]_{\epsilon = 0}
    = \frac{\partial \Psi}{\partial \boldsymbol{\chi}}
    \cdot \delta \boldsymbol{\chi}
  \end{align}
  In the case of an exact differential, the
  Vainberg's theorem \citep{Vainberg,Hjelmstad}
  implies that
  \begin{align}
    \Psi[\boldsymbol{\chi}] = \int_{0}^{1} \delta
    \Psi[\tau \boldsymbol{\chi};\boldsymbol{\chi}]d\tau 
  \end{align}
  where $\tau$ is a dummy variable
  introduced for integration.
  %------------------------------------;
  %  Item: Positive semi-definiteness  ;
  %------------------------------------;
\item \textit{Positive semi-definiteness.}
  The total power expended at the interface 
  should be \emph{physically} non-negative.
  This can be ensured by assuming $\Psi$
  to be a positive semi-definite functional.
  Mathematically, 
  \begin{align}
    \label{Eqn:positive_definiteness_of_Psi}
    \Psi[\boldsymbol{\chi}] \geq 0 
    \quad \forall \boldsymbol{\chi}
  \end{align}
  %============================;
  %  Item: Dependence of \Psi  ;
  %----------------------------;
\item \textit{Dependence of $\Psi$ on velocities.} We
  take the set of variables for the functional
  dependence of $\Psi$ as follows:
  \begin{align}
    \boldsymbol{\chi} = \{\overset{*}{\mathbf{v}}_{\mathrm{free}}(\mathbf{x}),
    \overset{*}{\mathbf{v}}_{\mathrm{por}}(\mathbf{x}),v_n(\mathbf{x})\}
  \end{align}
  where
  \begin{align}
  \label{Eqn:ICs_notation_for_vn}
    v_{n}(\mathbf{x}) := v^{(n)}_{\mathrm{free}}(\mathbf{x})
  \end{align}
  Recall that the tangential velocities
  have been defined in equation
  \eqref{Eqn:ICs_tangential_velocities}. 
  Since the true fluid densities in the porous
  and free flow regions are assumed to be the
  same, the balance of mass across the interface
  implies that 
  \begin{align}
    v_{n}(\mathbf{x}) = -v^{(n)}_{\mathrm{por}}(\mathbf{x})
  \end{align}
  The chosen functional dependence will imply that 
\begin{align}
  \delta \Psi =
  \frac{\partial \Psi}{\partial \overset{*}{\mathbf{v}}_{\mathrm{free}}}
  \cdot \delta \overset{*}{\mathbf{v}}_{\mathrm{free}}
  + \frac{\partial \Psi}{\partial \overset{*}{\mathbf{v}}_{\mathrm{por}}}
  \cdot \delta \overset{*}{\mathbf{v}}_{\mathrm{por}}
  + \frac{\partial \Psi}{\partial v_n} \cdot \delta v_n
\end{align}
Noting that $\delta \mathbf{v}_{\mathrm{free}}$ and
$\delta \mathbf{v}_{\mathrm{por}}$ are relative velocities
with respect to the rigid porous solid, they vanish
under a rigid body motion of the entire domain.
Hence, $\delta \Psi$ vanishes under a rigid body
motion of the virtual velocities. This point
is important to satisfy the statement (P2)
under the proposed principle of virtual power. 
  %====================;
  %  Item: Invariance  ;
  %--------------------;
\item \textit{Invariance.}
  We require the constitutive relations emanating
  from the functional $\Psi$ to satisfy the principle
  of material frame-indifference. Following
  \citep{leigh1968nonlinear,svendsen1999frame,bertram2001material},
  this amounts to enforcing form invariance on
  the functional and invariance under a 
  Euclidean transformation between frames
  (i.e., observers).
  Before proceeding further, we will first recall
  that $\mathbf{v}_{\mathrm{free}}(\mathbf{x})$
  and $\mathbf{v}_{\mathrm{por}}(\mathbf{x})$ are
  relative velocities with respect to the rigid
  porous solid, which is assumed to be at rest.
  It is important to realize that $\mathbf{v}_{\mathrm{free}}$
  and $\mathbf{v}_{\mathrm{por}}$ are relative velocities
  between two constituents at the same
  point in the space and they are not
  relative velocities (of the same
  constituent) between two different points in the
  space.
  Such a distinction is germane to TIC and
  is paramount to our discussion, as the
  former quantities are
  invariant under a Euclidean transformation
  between frames of reference, while the later
  ones are not. In fact, a relative velocity
  between two points in the space is not an
  invariant even under a Galilean transformation
  between frames.
  
  Now consider two frames of reference,
  $(\mathbf{x}^{'},t^{'})$ and $(\mathbf{x},t)$,
  which differ by a Euclidean transformation: 
  %-------------------------------------;
  %  Equation: Galilean transformation  ;
  %-------------------------------------;
  \begin{align}
    &\mathbf{x}^{'} \leftarrow \mathbf{Q}(t) \mathbf{x} 
    + \mathbf{c}(t) 
    \quad \mathrm{and} \quad 
    t^{'} \leftarrow t + t_0 
  \end{align}
  where $\mathbf{c}(t)$ is a translation vector,
  and $\mathbf{Q}(t) \in \mathrm{SO(3)}$ is
  a rotation for each $t$.\footnote{Also
  see footnote \ref{FN:ICs_RBM}.}  
  Under this Euclidean transformation, the
  tangential and the normal components of
  these (relative) velocity fields satisfy:
  \begin{align}
    \label{Eqn:ICs_relative_velocity_components}
    {\mathop{\mathbf{v}}^{*}}_{\mathrm{free}}^{'}
    = \mathbf{Q}(t) {\mathop{\mathbf{v}}^{*}}_{\mathrm{free}}, \;
    {\mathop{\mathbf{v}}^{*}}_{\mathrm{por}}^{'}
    = \mathbf{Q}(t) {\mathop{\mathbf{v}}^{*}}_{\mathrm{por}}
    \; \mathrm{and} \;
    v_n^{'} = v_n 
  \end{align}
  where the quantities with a prime are
  under $\mathbf{x}^{'}$ frame of reference.
  The above expressions
  \eqref{Eqn:ICs_relative_velocity_components}
  and the form invariance of the function
  together imply that 
  \begin{align}
    \Psi^{'}[{\mathop{\mathbf{v}}^{*}}_{\mathrm{free}}^{'},
      {\mathop{\mathbf{v}}^{*}}_{\mathrm{por}}^{'},v_n^{'}]
    = \Psi[\mathbf{Q}(t) {\mathop{\mathbf{v}}^{*}}_{\mathrm{free}},
      \mathbf{Q}(t) {\mathop{\mathbf{v}}^{*}}_{\mathrm{por}},v_n]
    = \Psi[{\mathop{\mathbf{v}}^{*}}_{\mathrm{free}},
      {\mathop{\mathbf{v}}^{*}}_{\mathrm{por}},v_n]
    \quad \forall \mathbf{Q}(t) \in \mathrm{SO(3)}
  \end{align}
  which implies that $\Psi$ is an isotropic
  functional of its arguments. 
  From the representation theory, we further
  assert that $\Psi$ can depend only on the
  following individual and joint invariants
  \citep{smith1971isotropic}:
  \[
    {\mathop{\mathbf{v}}^{*}}_{\mathrm{free}} \cdot
    {\mathop{\mathbf{v}}^{*}}_{\mathrm{free}}, \;
    {\mathop{\mathbf{v}}^{*}}_{\mathrm{por}} \cdot
    {\mathop{\mathbf{v}}^{*}}_{\mathrm{por}}, \;
    {\mathop{\mathbf{v}}^{*}}_{\mathrm{free}} \cdot
    {\mathop{\mathbf{v}}^{*}}_{\mathrm{por}}
    \; \mathrm{and} \;
    v_n
  \]
\end{enumerate}

%*****************************************************;
%                                                     ;
%  NAME                                               ;
%    S4_ICs_Derivation.tex                            ;
%                                                     ;
%  WRITTEN BY                                         ;
%      Kalyana Babu Nakshatrala  	              ;
%      Mohammad S. Joshaghani                         ;
%                                                     ;
%*****************************************************;
\section{DERIVATION OF INTERFACE CONDITIONS AND FIELD EQUATIONS}
\label{Sec:S4_ICs_Derivation}
Under a rigid body motion,
$\mathscr{P}^{\mathrm{(internal)}}_{\mathrm{int}} = 0$
and $\mathscr{P}^{\mathrm{(internal)}}_{\mathrm{por,\;interactions}}
= 0$, as they (linearly) depend on the relative virtual velocities.
Recall that under a rigid body motion
$\mathrm{grad}[\mathbf{w}_{\mathrm{free}}]$
and $\mathrm{grad}[\mathbf{w}_{\mathrm{por}}]$
are skew symmetric tensor fields. Thus, the main consequence
of the statement (P2) is the symmetry of the Cauchy
stresses in both the regions, which is equivalent
to the balance of angular momentum. That is,
%-----------------------------------------;
%  Equation: Symmetry of Cauchy stresses  ;
%-----------------------------------------;
\begin{align}
  \label{Eqn:ICs_symmetry_Cauchy_stresses}
  \mathbf{T}_{\mathrm{free}}(\mathbf{x}) 
  = \mathbf{T}^{\mathrm{T}}_{\mathrm{free}}(\mathbf{x})
  \quad \forall \mathbf{x} \in \mathcal{K}_{\mathrm{free}}
  \quad \mathrm{and} \quad
  \mathbf{T}_{\mathrm{por}}(\mathbf{x}) 
  = \mathbf{T}^{\mathrm{T}}_{\mathrm{por}}(\mathbf{x}) 
  \quad \forall \mathbf{x} \in \mathcal{K}_{\mathrm{por}}
\end{align}
%=============================================;
%  Subsection: Handling internal constraints  ;
%=============================================;
\subsection{Handling internal constraints}
We extend the approach put-forth by
\citet{carlson2004geometrically} for
handling internal constraints to the
case of flows in coupled free-porous media. 
Consider a constraint manifold for the
motion in $\mathcal{K}_{\mathrm{free}}$:
%---------------------------------;
%  Equation: Constraint manifold  ;
%---------------------------------;
\begin{align}
  \mathscr{C}_{\mathrm{free}} := \left\{
  \mathbf{L}_{\mathrm{free}} \; \big| \;
  \Upsilon_{\mathrm{free}}(\mathbf{L}_{\mathrm{free}}) 
  = 0 \; \mathrm{in} \; \mathcal{K}_{\mathrm{free}}
  \right\}
\end{align}
where the constraint function is:
%-------------------------------------------;
%  Equation: Free flow constraint function  ;
%-------------------------------------------;
\begin{align}
  \Upsilon_{\mathrm{free}} :
  \mathrm{Lin}(\mathcal{K}_{\mathrm{free}})
  \rightarrow \mathbb{R}
\end{align}
Herein, we have employed the standard notation
for $\mathrm{Lin}(\mathcal{K})$ to denote the
linear space of all (second-order) tensors
defined over $\mathcal{K}$.
The normal space to $\mathscr{C}_{\mathrm{free}}$ at
$\mathbf{L}_{\mathrm{free}} \in \mathrm{Lin}(\mathcal{K}_{\mathrm{free}})$
can be written as follows\footnote{$\mathrm{Grad}
  [\Upsilon_{\mathrm{free}}(\mathbf{L}_{\mathrm{free}})]$ means
  gradient of $\Upsilon_{\mathrm{free}}$ with respect
  to its argument $\mathbf{L}_{\mathrm{free}}$.}:
%--------------------------;
%  Equation: Normal space  ;
%--------------------------;
\begin{align}
  \mathrm{Norm}(\mathscr{C}_{\mathrm{free}}) := \mathrm{Lsp}
  \left\{\mathrm{Grad}[
    \Upsilon_{\mathrm{free}}(\mathbf{L}_{\mathrm{free}})]\right\} 
\end{align}
where $\mathrm{Lsp}\{\cdot\}$ denotes the
linear space spanned by its argument.
It is easy to check that $\mathrm{Norm}
(\mathscr{C}_{\mathrm{free}})$ is a subspace
of $\mathrm{Lin}(\mathcal{K}_{\mathrm{free}})$. 
Then the orthogonal complement of the normal
space (which is commonly referred to as the
tangent space) at $\mathbf{L}_{\mathrm{free}}$ 
can be defined as follows: 
%---------------------------;
%  Equation: Tangent space  ;
%---------------------------;
\begin{align}
  \mathrm{Tan}(\mathscr{C}_{\mathrm{free}}) =
  \left(\mathrm{Norm}(\mathscr{C}_{\mathrm{free}})
  \right)^{\perp}
  := \{\mathbf{A}_{\mathrm{free}} \in
  \mathrm{Lin}(\mathcal{K}_{\mathrm{free}})
  \; \big| \; \mathbf{A}_{\mathrm{free}}
  \cdot \mathbf{B}_{\mathrm{free}} = 0
  \; \forall \mathbf{B}_{\mathrm{free}}
  \in \mathrm{Norm}(\mathscr{C}_{\mathrm{free}})\} 
\end{align}
The projection theorem implies the following
direct sum decomposition:
%--------------------------------;
%  Equation: Projection theorem  ;
%--------------------------------;
\begin{align}
  \mathrm{Lin}(\mathcal{K}_{\mathrm{free}}) =
  \mathrm{Norm}(\mathscr{C}_{\mathrm{free}}) \oplus
  \mathrm{Tan}(\mathscr{C}_{\mathrm{free}}) 
\end{align}
This implies that for each $\mathbf{A}_{\mathrm{free}}
\in \mathrm{Lin}(\mathcal{K}_{\mathrm{free}})$ we have
\begin{align}
  \mathbf{A}_{\mathrm{free}} =  \mathbf{A}_{\mathrm{free}}^{\perp}
  + \mathbf{A}_{\mathrm{free}}^{\parallel}
\end{align}
where $\mathbf{A}_{\mathrm{free}}^{\perp} \in
\mathrm{Norm}(\mathscr{C}_{\mathrm{free}})$
and $\mathbf{A}_{\mathrm{free}}^{\parallel} \in
\mathrm{Tan}(\mathscr{C}_{\mathrm{free}})$
are, respectively, the active and
reactive components of $\mathbf{A}_{\mathrm{free}}$.
Similarly, one can define the constraint
manifold $\mathscr{C}_{\mathrm{por}}$ in
terms of $\mathbf{L}_{\mathrm{por}}$ for
the region $\mathcal{K}_{\mathrm{por}}$
and the corresponding
$\mathrm{Norm}(\mathscr{C}_{\mathrm{por}})$
and $\mathrm{Tan}(\mathscr{C}_{\mathrm{por}})$
subspaces of $\mathrm{Lin}(\mathcal{K}_{\mathrm{por}})$. 

Specifically in our case, the
constraint functions are: 
\begin{align}
  \Upsilon_{\mathrm{free}}(\mathbf{L}_{\mathrm{free}})
  = \mathrm{tr}[\mathbf{L}_{\mathrm{free}}] = 0
  \quad \mathrm{and} \quad
  \Upsilon_{\mathrm{por}}(\mathbf{L}_{\mathrm{por}})
  = \mathrm{tr}[\mathbf{L}_{\mathrm{por}}] = 0
\end{align}
where $\mathrm{tr}[\cdot]$ denotes the standard
trace of second-order tensors. The corresponding
normal spaces take the following form:
\begin{align}
  \mathrm{Norm}(\mathscr{C}_{\mathrm{free}})
  = \mathrm{Lsp}\{\mathbf{I}\}
  \quad \mathrm{and} \quad 
  \mathrm{Norm}(\mathscr{C}_{\mathrm{por}})
  = \mathrm{Lsp}\{\mathbf{I}\}
\end{align}
The direct sum decomposition form the
projection theorem implies that the Cauchy
stresses under the constrained motion
due to internal constraints can be
written as follows\footnote{The minus sign is
  introduced for convenience so that
  $p_{\mathrm{free}}$ and $p_{\mathrm{por}}$
  will be the mechanical pressures.} :
%----------------------------;
%  Equation: Extra stresses  ;
%----------------------------;
\begin{subequations}
  \label{Eqn:ICs_decomposition_of_Cauchy}
  \begin{align}
    \label{Eqn:ICs_decomposition_Tfree}
    &\mathbf{T}_{\mathrm{free}}(\mathbf{x}) =
    -p_{\mathrm{free}}(\mathbf{x}) \mathbf{I}
    + \mathbf{T}_{\mathrm{free}}^{\mathrm{extra}}(\mathbf{x}) \\
    \label{Eqn:ICs_decomposition_Tpor}
    &\mathbf{T}_{\mathrm{por}}(\mathbf{x}) =
    -p_{\mathrm{por}}(\mathbf{x}) \mathbf{I}
    + \mathbf{T}_{\mathrm{por}}^{\mathrm{extra}}(\mathbf{x}) 
  \end{align}
\end{subequations}
where the extra stresses,
$\mathbf{T}_{\mathrm{free}}^{\mathrm{extra}}$
and $\mathbf{T}_{\mathrm{free}}^{\mathrm{extra}}$,
belong to the tangent spaces and should be
prescribed through constitutive specifications
\footnote{See \citep{o2001decomposition} for an
  insightful discussion on a related issue in the
  context of particle dynamics. They discussed active
  and reactive components due to a constraint,
  what flexibility a dynamicist will have as
  a part of constitutive specifications, and
  the relation between the prescription for
  the reactive component and the Gauss's
  principle of least constraint.}. 
Moreover, the no-work by the active
components will be a trivial mathematical
consequence. To wit,
%---------------------;
%  Equation: No-work  ;
%---------------------;
\begin{align}
  \mathbf{T}_{\mathrm{free}}^{\perp} \cdot
  \mathbf{L}_{\mathrm{free}} = -p_{\mathrm{free}} \mathbf{I}
  \cdot \mathbf{L}_{\mathrm{free}} = -p_{\mathrm{free}}
  \mathrm{tr}[\mathbf{L}_{\mathrm{free}}] = 0  
\end{align}
A similar reasoning holds for $\mathbf{T}_{\mathrm{por}}^{\perp}$.
The following relations will also be
mathematical consequences:
\begin{align}
  \mathbf{T}_{\mathrm{free}} \cdot \mathbf{L}_{\mathrm{free}} 
  = \mathbf{T}_{\mathrm{free}}^{\parallel} \cdot
  \mathbf{L}_{\mathrm{free}}
  = \mathbf{T}_{\mathrm{free}}^{\mathrm{extra}} \cdot
  \mathbf{L}_{\mathrm{free}} 
\end{align}

%==============================================;
%  Subsection: Consequences of (P1) statement  ; 
%==============================================;
\subsection{Consequences of (P1) statement}
Using Green's identity and noting that the virtual
velocity fields vanish on $\Gamma_{\mathrm{free}}^{v}$
and $\Gamma_{\mathrm{por}}^{v}$, the (P1) statement
\eqref{Eqn:ICs_proposed_PVP_P1} can be rewritten
as follows:
%======================;
%  Step 1: Derivation  ;
%----------------------;
\begin{align}
  \label{Eqn: First_optimality_condition}
  &\int_{\Gamma_{\mathrm{free}}^{t}}
  \mathbf{w}_{\mathrm{free}} \cdot
  \left\{\mathbf{T}_{\mathrm{free}}
  \widehat{\mathbf{n}}_{\mathrm{free}}^{\mathrm{ext}}
  - \mathbf{t}^{\mathrm{p}}_{\mathrm{free}}
  \right\} \mathrm{d}\Gamma
  -\int_{\mathcal{K}_{\mathrm{free}}} \mathbf{w}_{\mathrm{free}}
  \cdot \left\{\mathrm{div}[\mathbf{T}_{\mathrm{free}}]
  + \gamma \mathbf{b}_{\mathrm{free}}\right\}
  \mathrm{d} \Omega \notag \\
  &+ \int_{\Gamma_{\mathrm{por}}^{t}}
  \mathbf{w}_{\mathrm{por}} \cdot
  \left\{\mathbf{T}_{\mathrm{por}}
  \widehat{\mathbf{n}}_{\mathrm{por}}^{\mathrm{ext}}
  - \mathbf{t}^{\mathrm{p}}_{\mathrm{por}} \right\}
  \mathrm{d}\Gamma 
  -\int_{\mathcal{K}_{\mathrm{por}}} \mathbf{w}_{\mathrm{por}}
  \cdot \left\{\mathrm{div}[\mathbf{T}_{\mathrm{por}}]
  + \gamma \phi_{\mathrm{por}} 
  \mathbf{b}_{\mathrm{por}} - \mathbf{i}_{\mathrm{por}}
  \right\} \; \mathrm{d} \Omega \nonumber \\
  &+ \int_{\Gamma_{\mathrm{int}}} \left\{
  \mathbf{w}_{\mathrm{free}} \cdot
  \mathbf{T}_{\mathrm{free}} \widehat{\mathbf{n}}_{\mathrm{free}}
  + \mathbf{w}_{\mathrm{por}} \cdot
  \mathbf{T}_{\mathrm{por}} \widehat{\mathbf{n}}_{\mathrm{por}}
  + \overset{*}{\mathbf{w}}_{\mathrm{free}}
  \cdot \frac{\partial \Psi}{\partial
    \overset{*}{\mathbf{v}}_{\mathrm{free}}}
	+ \overset{*}{\mathbf{w}}_{\mathrm{por}} \cdot
        \frac{\partial \Psi}{\partial
          \overset{*}{\mathbf{v}}_{\mathrm{por}}}
        + w_n \cdot
        \frac{\partial \Psi}{\partial v_n}
	\right\} \mathrm{d}\Gamma = 0 \nonumber \\
	&\forall
        \left(\mathbf{w}_{\mathrm{free}},
        \mathbf{w}_{\mathrm{por}}\right)
        \in \widetilde{\mathcal{V}}
	\end{align}

We now invoke the arbitrariness of the fields
$\mathbf{w}_{\mathrm{free}}(\mathbf{x})$ and
$\mathbf{w}_{\mathrm{por}}(\mathbf{x})$ but 
respecting the requirements of kinematic
admissibility. 
The first two terms give rise to the
following governing equations for the
free flow region except along the part
of the boundary that shares with the
interface:
%----------------------------------------;
%  Equation: GE in the free flow region  ;
%----------------------------------------;
\begin{subequations}
  \begin{alignat}{2}
    \label{Eqn:BoLM_free_region}
    &\mathrm{div}[\mathbf{T}_{\mathrm{free}}]
    + \gamma \mathbf{b}_{\mathrm{free}}
    = \mathbf{0}
    && \quad \mathrm{in} \; \mathcal{K}_{\mathrm{free}} \\
    \label{Eqn:BoM_free_region}
    &\mathrm{div}[\mathbf{v}_{\mathrm{free}}]
    = 0
    && \quad \mathrm{in} \; \mathcal{K}_{\mathrm{free}} \\
    \label{Eqn:Traction_BC_free_region}
    &\mathbf{T}_{\mathrm{free}}
    \widehat{\mathbf{n}}_{\mathrm{free}}^{\mathrm{ext}}(\mathbf{x}) 
    = \mathbf{t}^{\mathrm{p}}_{\mathrm{free}}(\mathbf{x})
    && \quad \mathrm{on} \; \Gamma_{\mathrm{free}}^{t} \\
    \label{Eqn:vBC_free_region}
    &\mathbf{v}_{\mathrm{free}}(\mathbf{x}) 
    = \mathbf{v}^{\mathrm{p}}_{\mathrm{free}}(\mathbf{x})
    && \quad \mathrm{on} \; \Gamma_{\mathrm{free}}^{v} 
  \end{alignat}
\end{subequations}
The third and fourth terms give rise to the
following governing equations for the porous
region except along the part of the boundary
that shares with the interface:
%-------------------------------------;
%  Equation: GE in the porous region  ;
%-------------------------------------;
\begin{subequations}
  \begin{alignat}{2}
    \label{Eqn:BoLM_porous_region}
    &\mathrm{div}[\mathbf{T}_{\mathrm{por}}]
    + \gamma \phi_{\mathrm{por}} \mathbf{b}_{\mathrm{por}} 
    - \mathbf{i}_{\mathrm{por}} = \mathbf{0}
    && \quad \mathrm{in} \; \mathcal{K}_{\mathrm{por}} \\
    \label{Eqn:BoM_porous_region}
    &\mathrm{div}[\mathbf{v}_{\mathrm{por}}]
    = 0
    && \quad \mathrm{in} \; \mathcal{K}_{\mathrm{por}} \\
    \label{Eqn:Traction_BC_porous_region}
    &\mathbf{T}_{\mathrm{por}}
    \widehat{\mathbf{n}}_{\mathrm{por}}^{\mathrm{ext}}(\mathbf{x}) 
    = \mathbf{t}^{\mathrm{p}}_{\mathrm{por}}(\mathbf{x}) 
    && \quad \mathrm{on} \; \Gamma_{\mathrm{por}}^{t} \\
    \label{Eqn:vBC_porous_region}
    &\mathbf{v}_{\mathrm{por}}(\mathbf{x}) 
    = \mathbf{v}^{\mathrm{p}}_{\mathrm{por}}(\mathbf{x}) 
    && \quad \mathrm{on} \; \Gamma_{\mathrm{por}}^{v} 
  \end{alignat}
\end{subequations}
Noting the decomposition given in equation 
\eqref{Eqn:ICs_tangential_velocities}, the
fifth term gives rise to the following
interface conditions on $\Gamma_{\mathrm{int}}$:
%--------------------------------------------------;
%  Equation: Compact form of interface conditions  ;
%--------------------------------------------------;
\begin{tcolorbox}
  \begin{subequations}
  \label{Eqn:Compact_form_of_interface}
    \begin{align}
      \label{Eqn:vn_jump_condition}
      &v^{(n)}_{\mathrm{free}}(\mathbf{x}) + v^{(n)}_{\mathrm{por}}(\mathbf{x}) = 0 \\
      \label{Eqn:normal_traction_continuity}
      & \widehat{\mathbf{n}}_{\mathrm{free}}(\mathbf{x}) 
      \cdot \mathbf{T}_{\mathrm{free}}(\mathbf{x}) 
      \widehat{\mathbf{n}}_{\mathrm{free}}(\mathbf{x})
      + \frac{\partial \Psi}{\partial v_n}
      = \widehat{\mathbf{n}}_{\mathrm{por}}(\mathbf{x}) 
      \cdot \mathbf{T}_{\mathrm{por}}(\mathbf{x}) 
      \widehat{\mathbf{n}}_{\mathrm{por}}(\mathbf{x}) \\
      \label{Eqn:tangential_traction_free}
      &\widehat{\mathbf{s}}(\mathbf{x}) \cdot 
      \mathbf{T}_{\mathrm{free}}^{\mathrm{extra}}
      \widehat{\mathbf{n}}_{\mathrm{free}}(\mathbf{x})
      = -\widehat{\mathbf{s}}(\mathbf{x})
      \cdot \frac{\partial \Psi}{\partial
        \overset{*}{\mathbf{v}}_{\mathrm{free}}} \\
      \label{Eqn:tangential_traction_porous}
      &\widehat{\mathbf{s}}(\mathbf{x}) \cdot 
      \mathbf{T}_{\mathrm{por}}^{\mathrm{extra}}
      \widehat{\mathbf{n}}_{\mathrm{por}}(\mathbf{x})
      = -\widehat{\mathbf{s}}(\mathbf{x}) \cdot
      \frac{\partial \Psi}{\partial \overset{*}{\mathbf{v}}_{\mathrm{por}}} 
    \end{align}
  \end{subequations}
\end{tcolorbox}

Equation \eqref{Eqn:vn_jump_condition} is in fact
the jump condition corresponding to the balance of
mass (cf. equation \eqref{Eqn:ICs_JC_BoM_final} in
\textbf{\S\ref{Sec:App_ICs_JCs}}). The other three conditions 
are in general not jump conditions and they stem
from a constitutive specification in the form of a
prescription for the functional $\Psi$. If $\Psi$
is independent of $v_n$ (which is assumed in
\textbf{\S\ref{Sec:S5_ICs_Special_cases}} to
obtain some popular conditions like the BJ
and BJS conditions)
then the second condition \eqref{Eqn:normal_traction_continuity}
will reduce to the normal
component of the jump condition
for the balance of linear momentum\footnote{The second interface 
condition can be interpreted in a more familiar form using tractions, 
see Appendix \ref{App:ICs_Cauchy_theorem}.}. 
To summarize, the complete set of governing
equations for flows in coupled
free-porous media is:
\begin{itemize}
\item the equations in the free flow region
  along with the boundary conditions on
  the external boundary (not including
  $\Gamma_{\mathrm{int}}$) of the region 
  \eqref{Eqn:BoLM_free_region}--\eqref{Eqn:vBC_free_region},
\item the equations in the porous region
  along with the boundary conditions on
  the external boundary (not including
  $\Gamma_{\mathrm{int}}$) of the region
  \eqref{Eqn:BoLM_porous_region}--\eqref{Eqn:vBC_porous_region},
\item the symmetry of Cauchy stresses
  \eqref{Eqn:ICs_symmetry_Cauchy_stresses},
  \item the decomposition of Cauchy
  stresses 
  \eqref{Eqn:ICs_decomposition_Tfree}--\eqref{Eqn:ICs_decomposition_Tpor},
\item the interface conditions
  \eqref{Eqn:vn_jump_condition}--\eqref{Eqn:tangential_traction_porous}
  and 
\item the (prescribed) constitutive specifications
  for $\mathbf{T}_{\mathrm{free}}^{\mathrm{extra}}$,
  $\mathbf{T}_{\mathrm{por}}^{\mathrm{extra}}$,
  $\mathbf{i}_{\mathrm{por}}$ and $\Psi$.  
\end{itemize}
The solution fields will be
$\mathbf{v}_{\mathrm{free}}(\mathbf{x})$,
$\mathbf{v}_{\mathrm{por}}(\mathbf{x})$,
$p_{\mathrm{free}}(\mathbf{x})$ and
$p_{\mathrm{por}}(\mathbf{x})$.

%*****************************************************;
%                                                     ;
%  NAME                                               ;
%    S5_ICs_Special_cases.tex                         ;
%                                                     ;
%  WRITTEN BY                                         ;
%    Kalyana Babu Nakshatrala                         ;
%    Mohammad S. Joshaghani                           ;
%                                                     ;
%*****************************************************;
\section{SPECIAL CASES}
\label{Sec:S5_ICs_Special_cases}
We now show the BJ and BJS conditions,
and the no-slip condition (which is
commonly employed in the fluid mechanics
for free flows) are, respectively, special
cases and a limiting case of the proposed
framework.
The following assumptions will be common
to all the mentioned conditions:

\begin{enumerate}{}
\item[(A1)] The normal component of the velocity
  at the interface does not contribute
  towards the power expended density at
  the interface. That is, $\Psi$ is
  independent of $v_n$.
\item[(A2)] $\Psi$ is a quadratic functional
  of the tangential (relative) velocities,
  and the invariance requirements demand 
  that this functional has to be in terms
  of individual and joint invariants of
  the tangential (relative) velocities.
  Thus, mathematically, we write the
  functional as follows: 
  %---------------------------------------------;
  %  Equation: Specific dissipation functional  ;
  %---------------------------------------------;
  \begin{align}
    \label{Eqn:Specific_dissipation_functional}
    \Psi[\overset{*}{\mathbf{v}}_{\mathrm{free}},
      \overset{*}{\mathbf{v}}_{\mathrm{por}},v_n] = 
    \alpha_{11} {\overset{*}{\mathbf{v}}_{\mathrm{free}}
      \cdot \overset{*}{\mathbf{v}}_{\mathrm{free}}}
    + 2\alpha_{12} {\overset{*}{\mathbf{v}}_{\mathrm{free}}
      \cdot \overset{*}{\mathbf{v}}_{\mathrm{por}}}
    +\alpha_{22} {\overset{*}{\mathbf{v}}_{\mathrm{por}}
      \cdot \overset{*}{\mathbf{v}}_{\mathrm{por}}}
  \end{align}
  where $\alpha_{11}$, $\alpha_{12}$ and $\alpha_{22}$ are
  constants, and $\mathop{\mathbf{v}}_{\mathrm{free}}^{*}$
  and $\mathop{\mathbf{v}}_{\mathrm{por}}^{*}$ are the
  tangential velocities.
\item[(A3)] The non-negativity of $\Psi$ is
  enforced by assuming that
  %-----------------------------------;
  %  Equation: Non-negativity of Psi  ;
  %-----------------------------------;
  \begin{align}
    \label{Eqn:Specific_dissipation_functional_NN}
    \alpha_{11} \alpha_{22} \geq \alpha_{12}^{2}
  \end{align}
  \item[(A4)] The Stokes model is assumed 
  to describe the flow in the free flow region. 
  That is, the flow 
  in the free flow region is assumed to be a 
  creeping flow, which implies the following: 
  \begin{align}
  \mathbf{T}_{\mathrm{free}}^{\mathrm{extra}} = 
   2 \mu \mathbf{D}_{\mathrm{free}}
  \end{align}
\end{enumerate}

The above assumptions give rise to the
following interface conditions for the
tangential component of the tractions:
%--------------------------------------------;
%  Equation: Tangent component of tractions  ;
%--------------------------------------------;
\begin{subequations}
  \begin{alignat}{2}
    \label{Eqn:Interface_free_flow_specific_A}
    &\widehat{\mathbf{s}} \cdot
    \mathbf{T}_{\mathrm{free}}^{\mathrm{extra}}
    \widehat{\mathbf{n}}_{\mathrm{free}}
    = -\frac{\partial \Psi}{\partial
      \overset{*}{\mathbf{v}}_{\mathrm{free}}} \cdot \widehat{\mathbf{s}}
    = -2(\alpha_{11}\overset{*}{\mathbf{v}}_{\mathrm{free}}
    + \alpha_{12}\overset{*}{\mathbf{v}}_{\mathrm{por}})
    \cdot \widehat{\mathbf{s}}
    &&\qquad \mbox{on} \; \Gamma_{\mathrm{free}}\\
    \label{Eqn:Interface_free_flow_specific_B}
    &\widehat{\mathbf{s}} \cdot \mathbf{T}_{\mathrm{por}}^{\mathrm{extra}}\widehat{\mathbf{n}}_{\mathrm{por}}
    = -\frac{\partial \Psi}{\partial \overset{*}{\mathbf{v}}_{\mathrm{por}}} \cdot \widehat{\mathbf{s}}
    =  -2(\alpha_{12}\overset{*}{\mathbf{v}}_{\mathrm{free}}
    + \alpha_{22}\overset{*}{\mathbf{v}}_{\mathrm{por}})
    \cdot \widehat{\mathbf{s}}
    &&\qquad \mbox{on} \; \Gamma_{\mathrm{por}}
  \end{alignat}
\end{subequations}
where $\widehat{\mathbf{s}}(\mathbf{x})$
denotes an arbitrary unit tangent vector
field along the interface.

%========================================;
%  Subsection: Beavers-Joseph condition  ;
%========================================;
\subsection{Beavers-Joseph condition}
The BJ condition can be obtained by
further making the following choices: 
%------------------------------------------------;
%  Equation: Choices for alpha for BJ condition  ;
%------------------------------------------------;
\begin{align}
  \alpha_{11} = \alpha_{22} = 
  \frac{\alpha \mu \sqrt{3}}{2 \sqrt{\mathrm{tr}[\mathbf{K}]}}
  \quad \mathrm{and} \quad 
  \alpha_{12}=\frac{-\alpha \mu \sqrt{3}}{2 \sqrt{\mathrm{tr}[\mathbf{K}]}} 
\end{align}
where $\mathrm{tr}[\cdot]$ denotes
the trace of a second-order tensor. 
Then equation \eqref{Eqn:Interface_free_flow_specific_A}
will reduce to:
\begin{equation}
  \label{Eqn:Beavers_Joseph}
  \widehat{\mathbf{s}} \cdot (-2\mu \; \mathbf{D}_{\mathrm{free}})
  \widehat{\mathbf{n}}_{\mathrm{free}}
  = \frac{\alpha \mu \sqrt{3}}{\sqrt{\mathrm{tr}[\mathbf{K}]}}
  \widehat{\mathbf{s}} \cdot (\mathbf{v}_{\mathrm{free}}-\mathbf{v}_{\mathrm{por}})
\end{equation}
which is the ``boundary'' condition proposed in
\citep{1967_Beavers_JFM} for the free flow region
due to the presence of a pervious boundary.
By aligning the coordinate axes similar to the
one shown in \textbf{Fig.~\ref{Fig:ICs_BJ_velocity_profile}}
and by taking the x-component of $\mathbf{v}_{\mathrm{por}}$
to be $Q$, one will get an expression similar
to the one provided in \citep{1967_Beavers_JFM}
(cf. equation \eqref{Eqn:ICs_BJ_original_equation}).
It should be however noted that
\citeauthor{1967_Beavers_JFM}
do not provide a corresponding condition
for the flow in the porous media, which
lies on the other side of the interface.
 
On the other hand, using the proposed framework,
one can obtain a corresponding condition for
the flow on the other side of the interface
(i.e., the porous medium); which is needed
if one wants to simulate a coupled flow in
both free and porous regions. Using equation
\eqref{Eqn:Interface_free_flow_specific_B},
the interface condition on $\Gamma_{\mathrm{por}}$
can be written as follows: 
\begin{equation}
 \label{Eqn:BC_in_porous}
  \widehat{\mathbf{s}} \cdot 
  \mathbf{T}_{\mathrm{por}}^{\mathrm{extra}}
  \widehat{\mathbf{n}}_{\mathrm{por}}
  = \frac{\alpha \mu^{\prime} \sqrt{3}}{\sqrt{\mathrm{tr}[\mathbf{K}]}}
  \widehat{\mathbf{s}} \cdot (\mathbf{v}_{\mathrm{free}}-\mathbf{v}_{\mathrm{por}})
\end{equation}

%===================================================;
%  Subsubsection: A discussion on the BJ condition  ;
%---------------------------------------------------;
\subsubsection{A discussion on the BJ condition}
The velocity field in the porous region
  is assumed to be known \emph{a priori}. Moreover,
  the flow in the porous region is tacitly assumed
  to be uniform beyond a boundary layer (see
  \textbf{Fig.~\ref{Fig:ICs_BJ_velocity_profile}}).
  But the velocity field in the porous region
  is seldom known \emph{a priori} and this is
  particularly true in the case of flows in
  coupled free-porous media.
  Even if the velocity field
  in the porous region is known, this field
  will not be uniform due to
  spatial heterogeneity of medium properties
  (e.g., permeability). (Heterogeneity is
  inherent to the two application problems
  that we discussed in the introduction.)
  This will create an ambiguity in assigning
  a value to $Q$ (cf. equation
  \eqref{Eqn:ICs_BJ_original_equation}).
  Specifically, at what depth one has to
  sample the (horizontal or tangential)
  velocity to specify $Q$ (cf. \textbf{Fig.~\ref{Fig:ICs_BJ_velocity_profile}}).

  Last but not least, the BJ condition may not
  be compatible with all porous media model.
  For example, if the flow in the porous
  region is modeled using the Darcy model,
  for which, $\mathbf{T}_{\mathrm{por}}^{\mathrm{(extra)}}
  = \mathbf{0}$. Equation \eqref{Eqn:BC_in_porous} 
  will then imply that
  \begin{align*}
    \widehat{\mathbf{s}} \cdot
    (\mathbf{v}_{\mathrm{free}}-\mathbf{v}_{\mathrm{por}})
    = 0 
  \end{align*}
  which, based on the BJ condition
  \eqref{Eqn:Beavers_Joseph}, will
  further imply that
  \begin{align*}
    \widehat{\mathbf{s}} \cdot \mathbf{D}_{\mathrm{free}}
    \widehat{\mathbf{n}} = 0 
  \end{align*}
  But this condition will not be met in general,
  as, for example, the horizontal velocity
  can depend on the $y$-coordinate or the
  vertical velocity can depend on the
  $x$-coordinate. 
    
%================================================;
%  Subsection: Beavers-Joseph-Saffman condition  ;
%================================================;
\subsection{Beavers-Joseph-Saffman condition} 
In addition to the aforementioned four assumptions 
(A1)--(A4), we make the following choices
to obtain the BJS condition: 
\begin{align}
  \alpha_{11} = \alpha_{22} = 
  \frac{\alpha \mu \sqrt{3}}{2 \sqrt{\mathrm{tr}[\mathbf{K}]}}
  \quad \mathrm{and} \quad  
  \alpha_{12} = 0 
\end{align}
Then, using equation
\eqref{Eqn:Interface_free_flow_specific_A},
the \emph{boundary} condition at
$\Gamma_{\mathrm{free}}$ for the flow
in the free region due to a juxtaposed
porous region takes the following form: 
\begin{equation}
  \label{Eqn:Beavers_Joseph_saffman}
  \widehat{\mathbf{s}} \cdot (-2\mu \; \mathbf{D}_{\mathrm{free}})
  \widehat{\mathbf{n}}_{\mathrm{free}}
  = \frac{\alpha \mu \sqrt{3}}{\sqrt{\mathrm{tr}[\mathbf{K}]}}
  \widehat{\mathbf{s}} \cdot \mathbf{v}_{\mathrm{free}}
\end{equation}

Using equation \eqref{Eqn:Interface_free_flow_specific_B},
the interface condition on $\Gamma_{\mathrm{por}}$ takes the
following form:
\begin{align}
  \label{Eqn:Beavers_Joseph_saffman_por_side}
  \widehat{\mathbf{s}} \cdot \mathbf{T}_{\mathrm{free}}^{\mathrm{extra}}
  \widehat{\mathbf{n}}_{\mathrm{free}}
  = \frac{\alpha \mu \sqrt{3}}{\sqrt{\mathrm{tr}[\mathbf{K}]}}
  \widehat{\mathbf{s}} \cdot \mathbf{v}_{\mathrm{por}}
\end{align}

%====================================;
%  Subsubsection: Discussion on BJS  ;
%------------------------------------;
\subsubsection{A discussion on the BJS condition}
Since the BJS condition \eqref{Eqn:Beavers_Joseph_saffman}
does not contain $Q$ (the mean velocity in the porous
region beyond the boundary layer), it does not
assume the velocity field in the porous region
is neither known \emph{a priori} nor uniform.
However, the BJS condition need not be compatible
with all porous media models. If one again considers
the Darcy model to describe the flow in the porous
region, equation \eqref{Eqn:Beavers_Joseph_saffman_por_side}
implies that $\mathbf{v}_{\mathrm{por}} = \mathbf{0}$---the no-slip boundary condition for the porous
region along the interface---which is not what
has been observed in the experiments \citep{1967_Beavers_JFM}.

On the other hand, if one uses the Darcy-Brinkman model,
for which $\mathbf{T}_{\mathrm{por}}^{\mathrm{extra}} =
2 \mu \mathbf{D}_{\mathrm{por}}$, the BJS condition
will be compatible with the chosen model. Saffman did
recognize that his condition is actually compatible
with the Darcy-Brinkman model and not the Darcy
model\footnote{See \citep[equation (2.18)]{1971_Saffman_SAM} and the text below that 
  equation.}. However, by using asymptotic analysis,
he argued that solutions from the Darcy-Brinkman model
and the Darcy model do not differ significantly outside
the boundary layer, and the size of the boundary layer
is in the order of the square-root of the (trace of)
permeability.

%=================================;
%  Subsection: No-slip condition  ;
%=================================;
\subsection{No-slip condition}
The classical no-slip condition can be
obtained by making the following choices
for the constants: 
\begin{align}
  \label{Eqn:ICs_slip_choices}
  \alpha_{11} = \frac{\alpha}{2\sqrt{\mathrm{tr}[\mathbf{K}]}},
  \alpha_{22} = 0 \quad \mathrm{and} \quad
  \alpha_{12} = 0 
\end{align}
and then by letting $\mathrm{tr}[\mathbf{K}]
\rightarrow 0$. To wit, based on the choices
made in equation \eqref{Eqn:ICs_slip_choices},
the interface condition
\eqref{Eqn:Interface_free_flow_specific_A}
reduces to the following: 
\begin{align}
  \widehat{\mathbf{s}} \cdot
  {\mathop{\mathbf{v}}^{*}}_{\mathrm{free}} 
  = - \left(\frac{\sqrt{\mathrm{tr}[\mathbf{K}]}}{\alpha}\right)
  \widehat{\mathbf{s}} \cdot \mathbf{T}_{\mathrm{free}}^{\mathrm{extra}}
  \widehat{\mathbf{n}}_{\mathrm{free}}
\end{align}
By letting $\mathrm{tr}[\mathbf{K}] \rightarrow 0$
and noting that $\widehat{\mathbf{s}}$ is an arbitrary
tangent vector along the interface, one can conclude
that $\overset{*}{\mathbf{v}}_{\mathrm{free}} =
\mathbf{0}$ on $\Gamma_{\mathrm{free}}$, which is the
no-slip condition. 
Note that $\mathrm{tr}[\mathbf{K}] \rightarrow 0$
basically implies that the boundary is impervious,
and the no-slip boundary condition is typically
enforced at an impervious boundary in an uncoupled
free flow.

%*****************************************************;
%                                                     ;
%  NAME                                               ;
%    S6_ICs_MPT.tex                                   ;
%                                                     ;
%  WRITTEN BY                                         ;
%      Kalyana Babu Nakshatrala  	              ;
%      Mohammad S. Joshaghani                         ;
%                                                     ;
%*****************************************************;
\section{MINIMUM POWER THEOREM FOR
  A CLASS OF COUPLED FLOWS}
\label{Sec:S6_ICs_MPT}
It is well-known that an uncoupled creeping
flow, which is governed by the incompressible
Stokes equations, enjoys a minimum power
theorem \citep{Guazzelli_Morris}. It has
also been established that an uncoupled
flow through porous media based on either
Darcy equations or Darcy-Brinkman equations
enjoys a minimum power theorem
\citep{2016_Shabouei_CICP}.
\emph{It is thus natural to ask whether a
  flow in coupled free-porous media enjoys
  a minimum power theorem.}

We now show that the answer to this question
is affirmative for a class of coupled flows. 
This class of flows is characterized
by these two requirements:
\begin{enumerate}
  \item[(R1)] There exists two potentials, 
    $\Phi_{\mathrm{free}}$ and $\Phi_{\mathrm{por}}$,
    with the following properties: 
    \begin{enumerate}[(i)]
    \item They satisfy the form-invariance and the
      invariance under a Euclidean transformation
      (i.e., they satisfy the principle of material
      frame indifference). Specifically these potentials
      can be expressed as
      $\Phi_{\mathrm{free}}[\mathbf{D}_{\mathrm{free}}]$
      and $\Phi_{\mathrm{por}}[\mathbf{D}_{\mathrm{por}},
      \mathbf{v}_{\mathrm{por}}]$.\footnote{$\mathbf{v}_{\mathrm{por}}$
      should be interpreted with respect to the velocity
      of the porous solid, and hence it is objective
      under a Euclidean transformation.}
    \item They provide the constitutive relations
      of the following form for the extra
      Cauchy stresses and the interaction
      term:
    \begin{subequations}
      \begin{align}
        \label{Eqn:ICs_T_free_extra_Phi_free}
        &\mathbf{T}_{\mathrm{free}}^{\mathrm{extra}} =
        \ddfrac{\partial \Phi_{\mathrm{free}}}{\partial
          \mathbf{D}_{\mathrm{free}}} \\
        \label{Eqn:ICs_T_por_extra_Phi_por}
        &\mathbf{T}_{\mathrm{por}}^{\mathrm{extra}} =
        \ddfrac{\partial \Phi_{\mathrm{por}}}{\partial
          \mathbf{D}_{\mathrm{por}}} \\
        \label{Eqn:ICs_i_por_Phi_por}
        &\mathbf{i}_{\mathrm{por}} = 
        \ddfrac{\partial \Phi_{\mathrm{por}}}{\partial
          \mathbf{v}_{\mathrm{por}}}
      \end{align}
    \end{subequations}
  \item Each of the potentials has a positive
    definite Hessian\footnote{The
      Hessian of a functional is the Jacobian
      matrix containing the second derivatives
      of the functional with respect to its
      input arguments. A positive definite
      Hessian means that the Jacobian matrix
      is positive definite. In other words,
      the second variation of the functional
      is positive under all non-zero variations
      of its input arguments.}.
    \end{enumerate}
  \item[(R2)] The functional $\Psi$ has a
    positive definite Hessian. 
\end{enumerate}
The requirement (R2) is in addition to
the properties that outlined in
\textbf{\S\ref{Sec:S3_ICs_Proposed_framework}}
for $\Psi$ to satisfy.
It is easy to construct $\Psi$ to
have a positive definite Hessian; the functional
\eqref{Eqn:Specific_dissipation_functional}
satisfying the condition
\eqref{Eqn:Specific_dissipation_functional_NN}
is one such example.

%=================================================;
%  Subsection: On construction of the potentials  ;
%=================================================;
\subsection{On construction of the potentials}
For many popular uncoupled free flow models
(e.g., Stokes equations) and porous media
models (e.g., Darcy equations, Darcy-Brinkman
equations), the rate of internal dissipation
density satisfies the conditions
\eqref{Eqn:ICs_T_free_extra_Phi_free}--\eqref{Eqn:ICs_i_por_Phi_por}.
One can take the same approach to construct
the potentials $\Phi_{\mathrm{free}}$ and
$\Phi_{\mathrm{por}}$ even for the case of
coupled flows. This approach can be
best illustrated by the following 
examples. 

Under the Stokes model, the Cauchy stress
and the extra Cauchy stress are given by
%----------------------------------------------;
%  Equation: Cauchy stress under Stokes model  ;
%----------------------------------------------;
\begin{align}
  \mathbf{T}_{\mathrm{free}} = -p_{\mathrm{free}} \mathbf{I}
  + 2 \mu \mathbf{D}_{\mathrm{free}} 
  = -p_{\mathrm{free}} \mathbf{I}
  + \mathbf{T}_{\mathrm{free}}^{\mathrm{extra}}
\end{align}
and the rate of internal dissipation
density is given by
%-----------------------------------------------------;
%  Equation: Internal dissipation under Stokes model  ;
%-----------------------------------------------------;
\[
  2 \mu \mathbf{D}_{\mathrm{free}} \cdot
  \mathbf{D}_{\mathrm{free}}  
\]
Clearly, by choosing the potential
$\Phi_{\mathrm{free}}$ to be
%---------------------------------------;
%  Equation: Phi_free for Stokes model  ;
%---------------------------------------;
\begin{align}
  \label{Eqn:ICs_Stokes_Phi_free_potential}
  2 \Phi_{\mathrm{free}}[\mathbf{v}_{\mathrm{free}}] =
  2 \mu \mathbf{D}_{\mathrm{free}} \cdot
  \mathbf{D}_{\mathrm{free}}
\end{align}
one can satisfy the requirement
\eqref{Eqn:ICs_T_free_extra_Phi_free}.
Similarly, under the Darcy model, the
extra Cauchy stress and interaction
term are, respectively, given by
%------------------------------------------------------------------;
%  Equation: Cauchy stress and interaction term under Darcy model  ;
%------------------------------------------------------------------;
\begin{align}
  \mathbf{T}_{\mathrm{por}}^{\mathrm{extra}} = \mathbf{0}
  \quad \mathrm{and} \quad
  \mathbf{i}_{\mathrm{por}} = \mu \mathbf{K}^{-1}
  \mathbf{v}_{\mathrm{por}}(\mathbf{x}) 
\end{align}
By choosing the potential
$\Phi_{\mathrm{por}}$ to be
%-------------------------------------;
%  Equation: Phi_por for Darcy model  ;
%-------------------------------------;
\begin{align}
  \label{Eqn:ICs_Darcy_Phi_por_potential}
  2 \Phi_{\mathrm{por}}[\mathbf{v}_{\mathrm{por}}] =
  \underbrace{\mu \mathbf{K}^{-1} \mathbf{v}_{\mathrm{por}}(\mathbf{x})
    \cdot \mathbf{v}_{\mathrm{por}}(\mathbf{x})}_{\mbox{rate of internal dissipation density}}
\end{align}
one can satisfy the requirements
\eqref{Eqn:ICs_T_por_extra_Phi_por}
and \eqref{Eqn:ICs_i_por_Phi_por}.
Under the Darcy-Brinkman model, the extra
Cauchy stress and interaction term are,
respectively, given by
%---------------------------------------------------------------;
%  Equation: Cauchy stress and interaction term under DB model  ;
%---------------------------------------------------------------;
\begin{align}
  \mathbf{T}_{\mathrm{por}}^{\mathrm{extra}}
  = 2 \mu \mathbf{D}_{\mathrm{por}}
  \cdot \mathbf{D}_{\mathrm{por}}
  \quad \mathrm{and} \quad
  \mathbf{i}_{\mathrm{por}} = \mu \mathbf{K}^{-1}
  \mathbf{v}_{\mathrm{por}}(\mathbf{x}) 
\end{align}
By choosing the potential
$\Phi_{\mathrm{por}}$ to be
%----------------------------------;
%  Equation: Phi_por for DB model  ;
%----------------------------------;
\begin{align}
  \label{Eqn:ICs_DB_Phi_por_potential}
  2 \Phi_{\mathrm{por}}[\mathbf{v}_{\mathrm{por}}] =
  \underbrace{2 \mu \mathbf{D}_{\mathrm{por}} 
    \cdot \mathbf{D}_{\mathrm{por}}
    + \mu \mathbf{K}^{-1} \mathbf{v}_{\mathrm{por}}(\mathbf{x})
    \cdot \mathbf{v}_{\mathrm{por}}(\mathbf{x})}_{\mbox{rate of internal dissipation density}}
\end{align}
one can satisfy the requirements
\eqref{Eqn:ICs_T_por_extra_Phi_por}
and \eqref{Eqn:ICs_i_por_Phi_por}.

If the coupled flow is modeled based
on Stokes-Darcy equations (i.e., Stokes
model is used for the free flow region,
and Darcy model is used for the porous
region), then the two potentials for
the coupled flow can be chosen based
on equations \eqref{Eqn:ICs_Stokes_Phi_free_potential}
and \eqref{Eqn:ICs_Darcy_Phi_por_potential},
which are for uncoupled flows.
Similarly, if the coupled flow is based
on Stokes-Darcy-Brinkman equations (i.e.,
Stokes model is used for the free flow
region and Darcy-Brinkman model is used
for the porous region), then the two potentials for the
coupled flow can be chosen based on equations
\eqref{Eqn:ICs_Stokes_Phi_free_potential}
and \eqref{Eqn:ICs_DB_Phi_por_potential}.

%=====================================;
%  Subsection: Minimum power theorem  ;
%=====================================;
\subsection{Minimum power theorem}
We define the total mechanical power
functional as follows:
%-----------------------------------------------;
%  Equation: Total mechanical power functional  ;
%-----------------------------------------------;
\begin{align}
  \label{Eqn:Minimum_total_mechanical_power_Statement}
  \mathcal{P}_{\mathrm{coupled}}[\mathbf{z}_{\mathrm{free}}
    (\mathbf{x}),\mathbf{z}_{\mathrm{por}}(\mathbf{x})] 
  :&= \int_{\mathcal{K}_{\mathrm{free}}}
  \Phi_{\mathrm{free}}[\mathbf{z}_{\mathrm{free}}(\mathbf{x})]
  \; \mathrm{d} \Omega 
  + \int_{\mathcal{K}_{\mathrm{por}}}
    \Phi_{\mathrm{por}}[\mathbf{z}_{\mathrm{por}}(\mathbf{x})]
    \; \mathrm{d} \Omega \notag \\
  &+ \int_{\Gamma_{\mathrm{int}}} \Psi[\overset{*}{\mathbf{z}}_{\mathrm{free}}(\mathbf{x}),
    \overset{*}{\mathbf{z}}_{\mathrm{por}}(\mathbf{x}),z_n(\mathbf{x})]
  \; \mathrm{d} \Gamma
  \nonumber \\
  &- \int_{\mathcal{K}_{\mathrm{free}}} \gamma \mathbf{b}_{\mathrm{free}}(\mathbf{x})
  \cdot \mathbf{z}_{\mathrm{free}}(\mathbf{x}) \; \mathrm{d}\Omega 
  -\int_{\Gamma^{t}_{\mathrm{free}}} \mathbf{t}^{\mathrm{p}}_{\mathrm{free}}(\mathbf{x})
  \cdot \mathbf{z}_{\mathrm{free}}(\mathbf{x}) \; \mathrm{d}\Gamma \nonumber\\
  &- \int_{\mathcal{K}_{\mathrm{por}}} \gamma \phi_{\mathrm{por}}(\mathbf{x}) 
  \mathbf{b}_{\mathrm{por}}(\mathbf{x})
  \cdot \mathbf{z}_{\mathrm{por}}(\mathbf{x}) \; \mathrm{d}\Omega 
  -\int_{\Gamma^{t}_{\mathrm{por}}} \mathbf{t}^{\mathrm{p}}_{\mathrm{por}}(\mathbf{x})
  \cdot \mathbf{z}_{\mathrm{por}}(\mathbf{x}) \; \mathrm{d}\Gamma
\end{align}
where $\mathbf{z}_{\mathrm{free}}:\mathcal{K}_{\mathrm{free}}
\rightarrow \mathbb{R}^{nd}$ and $\mathbf{z}_{\mathrm{por}}:
\mathcal{K}_{\mathrm{por}} \rightarrow \mathbb{R}^{nd}$ are
vector fields;
${\mathop{\mathbf{z}}^{*}}_{\mathrm{free}}$ and
${\mathop{\mathbf{z}}^{*}}_{\mathrm{por}}$
denote, respectively the tangential
components of $\mathbf{z}_{\mathrm{free}}$
and $\mathbf{z}_{\mathrm{por}}$; and 
\[
z_n(\mathbf{x}) := \mathbf{z}_{\mathrm{free}}(\mathbf{x})
\cdot \widehat{\mathbf{n}}_{\mathrm{free}}(\mathbf{x})
\]
We then establish the following result with a
proof provided in Appendix \ref{Sec:ICs_Supplementary}. 

%====================================================;
%  Theorem: Minimum power theorem for coupled flows  ; 
%----------------------------------------------------;
\begin{theorem}[Minimum power theorem for coupled flows]
  For the class of coupled flows satisfying
  the requirements (R1)--(R2), any pair of
  kinematically admissible vector fields
  $(\widetilde{\mathbf{v}}_{\mathrm{free}}(\mathbf{x}),
    \widetilde{\mathbf{v}}_{\mathrm{por}}(\mathbf{x}))$ satisfies  
    %----------------------------------------------;
    %  Equation: Total mechanical power principle  ;
    %----------------------------------------------;
    \begin{align}
      \label{Eqn:Minimum_total_mechanical_power_inequality}
      \mathcal{P}_{\mathrm{coupled}}[\mathbf{v}_{\mathrm{free}}(\mathbf{x}),
        \mathbf{v}_{\mathrm{por}}(\mathbf{x})]
      \leq
      \mathcal{P}_{\mathrm{coupled}}[\widetilde{\mathbf{v}}_{\mathrm{free}}
        (\mathbf{x}),\widetilde{\mathbf{v}}_{\mathrm{por}}(\mathbf{x})]
    \end{align}
    in which $\mathbf{v}_{\mathrm{free}}(\mathbf{x})$
    is the velocity field in the free flow region
    and $\mathbf{v}_{\mathrm{por}}(\mathbf{x})$ is
    the velocity field in the porous region. 
\end{theorem}

%************************************************;
%                                                ;
%  NAME                                          ;
%    S7_ICs_Uniqueness                	         ;
%                                                ;
%  WRITTEN BY                                    ;
%    Kalyana Babu Nakshatrala   		 ;
%    Mohammad S. Joshaghani                      ;
%                                                ;
%************************************************;
\section{UNIQUENESS OF SOLUTIONS}
\label{Sec:S7_ICs_Uniqueness}
We will use the minimum power theorem
to establish the uniqueness of solutions under
the proposed interface conditions. For brevity,
we will show for the case of coupled
Stokes-Darcy-Brinkman equations; however, with
straightforward alterations, one can show for
the case of Darcy equations coupled with the
Stokes equations.
We establish the uniqueness under the
following functional form for $\Psi$,
which is (slightly) more general than
the one considered in
\textbf{\S\ref{Sec:S5_ICs_Special_cases}}:
%-------------------------------------------;
%  Equation: Psi functional for uniqueness  ;
%-------------------------------------------;
\begin{align}
  \label{Eqn:ICs_Psi_functional_uniqueness}
  \Psi[{\mathop{\mathbf{v}}^{*}}_{\mathrm{free}}(\mathbf{x}),
    {\mathop{\mathbf{v}}^{*}}_{\mathrm{por}}(\mathbf{x}),
  v_n(\mathbf{x})] = \alpha_{11}
      {\mathop{\mathbf{v}}^{*}}_{\mathrm{free}}(\mathbf{x})
      \cdot {\mathop{\mathbf{v}}^{*}}_{\mathrm{free}}(\mathbf{x})
      &+ 2 \alpha_{12}
      {\mathop{\mathbf{v}}^{*}}_{\mathrm{free}}(\mathbf{x})
      \cdot {\mathop{\mathbf{v}}^{*}}_{\mathrm{por}}(\mathbf{x})
      \notag \\
      &+ \alpha_{22}
      {\mathop{\mathbf{v}}^{*}}_{\mathrm{por}}(\mathbf{x})
      \cdot {\mathop{\mathbf{v}}^{*}}_{\mathrm{por}}(\mathbf{x})
      + \beta v_{n}(\mathbf{x}) \cdot v_{n}(\mathbf{x}) 
\end{align}
with
%--------------------------;
%  Equation: Coefficients  ;
%--------------------------;
\begin{align}
  \alpha_{11} \alpha_{22} \geq \alpha_{12}^2
  \quad \mathrm{and} \quad \beta \geq 0 
\end{align}
To establish uniqueness under more general
conditions (e.g., a more general functional
form for $\Psi$), one needs to resort to
techniques from functional analysis, which
is beyond the scope of this paper. We establish
the following theorem with a proof provided in 
Appendix \ref{Sec:ICs_Supplementary}. 

%=======================;
%  Theorem: Uniqueness  ; 
%-----------------------;
\begin{theorem}[Uniqueness]
  \label{Thm:ICs_uniqueness_theorem}
  Under the prescribed data given by $\mathbf{b}_{\mathrm{free}}
  (\mathbf{x})$, $\mathbf{b}_{\mathrm{por}}(\mathbf{x})$, 
  $\mathbf{v}_{\mathrm{free}}^{\mathrm{p}}(\mathbf{x})$,
  $\mathbf{v}_{\mathrm{por}}^{\mathrm{p}}(\mathbf{x})$, 
  $\mathbf{t}_{\mathrm{free}}^{\mathrm{p}}(\mathbf{x})$
  and $\mathbf{t}_{\mathrm{por}}^{\mathrm{p}}(\mathbf{x})$;
  and under $\Psi$ given by equation
  \eqref{Eqn:ICs_Psi_functional_uniqueness}; 
  the solution to the coupled Stokes-Darcy-Brinkman
  equations is unique up to an arbitrary constant
  for the pressures. 
\end{theorem}

%*****************************************************;
%                                                     ;
%  NAME                                               ;
%    S8_ICs_CR.tex                          	      ;
%                                                     ;
%  WRITTEN BY                                         ;
%    Kalyana Babu Nakshatrala                         ;
%    Mohammad S. Joshaghani                           ;
%                                                     ;
%*****************************************************;
\section{CONCLUDING REMARKS}
\label{Sec:S8_ICs_CR}
We have considered the flows of incompressible fluids
in coupled free-porous media. We have presented a
theoretical framework to obtain a complete set of
self-consistent conditions, which describes the flow
dynamics at the interface of free flow and porous
regions. The interface conditions are essential
for the closure of the mathematical model. 
The framework is primarily built upon the
principle of virtual power, theory of interacting
continua, and a geometric argument for
enforcing internal constraints, which
in our case is the incompressibility
of the fluid. 
The central idea in the proposed principle
of virtual power is to account for the power
expended at the interface and thereby making it
possible to circumvent the need to estimate
the partial stress in the porous solid.  

Under the proposed framework, the set of interface
conditions is a combination of jump conditions and
a constitutive specification, which is provided by
prescribing the physically meaningful power expended
density at the interface.
We have also shown that the jump conditions
by themselves do not provide a workable set
of conditions, which is because of the
inability to quantify the traction taken
by the rigid porous solid under the theory
of interacting continua. 
The salient features of the proposed
framework of obtaining interface
conditions are:
(i) The framework enjoys a strong theoretical
  underpinning. 
(ii) The resulting interface conditions
  make the resulting mathematical model
  well-posed. Specifically, we have shown
  that the resulting mathematical model
  has a unique solution.
  (iii) The framework is amenable to generalizations,
  and the resulting interface conditions are valid
  for a wide variety of porous media models.
  (iv) Several popular conditions in the literature
  are special cases of the proposed framework.
  (v) Similar
  to uncoupled free flows and uncoupled
  flows in porous media, the flows in
  coupled free-porous media under the
  proposed interface conditions also enjoy
  a minimum power theorem.
  
  % It is possible to construct a weak formulation 
  % based on the minimum power theorem, which 
  % can be utilized in a numerical implementation
  % of the mathematical model under the finite
  % element method. Assessing the efficacy of
  % such a weak formulation is worthy of a
  % future investigation. 
  
  \emph{In closure, the proposed principle of
    virtual power for flows in coupled free-porous
    media encapsulates the balance of linear momentum,
    the balance of angular momentum, internal constraints,
    Cauchy's fundamental theorem for the stress, and
    interface conditions!}

\appendix

%*****************************************************;
%                                                     ;
%  NAME                                               ;
%    ICs_Appendix_JCs.tex                            ;
%                                                     ;
%  WRITTEN BY                                         ;
%      Kalyana Babu Nakshatrala  	              ;
%                                                     ;
%*****************************************************;
\section{ON JUMP CONDITIONS}
\label{Sec:App_ICs_JCs}
It can be tempting to treat the interface
as a singular surface, obtain the jump
conditions across the singular surface
and consider them as an appropriate set
of interface conditions.
We will now show why the jump conditions
will not render a useful set of conditions
at the interface for flows in coupled
free-porous media, especially when the
porous solid is assumed to be rigid. 

The jump conditions (which are the balance
laws across a singular surface) in the
context of a single constituent can be
found in many standard texts on continuum
mechanics (e.g., \citep{2012_Chadwick_Book,
  liu2013continuum}).
But the problem central to this paper
involves a porous medium, which is not
a single constituent.
A jump condition for a mixture (i.e., a 
continuum with multiple constituents) 
will be a bit more than the balance laws,
as one need to make additional assumptions 
on defining quantities for the mixture on 
the whole in terms of the corresponding 
quantities of its constituents. 
We first present the jump conditions in
the most familiar setting of a single
constituent and then extend to the
case of multiple constituents using 
TIC. 
Only the jump conditions pertaining to
the balance of mass and the balance of
linear momentum will be relevant here.

%------------------------------;
%  Figure 4: singular surface  ;
%------------------------------;
\begin{figure}
  \centering
  \includegraphics[scale=0.8]{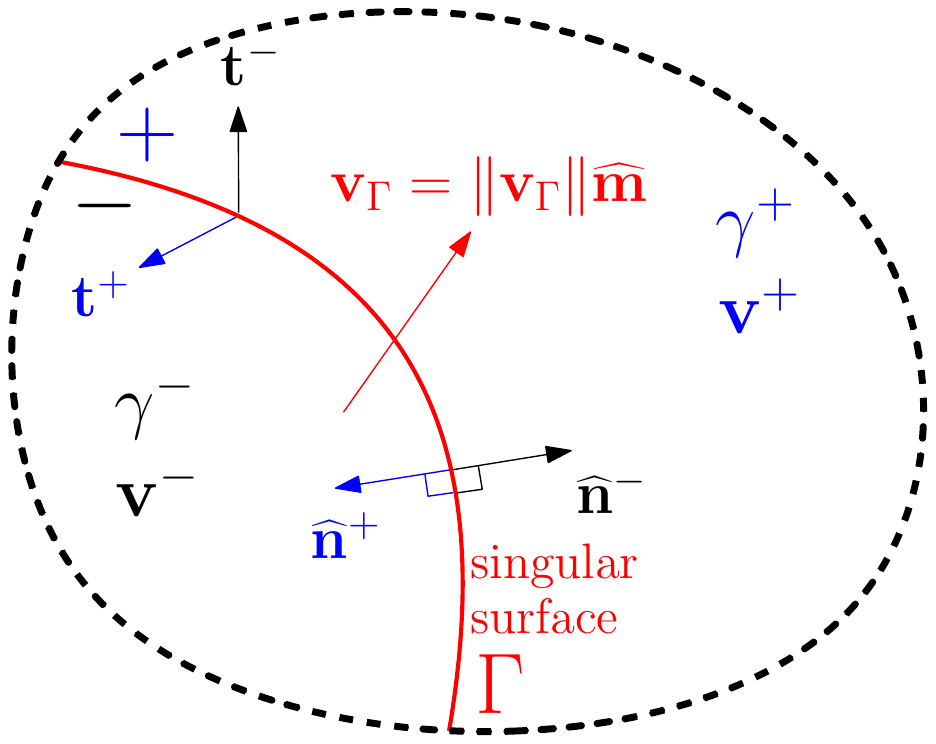}
  \caption{A singular surface $\Gamma$ evolves
    with a velocity $\mathbf{v}_{\Gamma}$ along
    the direction given by the unit vector
    $\widehat{\mathbf{m}}$. The regions on the
    either side of $\Gamma$ are
    denoted by ``$+$'' and ``$-$''.
    The corresponding quantities are
    denoted using these signs as
    superscripts. The tractions are
    denoted by $\mathbf{t}^{+}$ and
    $\mathbf{t}^{-}$, and the unit
    outward normals are denoted by
    $\widehat{\mathbf{n}}^{+}$ and
    $\widehat{\mathbf{n}}^{-}$. The
    jump conditions are balance laws
    across such a singular surface. 
    \label{Fig:Jump_conditions}}
\end{figure}

Consider a singular surface $\Gamma$
which evolves with a velocity vector
$\mathbf{v}_{\Gamma}$. The regions on
the either sides of $\Gamma$ and the
corresponding quantities are indicated
by ``$+$'' and ``$-$'' (see
\textbf{Fig.~\ref{Fig:Jump_conditions}}). 
The velocity vector of the interface,
in general, need not be along the
normal to the interface. That is,
the unit vector $\widehat{\mathbf{m}}$
need not be parallel to
$\widehat{\mathbf{n}}^{+}$
or $\widehat{\mathbf{n}}^{-}$. However,
only the normal component of the interface
velocity manifests in the jump conditions.
To this end, without loss of generality, we
define the normal component of the interface
velocity as follows:
%-----------------------------------------;
%  Equation: Normal component of v_Gamma  ;
%-----------------------------------------;
\begin{align}
  \mathbb{V}_{\Gamma} := \mathbf{v}_{\Gamma}
  \cdot \widehat{\mathbf{n}}^{-}
\end{align}
We define the jump operator acting
on a quantity $\eta$ as follows:
%---------------------------;
%  Equation: Jump operator  ;
%---------------------------;
\begin{align}
  \llbracket \eta \rrbracket =
  \eta^{+} - \eta^{-}
\end{align}

%============================================;
%  Subsection: JCs for a single constituent  ;
%============================================;
\subsection{Jump conditions for a single constituent}
The jump condition for the balance
of mass across $\Gamma$ reads:
%--------------------------------------------------------;
%  Equation: Jump condition for single constituent BoM  ; 
%--------------------------------------------------------;
\begin{align}
  \llbracket \gamma (\mathbb{V}_{\Gamma}
  - \mathbf{v} \cdot \widehat{\mathbf{n}}
  )\rrbracket = 0
\end{align}
which when expanded reads as follows: 
%---------------------------------------------;
%  Equation: Expanded BoM single constituent  ;
%---------------------------------------------;
\begin{align}
  (\gamma^{+} - \gamma^{-}) \mathbb{V}_{\Gamma}
  + \gamma^{+} \mathbf{v}^{+} \cdot
  \widehat{\mathbf{n}}^{+} 
  + \gamma^{-} \mathbf{v}^{-} \cdot
  \widehat{\mathbf{n}}^{-} = 0 
\end{align}
The jump condition for the balance of
linear momentum across $\Gamma$ reads:
%---------------------------------------------------------;
%  Equation: Jump condition for single constituent BoLM  ; 
%---------------------------------------------------------;
\begin{align}
  \llbracket \gamma (\mathbb{V}_{\Gamma}
  - \mathbf{v} \cdot \widehat{\mathbf{n}})
  \mathbf{v} \rrbracket
  + \mathbf{t}^{-} + \mathbf{t}^{+} = \mathbf{0} 
\end{align}
where $\mathbf{t}^{-}$ and $\mathbf{t}^{+}$
denote the tractions on the either side
of the singular surface. 

%=====================================;
%  Subsection: Multiple constituents  ;
%=====================================;
\subsection{Multiple constituents}
For the coupled free-porous media, we associate,
without loss of generality, the ``$-$'' region
with the free flow region and the ``$+$'' region
with the porous region. 
The jump condition for the balance of the mass 
for the fluid takes the following form:
%-------------------------------------------------;
%  Equation: Jump conditions for balance of mass  ;
%-------------------------------------------------;
\begin{align}
  \label{Eqn:Jump_conditions_BoM}
  (\gamma_{\mathrm{por}} - \gamma_{\mathrm{free}})\mathbb{V}_{\Gamma}
  + \gamma_{\mathrm{free}} v_{\mathrm{free}}^{(n)}(\mathbf{x}) 
  + \gamma_{\mathrm{por}} v_{\mathrm{por}}^{(n)}(\mathbf{x}) 
  = 0 
\end{align} 
where $v^{(n)}_{\mathrm{por}}$ is the normal 
component of the discharge velocity, which is 
equal to the product of the (surface) porosity 
and the seepage velocity. 

\begin{figure}
  \centering
  \includegraphics[scale=0.72]{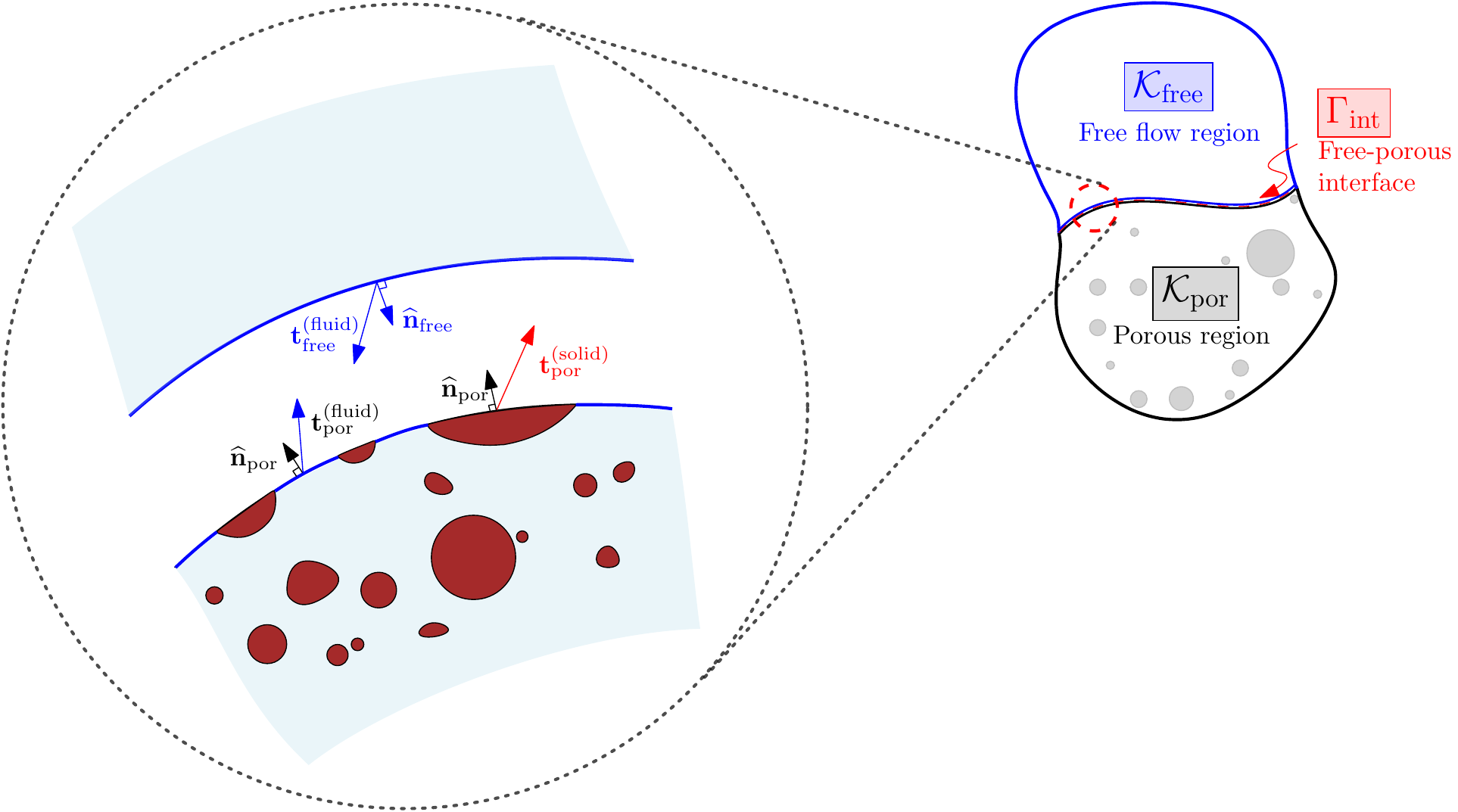}
  \caption{A conceptual visualization of the tractions
    at the interface of free-porous media. The
    interface traction of free flow domain is
    denoted by $\mathbf{t}_{\mathrm{free}}$ and
    the interface traction of fluid and solid
    constituents of porous domain are, respectively,
    denoted by $\mathbf{t}_{\mathrm{por}}^{(\mathrm{fluid})}$
    and $\mathbf{t}_{\mathrm{por}}^{(\mathrm{solid})}$.
    \label{Fig:Tractions_at_interface}}
\end{figure}

In order to write the jump condition
for the balance of linear momentum,
the multi-constituent nature of the
porous medium needs to be considered
and an additional assumption on the
total traction of the mixture needs
to be made.
Even in the simplest case as considered in
this paper, a porous medium consists of
two constituents; one of them being the
porous solid and the other one is the fluid
in the pores.
Although different definitions are employed
under TIC to define a quantity of a mixture in terms
of the corresponding quantities of its
constituents \citep{hansen1991some}, 
it is however common to assume that the total
traction of a mixture is the sum of the
partial tractions of its constituents.
Thus, the total traction in the porous
medium (consisting of a fluid and a solid
constituents) is taken as  
\[
\mathbf{t}_{\mathrm{por}}^{\mathrm{(fluid)}} +
\mathbf{t}_{\mathrm{por}}^{\mathrm{(solid)}} 
\]
where $\mathbf{t}_{\mathrm{por}}^{\mathrm{(fluid)}}$
and $\mathbf{t}_{\mathrm{por}}^{\mathrm{(solid)}}$
are, respectively, the partial tractions
in the fluid and solid constituents; see Figure \ref{Fig:Tractions_at_interface}.
The jump condition for the balance of
linear momentum for the entire mixture
(i.e., all the constituents) across
$\Gamma$ can be written as follows:
%---------------------------------;
%  Equation: BoLM jump condition  ;
%---------------------------------;
\begin{align}
  \label{Eqn:ICS_JC_BoLM_mixture_general}
  \gamma_{\mathrm{free}} \left(\mathbb{V}_{\Gamma}
  - \mathbf{v}_{\mathrm{free}} 
  \cdot \widehat{\mathbf{n}}_{\mathrm{free}}\right)
  \mathbf{v}_{\mathrm{free}}
  + \gamma_{\mathrm{por}} \left(\mathbb{V}_{\Gamma}
  - \mathbf{v}_{\mathrm{por}} 
  \cdot \widehat{\mathbf{n}}_{\mathrm{por}}\right)
  \mathbf{v}_{\mathrm{por}}
  + \mathbf{t}_{\mathrm{free}}
  + \left(\mathbf{t}_{\mathrm{por}}^{\mathrm{(fluid)}} +
  \mathbf{t}_{\mathrm{por}}^{\mathrm{(solid)}}\right)
  = \mathbf{0}
\end{align}

We now specialize to the case when the
singular surface is stationary (which
implies $\mathbb{V}_{\Gamma} = 0$) and
the true density of the fluid across
the singular surface
is the same (i.e., $\gamma_{\mathrm{free}}
= \gamma_{\mathrm{por}}$). The jump conditions
for the balance of mass and the balance of
linear momentum can be compactly written
as follows: 
%---------------------------------------------;
%  Equation: Complete set of jump conditions  ;
%---------------------------------------------;
\begin{subequations}
  \begin{align}
    \label{Eqn:ICs_JC_BoM_final}
    &v_{\mathrm{free}}^{(n)}(\mathbf{x}) + 
    v_{\mathrm{por}}^{(n)}(\mathbf{x}) = 0 \\
    \label{Eqn:ICs_JC_BoLM_normal_final}
    &\mathbf{t}_{\mathrm{free}} \cdot
    \widehat{\mathbf{n}}_{\mathrm{free}}
    = \left(\mathbf{t}_{\mathrm{por}}^{\mathrm{(fluid)}} +
    \mathbf{t}_{\mathrm{por}}^{\mathrm{(solid)}}\right)
    \cdot \widehat{\mathbf{n}}_{\mathrm{por}} \\
    \label{Eqn:ICs_JC_BoLM_tangential_final}
    &\gamma v_n (\mathbf{v}_{\mathrm{free}} -
    \mathbf{v}_{\mathrm{por}})\cdot \widehat{\mathbf{s}}
    = \mathbf{t}_{\mathrm{free}} \cdot
    \widehat{\mathbf{s}}
    + \left(\mathbf{t}_{\mathrm{por}}^{\mathrm{(fluid)}} +
    \mathbf{t}_{\mathrm{por}}^{\mathrm{(solid)}}\right)
    \cdot \widehat{\mathbf{s}}
  \end{align}
\end{subequations}
Equations \eqref{Eqn:ICs_JC_BoLM_normal_final}
and \eqref{Eqn:ICs_JC_BoLM_tangential_final}
are, respectively, the normal and tangential
components of equation \eqref{Eqn:ICS_JC_BoLM_mixture_general}.
Equation \eqref{Eqn:ICs_JC_BoM_final} has been
invoked in obtaining equations \eqref{Eqn:ICs_JC_BoLM_normal_final}
and \eqref{Eqn:ICs_JC_BoLM_tangential_final}.

%==========================;
%  Subsection: Discussion  ;
%==========================;
\subsection{Discussion} 
We now compare the above set of three jump conditions
with the set of four interface conditions
\eqref{Eqn:vn_jump_condition}--\eqref{Eqn:tangential_traction_porous}.
The following are the similarities and 
the notable differences:
\begin{enumerate}[(a)]
\item The jump condition pertaining to the
  balance of mass \eqref{Eqn:ICs_JC_BoM_final}
  is exactly the same as the first interface
  condition \eqref{Eqn:vn_jump_condition}, which
  is the reason why we mentioned earlier that
  the interface condition
  \eqref{Eqn:vn_jump_condition} stems from
  the jump conditions.
\item There is only one jump condition
  involving the tangential part of the
  tractions. On the other hand, two interface
  conditions are related to the tangential
  components of the tractions.
\item The jump conditions
  \eqref{Eqn:ICs_JC_BoLM_normal_final}--\eqref{Eqn:ICs_JC_BoLM_tangential_final}
  involve $\mathbf{t}^{\mathrm{(solid)}}_{\mathrm{por}}$
  but the interface conditions \eqref{Eqn:tangential_traction_free}--\eqref{Eqn:tangential_traction_porous}
  involve
  the functional $\Psi$ instead.
\end{enumerate}
    
Let us now focus on equation \eqref{Eqn:ICs_JC_BoLM_normal_final}.
The total traction in the porous medium is
distributed among these two constituents:
the porous solid and the fluid in the pores.
If the porous solid is rigid, one cannot
estimate what part of the total traction
is taken up by the porous solid, and hence
one will not be able to find the traction
taken by the fluid in the pores. 
A similar case exists even with the condition
\eqref{Eqn:ICs_JC_BoLM_tangential_final}.
Thus, the jump condition related to the
balance of linear momentum does not provide
a workable condition. 
This type of difficulty (i.e., finding the partial 
tractions of the individual constituents from 
the total traction) is inherent to porous media 
models which are based on TIC and is not just limited 
to the case when one of the constituents 
is rigid \citep{rajagopal1995mechanics}.

Since we do not deal with the partial traction of the 
porous solid in the rest of this paper, our usage 
$\mathbf{t}_{\mathrm{por}}$ (instead of 
$\mathbf{t}_{\mathrm{por}}^{\mathrm{(fluid)}}$) in 
the main text to 
denote
the partial traction of the fluid in the porous region 
should not cause any confusion. Similarly, 
$\mathbf{t}_{\mathrm{por}}^{\mathrm{p}}$ will be
used to denote the prescribed traction for
the fluid in the porous region. 

%------------------------------------------------------;
%  Figure 5: Jump conditions for interface conditions  ;
%------------------------------------------------------;

%*****************************************************;
%                                                     ;
%  NAME                                               ;
%    ICs_Appendix_Cauchy_theorem.tex                    ;
%                                                     ;
%  WRITTEN BY                                         ;
%      Kalyana Babu Nakshatrala  	              ;
%                                                     ;
%*****************************************************;
\section{RECOVERING CAUCHY'S FUNDAMENTAL THEOREM}
\label{App:ICs_Cauchy_theorem}
To recover the Cauchy's fundamental theorem for the stress, 
one need to enforce the balance of virtual power on arbitrary 
subsets of the domain. To this end, we replace statement
(P1) in the principle of virtual power
\eqref{Eqn:ICs_proposed_PVP_P1} with the following: 
%----------------------------------;
%  Equation: P1' statement of PVP  ;
%----------------------------------;
\begin{align}
  \mathrm{(P1')} \qquad \qquad \qquad 
  \mathscr{P}^{\mathrm{(internal)}}(\mathcal{B})
  = \mathscr{P}^{\mathrm{(external)}}(\mathcal{B})
  \qquad \forall (\mathbf{w}_{\mathrm{free}},
  \mathbf{w}_{\mathrm{por}}) \in
  \widetilde{\mathcal{W}}
  \; \mathrm{and} \;
  \forall \mathcal{B} \subseteq \Omega 
\end{align}
where $\mathcal{B}$ is an arbitrary
subset of the domain $\Omega$ and 
%-------------------------------------;
%  Equation: Internal power expended  ;
%-------------------------------------;
\begin{align}
  \mathscr{P}^{\mathrm{(internal)}}(\mathcal{B})
  &:= \int_{\mathcal{K}_{\mathrm{free}} \cap \mathcal{B}}
  \mathbf{T}_{\mathrm{free}} \cdot
  \mathrm{grad}[\mathbf{w}_{\mathrm{free}}] 
  \; \mathrm{d} \Omega 
  + \int_{\mathcal{K}_{\mathrm{por}} \cap \mathcal{B}}
  \mathbf{T}_{\mathrm{por}} \cdot
  \mathrm{grad}[\mathbf{w}_{\mathrm{por}}] 
  \; \mathrm{d} \Omega \notag \\
  &+ \int_{\mathcal{K}_{\mathrm{por}} \cap \mathcal{B}}
  \mathbf{i}_{\mathrm{por}} \cdot
  \mathbf{w}_{\mathrm{por}} 
  \; \mathrm{d} \Omega 
  + \int_{\Gamma_{\mathrm{int}} \cap \mathcal{B}}
  \delta \Psi \; \mathrm{d} \Gamma
\end{align}
%-------------------------------------;
%  Equation: External power expended  ;
%-------------------------------------;
\begin{align}
  \mathscr{P}^{\mathrm{(external)}}(\mathcal{B}) 
  &:= \int_{\partial \mathcal{K}_{\mathrm{free}} \cap \mathcal{B}}
  \mathbf{t}_{\mathrm{free}} \cdot \mathbf{w}_{\mathrm{free}} 
  \; \mathrm{d} \Gamma
  + \int_{\mathcal{K}_{\mathrm{free}} \cap \mathcal{B}}
  \gamma \mathbf{b}_{\mathrm{free}} \cdot
  \mathbf{w}_{\mathrm{free}} 
  \; \mathrm{d} \Omega \notag \\
  &+ \int_{\partial \mathcal{K}_{\mathrm{por}} \cap \mathcal{B}}
  \mathbf{t}_{\mathrm{por}} \cdot \mathbf{w}_{\mathrm{por}} 
  \; \mathrm{d} \Gamma
  + \int_{\mathcal{K}_{\mathrm{por}} \cap \mathcal{B}}
  \gamma \phi_{\mathrm{por}}\mathbf{b}_{\mathrm{por}}
  \cdot \mathbf{w}_{\mathrm{por}} 
  \mathrm{d} \Omega 
\end{align}
In the above expression, $\mathbf{t}_{\mathrm{free}}$
and $\mathbf{t}_{\mathrm{por}}$ denote the tractions,
respectively, on $\partial \mathcal{K}_{\mathrm{free}}$
and $\partial \mathcal{K}_{\mathrm{por}}$.

By taking the subset $\mathcal{B}$ to be
entirely within $\mathcal{K}_{\mathrm{free}}$
and by using a similar approach taken in
the previous sections (e.g., Green's identity,
the fundamental lemma of calculus of variations),
one can establish:
%------------------------------------;
%  Equation: Cauchy theorem in K_free  ;
%------------------------------------;
\begin{align}
  \label{Eqn:ICs_Cauchy_theorem_Kfree}
  \mathbf{t}_{\mathrm{free}}(\mathbf{x}) =
  \mathbf{T}_{\mathrm{free}}(\mathbf{x})
  \widehat{\mathbf{n}}(\mathbf{x})
\end{align}
on any surface in the free flow region 
($\mathcal{K}_{\mathrm{free}} \cup \partial 
\mathcal{K}_{\mathrm{free}}$) 
with the unit outward normal 
$\widehat{\mathbf{n}}(\mathbf{x})$.
Similarly, by taking the subset $\mathcal{B}$
to be entirely within $\mathcal{K}_{\mathrm{por}}$,
one can establish:
%-----------------------------------;
%  Equation: Cauchy theorem in K_por  ;
%-----------------------------------;
\begin{align}
  \label{Eqn:ICs_Cauchy_theorem_Kpor}
  \mathbf{t}_{\mathrm{por}}(\mathbf{x}) =
  \mathbf{T}_{\mathrm{por}}(\mathbf{x})
  \widehat{\mathbf{n}}(\mathbf{x})
\end{align}
on any surface in the porous region 
($\mathcal{K}_{\mathrm{por}} \cup \partial 
\mathcal{K}_{\mathrm{por}}$) 
with the unit outward normal 
$\widehat{\mathbf{n}}(\mathbf{x})$.
The relations \eqref{Eqn:ICs_Cauchy_theorem_Kfree}
and \eqref{Eqn:ICs_Cauchy_theorem_Kpor}, respectively, 
represent the Cauchy's fundamental theorem for the stress 
for the free flow region and the porous region. 
Using the traction-stress relations,
the second interface condition
\eqref{Eqn:normal_traction_continuity}
takes the following more familiar form: 
%-------------------------------------------------;
%  Equation: Modified normal interface condition  ;
%-------------------------------------------------;
\begin{align}
  \mathbf{t}_{\mathrm{free}}(\mathbf{x}) \cdot
  \widehat{\mathbf{n}}_{\mathrm{free}}(\mathbf{x})
  + \frac{\partial \Psi}{\partial v_n}
  = \mathbf{t}_{\mathrm{por}}(\mathbf{x}) 
  \cdot \widehat{\mathbf{n}}_{\mathrm{por}}(\mathbf{x})
  \qquad \forall \mathbf{x} \in
  \Gamma_{\mathrm{int}}
\end{align}

%%============================;
%%  Section: Acknowledgments  ;
%%============================;
\section*{ACKNOWLEDGMENTS}
The authors acknowledge the support through the 
\emph{High Priority Area Research Seed Grant} 
from the Division of Research, University of Houston.

%==============================;
%  Include all the references  ;
%==============================;
\bibliographystyle{plainnat}
\bibliography{References,References.bib}

\clearpage
\newpage
%************************************************;
%                                                ;
%  NAME                                          ;
%    ICs_Supplementary.tex           	         ;
%                                                ;
%  WRITTEN BY                                    ;
%    Kalyana Babu Nakshatrala   		 ;
%    Mohammad S. Joshaghani                      ;
%                                                ;
%************************************************;
\section{MATHEMATICAL PROOFS}
\label{Sec:ICs_Supplementary}

%==============================;
%  Subsection: A proof of MPT  ;
%==============================;
\subsection{A proof of the minimum power theorem}

Based on the first-order optimality
condition it will suffice to show that
  %--------------------;
  %  Equation: Step 1  ;
  %--------------------;
  \begin{align}
    \delta\mathcal{P}_{\mathrm{coupled}}[\mathbf{v}_{\mathrm{free}},
      \mathbf{v}_{\mathrm{por}};
      \delta\mathbf{v}_{\mathrm{free}},\delta\mathbf{v}_{\mathrm{por}}]
    := \left[\frac{d}{d\epsilon}
      \mathcal{P}_{\mathrm{coupled}}[\mathbf{v}_{\mathrm{free}}
      + \epsilon \delta\mathbf{v}_{\mathrm{free}} 
      ,\mathbf{v}_{\mathrm{por}} + \epsilon
      \delta \mathbf{v}_{\mathrm{por}}]
      \right]_{\epsilon = 0} = 0
    \notag \\
    \forall
    (\delta\mathbf{v}_{\mathrm{free}},\delta\mathbf{v}_{\mathrm{por}})
    \in \mathcal{W}
  \end{align}
  The positive definite Hessians will
  ensure that the extremum is in fact
  a minimum. The G\^ateaux variation
  can be written as
  follows\footnote{$\delta \mathbf{D}_{\mathrm{free}}
    := \frac{1}{2} (\mathrm{grad}[\delta \mathbf{v}_{\mathrm{free}}]
    + \mathrm{grad}[\delta \mathbf{v}_{\mathrm{free}}]^{\mathrm{T}})$
    and
    $\delta \mathbf{D}_{\mathrm{por}}
    := \frac{1}{2} (\mathrm{grad}[\delta \mathbf{v}_{\mathrm{por}}]
    + \mathrm{grad}[\delta \mathbf{v}_{\mathrm{por}}]^{\mathrm{T}})$}:
  %--------------------;
  %  Equation: Step 2  ;
  %--------------------;
  {\small
    \begin{align}
      \delta\mathcal{P}_{\mathrm{coupled}}[\mathbf{v}_{\mathrm{free}},
      \mathbf{v}_{\mathrm{por}};
      \delta\mathbf{v}_{\mathrm{free}},\delta\mathbf{v}_{\mathrm{por}}]
    &= \int_{\mathcal{K}_{\mathrm{free}}}
    \ddfrac{\partial \Phi_{\mathrm{free}}}{\partial \mathbf{D}_{\mathrm{free}}}
    \cdot \delta \mathbf{D}_{\mathrm{free}} \; \mathrm{d} \Omega
    + \int_{\mathcal{K}_{\mathrm{por}}}
    \left(\ddfrac{\partial \Phi_{\mathrm{por}}}{\partial \mathbf{v}_{\mathrm{por}}}
    \cdot \delta \mathbf{v}_{\mathrm{por}} 
    + \ddfrac{\partial \Phi_{\mathrm{por}}}{\partial \mathbf{D}_{\mathrm{por}}}
    \cdot \delta \mathbf{D}_{\mathrm{por}} \right) \mathrm{d} \Omega
    \notag \\
    &+ \int_{\Gamma_{\mathrm{int}}} \left(
    \ddfrac{\partial \Psi}{\partial {\mathop{\mathbf{v}}^{*}}_{\mathrm{free}}}
    \cdot \delta {\mathop{\mathbf{v}}^{*}}_{\mathrm{free}} 
    + \ddfrac{\partial \Psi}{\partial {\mathop{\mathbf{v}}^{*}}_{\mathrm{por}}}
    \cdot \delta {\mathop{\mathbf{v}}^{*}}_{\mathrm{por}} 
    + \ddfrac{\partial \Psi}{\partial v_n}
    \cdot \delta v_n \right) \mathrm{d} \Gamma
    \notag \\
    &- \int_{\mathcal{K}_{\mathrm{free}}}
    \gamma \mathbf{b}_{\mathrm{free}} \cdot
    \delta \mathbf{v}_{\mathrm{free}} \; \mathrm{d} \Omega
    -\int_{\Gamma^{t}_{\mathrm{free}}} \mathbf{t}^{\mathrm{p}}_{\mathrm{free}}(\mathbf{x})
    \cdot \delta \mathbf{v}_{\mathrm{free}}(\mathbf{x}) \; \mathrm{d}\Gamma 
    \notag \\
    &- \int_{\mathcal{K}_{\mathrm{por}}}
    \gamma \phi_{\mathrm{por}} \mathbf{b}_{\mathrm{por}} \cdot
    \delta \mathbf{v}_{\mathrm{por}} \; \mathrm{d} \Omega
    -\int_{\Gamma^{t}_{\mathrm{por}}} \mathbf{t}^{\mathrm{p}}_{\mathrm{por}}(\mathbf{x})
    \cdot \delta \mathbf{v}_{\mathrm{por}}(\mathbf{x}) \; \mathrm{d}\Gamma
  \end{align}}
  Using the conditions
  \eqref{Eqn:ICs_T_free_extra_Phi_free}--\eqref{Eqn:ICs_i_por_Phi_por}
    under the requirement (R1), we obtain the following:
  %--------------------;
  %  Equation: Step 3  ;
  %--------------------;
  {\small
    \begin{align}
      \delta\mathcal{P}_{\mathrm{coupled}}[\mathbf{v}_{\mathrm{free}},
      \mathbf{v}_{\mathrm{por}};
      \delta\mathbf{v}_{\mathrm{free}},\delta\mathbf{v}_{\mathrm{por}}]
    &= \int_{\mathcal{K}_{\mathrm{free}}}
    \mathbf{T}_{\mathrm{free}}^{\mathrm{extra}}
    \cdot \delta \mathbf{D}_{\mathrm{free}}
    \; \mathrm{d} \Omega 
    + \int_{\mathcal{K}_{\mathrm{por}}}
    \left(\mathbf{i}_{\mathrm{por}} \cdot
    \delta \mathbf{v}_{\mathrm{por}} 
    + \mathbf{T}_{\mathrm{por}}^{\mathrm{extra}}
    \cdot \delta \mathbf{D}_{\mathrm{por}} \right) \mathrm{d} \Omega
    \notag \\
    &+ \int_{\Gamma_{\mathrm{int}}} \left(
    \ddfrac{\partial \Psi}{\partial  {\mathop{\mathbf{v}}^{*}}_{\mathrm{free}}}
    \cdot \delta {\mathop{\mathbf{v}}^{*}}_{\mathrm{free}} 
    + \ddfrac{\partial \Psi}{\partial {\mathop{\mathbf{v}}^{*}}_{\mathrm{por}}}
    \cdot \delta {\mathop{\mathbf{v}}^{*}}_{\mathrm{por}} 
    + \ddfrac{\partial \Psi}{\partial v_n}
    \cdot \delta v_n \right) \mathrm{d} \Gamma
    \notag \\
    &- \int_{\mathcal{K}_{\mathrm{free}}}
    \gamma \mathbf{b}_{\mathrm{free}} \cdot
    \delta \mathbf{v}_{\mathrm{free}} \; \mathrm{d} \Omega
    -\int_{\Gamma^{t}_{\mathrm{free}}} \mathbf{t}^{\mathrm{p}}_{\mathrm{free}}(\mathbf{x})
    \cdot \delta \mathbf{v}_{\mathrm{free}}(\mathbf{x}) \; \mathrm{d}\Gamma 
    \notag \\
    &- \int_{\mathcal{K}_{\mathrm{por}}}
    \gamma \phi_{\mathrm{por}} \mathbf{b}_{\mathrm{por}} \cdot
    \delta \mathbf{v}_{\mathrm{por}} \; \mathrm{d} \Omega
    -\int_{\Gamma^{t}_{\mathrm{por}}} \mathbf{t}^{\mathrm{p}}_{\mathrm{por}}(\mathbf{x})
    \cdot \delta \mathbf{v}_{\mathrm{por}}(\mathbf{x}) \; \mathrm{d}\Gamma
  \end{align}}
  Noting the internal constraints
  \eqref{Eqn:BoM_free_region} and
  \eqref{Eqn:BoM_porous_region},
  utilizing the decomposition of
  the Cauchy stresses
  \eqref{Eqn:ICs_decomposition_of_Cauchy},
  and invoking the Green's identity, we
  obtain the following:
  %--------------------;
  %  Equation: Step 4  ;
  %--------------------;
  {\small
    \begin{align}
      \delta\mathcal{P}_{\mathrm{coupled}}[\mathbf{v}_{\mathrm{free}},
        \mathbf{v}_{\mathrm{por}};\delta\mathbf{v}_{\mathrm{free}},
        \delta\mathbf{v}_{\mathrm{por}}] &= 
    -\int_{\mathcal{K}_{\mathrm{free}}} \underbrace{\left(
    \mathrm{div}[\mathbf{T}_{\mathrm{free}}] 
    + \gamma \mathbf{b}_{\mathrm{free}}
    \right)}_{\mbox{= \textbf{0} due to \eqref{Eqn:BoLM_free_region}}}
    \cdot \delta \mathbf{v}_{\mathrm{free}} \; \mathrm{d} \Omega
    \notag \\
    &-\int_{\mathcal{K}_{\mathrm{por}}} \underbrace{\left(
    \mathrm{div}[\mathbf{T}_{\mathrm{por}}]
    + \gamma \phi_{\mathrm{por}} \mathbf{b}_{\mathrm{por}}
    - \mathbf{i}_{\mathrm{por}}
    \right)}_{\mbox{= \textbf{0} due to \eqref{Eqn:BoLM_porous_region}}}
    \cdot \delta \mathbf{v}_{\mathrm{por}} \; \mathrm{d} \Omega
    \notag \\
    &+\int_{\partial \mathcal{K}_{\mathrm{free}}}
    \left(\mathbf{T}_{\mathrm{free}}\widehat{\mathbf{n}}_{\mathrm{free}}\right)
    \cdot \delta \mathbf{v}_{\mathrm{free}}(\mathbf{x}) \; \mathrm{d}\Gamma
    -\int_{\Gamma^{t}_{\mathrm{free}}} \mathbf{t}^{\mathrm{p}}_{\mathrm{free}}(\mathbf{x})
    \cdot \delta \mathbf{v}_{\mathrm{free}}(\mathbf{x}) \; \mathrm{d}\Gamma
    \notag \\
    &+\int_{\partial \mathcal{K}_{\mathrm{por}}}
    \left(\mathbf{T}_{\mathrm{por}}\widehat{\mathbf{n}}_{\mathrm{por}}\right)
    \cdot \delta \mathbf{v}_{\mathrm{por}}(\mathbf{x}) \; \mathrm{d}\Gamma 
    -\int_{\Gamma^{t}_{\mathrm{por}}} \mathbf{t}^{\mathrm{p}}_{\mathrm{por}}(\mathbf{x})
    \cdot \delta \mathbf{v}_{\mathrm{por}}(\mathbf{x}) \; \mathrm{d}\Gamma
    \notag \\
    &+ \int_{\Gamma_{\mathrm{int}}} \left(
    \ddfrac{\partial \Psi}{\partial {\mathop{\mathbf{v}}^{*}}_{\mathrm{free}}}
    \cdot \delta {\mathop{\mathbf{v}}^{*}}_{\mathrm{free}} 
    + \ddfrac{\partial \Psi}{\partial {\mathop{\mathbf{v}}^{*}}_{\mathrm{por}}}
    \cdot \delta {\mathop{\mathbf{v}}^{*}}_{\mathrm{por}} 
    + \ddfrac{\partial \Psi}{\partial v_n}
    \cdot \delta v_n \right) \mathrm{d} \Gamma
  \end{align}
  }
  Noting the decomposition of the boundaries
  $\partial \mathcal{K}_{\mathrm{free}}$ and
  $\partial \mathcal{K}_{\mathrm{por}}$, given
  by equations
  \eqref{Eqn:ICs_decomposition_of_whole_boundary_Kfree}
  and \eqref{Eqn:ICs_decomposition_of_whole_boundary_Kpor},
  we obtain the following: 
  %--------------------;
  %  Equation: Step 5  ;
  %--------------------;
  {\small
    \begin{align}
      \delta\mathcal{P}_{\mathrm{coupled}}[\mathbf{v}_{\mathrm{free}},
        \mathbf{v}_{\mathrm{por}};\delta\mathbf{v}_{\mathrm{free}},
        \delta\mathbf{v}_{\mathrm{por}}] 
    &= \int_{\Gamma_{\mathrm{free}}^{t}}
    \underbrace{\left(\mathbf{T}_{\mathrm{free}}
    \widehat{\mathbf{n}}_{\mathrm{free}}^{\mathrm{ext}}
      - \mathbf{t}^{\mathrm{p}}_{\mathrm{free}}(\mathbf{x})
      \right)}_{\mbox{= \textbf{0} due to \eqref{Eqn:Traction_BC_free_region}}}
    \cdot \delta \mathbf{v}_{\mathrm{free}}(\mathbf{x}) \; \mathrm{d}\Gamma
    +\int_{\Gamma_{\mathrm{free}}^{v}}
    \left(\mathbf{T}_{\mathrm{free}}
    \widehat{\mathbf{n}}_{\mathrm{free}}^{\mathrm{ext}}
    \right)
    \cdot \delta \mathbf{v}_{\mathrm{free}}(\mathbf{x}) \; \mathrm{d}\Gamma
    \notag \\
    &+\int_{\Gamma_{\mathrm{por}}^{t}}
    \underbrace{\left(\mathbf{T}_{\mathrm{por}}
    \widehat{\mathbf{n}}_{\mathrm{por}}^{\mathrm{ext}}
      - \mathbf{t}^{\mathrm{p}}_{\mathrm{por}}(\mathbf{x})
      \right)}_{\mbox{= \textbf{0} due to \eqref{Eqn:Traction_BC_porous_region}}}
    \cdot \delta \mathbf{v}_{\mathrm{por}}(\mathbf{x}) \; \mathrm{d}\Gamma
    +\int_{\Gamma_{\mathrm{por}}^{v}}
    \left(\mathbf{T}_{\mathrm{por}}
    \widehat{\mathbf{n}}_{\mathrm{por}}^{\mathrm{ext}}
    \right)
    \cdot \delta \mathbf{v}_{\mathrm{por}}(\mathbf{x}) \; \mathrm{d}\Gamma
    \notag \\
    &+\int_{\Gamma_{\mathrm{int}}}
    \left(\mathbf{T}_{\mathrm{free}}\widehat{\mathbf{n}}_{\mathrm{free}}\right) 
    \cdot \delta \mathbf{v}_{\mathrm{free}}(\mathbf{x}) \; \mathrm{d}\Gamma 
    +\int_{\Gamma_{\mathrm{int}}}
    \left(\mathbf{T}_{\mathrm{por}}\widehat{\mathbf{n}}_{\mathrm{por}}\right)
    \cdot \delta \mathbf{v}_{\mathrm{por}}(\mathbf{x}) \; \mathrm{d}\Gamma
    \notag \\
    &+ \int_{\Gamma_{\mathrm{int}}} \left(
    \ddfrac{\partial \Psi}{\partial {\mathop{\mathbf{v}}^{*}}_{\mathrm{free}}}
    \cdot \delta {\mathop{\mathbf{v}}^{*}}_{\mathrm{free}} 
    + \ddfrac{\partial \Psi}{\partial {\mathop{\mathbf{v}}^{*}}_{\mathrm{por}}}
    \cdot \delta {\mathop{\mathbf{v}}^{*}}_{\mathrm{por}} 
    + \ddfrac{\partial \Psi}{\partial v_n}
    \cdot \delta v_n \right) \mathrm{d} \Gamma
  \end{align}
  }
  Invoking that $\delta \mathbf{v}_{\mathrm{free}}(\mathbf{x})$
  and $\delta \mathbf{v}_{\mathrm{por}}(\mathbf{x})$,
  respectively, vanish on $\Gamma_{\mathrm{free}}^{v}$
  and $\Gamma_{\mathrm{por}}^{v}$ (see
  \textbf{\S\ref{Subsec:ICs_kinematic_virtual}}), and
  using the first interface condition
  \eqref{Eqn:vn_jump_condition} and the
  notation introduced in
  \eqref{Eqn:ICs_notation_for_vn}, we
  obtain the following: 
  %--------------------;
  %  Equation: Step 6  ;
  %--------------------;
  {\small
    \begin{align}
      \delta\mathcal{P}_{\mathrm{coupled}}[\mathbf{v}_{\mathrm{free}},
        \mathbf{v}_{\mathrm{por}};\delta\mathbf{v}_{\mathrm{free}},
        \delta\mathbf{v}_{\mathrm{por}}] &= 
    \int_{\Gamma_{\mathrm{int}}}
    \left(\widehat{\mathbf{n}}_{\mathrm{free}} \cdot
    \mathbf{T}_{\mathrm{free}}\widehat{\mathbf{n}}_{\mathrm{free}}
    - \widehat{\mathbf{n}}_{\mathrm{por}} \cdot
    \mathbf{T}_{\mathrm{por}}\widehat{\mathbf{n}}_{\mathrm{por}}
    + \ddfrac{\partial \Psi}{\partial v_n}
    \right) \cdot \delta v_{n} \; \mathrm{d}\Gamma
    \notag \\
    &+\int_{\Gamma_{\mathrm{int}}}
    \left(\mathbf{T}_{\mathrm{free}}\widehat{\mathbf{n}}_{\mathrm{free}} 
    + \ddfrac{\partial \Psi}{\partial {\mathop{\mathbf{v}}^{*}}_{\mathrm{free}}}
    \right)\cdot \delta {\mathop{\mathbf{v}}^{*}}_{\mathrm{free}}
    \; \mathrm{d} \Gamma
    \notag \\
    &+ \int_{\Gamma_{\mathrm{int}}}
    \left(\mathbf{T}_{\mathrm{por}} \widehat{\mathbf{n}}_{\mathrm{por}}
    + \ddfrac{\partial \Psi}{\partial {\mathop{\mathbf{v}}^{*}}_{\mathrm{por}}}
    \right) \cdot \delta {\mathop{\mathbf{v}}^{*}}_{\mathrm{por}} 
    \; \mathrm{d} \Gamma
  \end{align}
  }
  Finally, by utilizing the interface
  conditions
  \eqref{Eqn:normal_traction_continuity}--\eqref{Eqn:tangential_traction_porous} we have 
  established that the first variation of $\mathcal{P}_{\mathrm{coupled}}$ vanishes.

%=============================================;
%  Subsection: A proof of uniqueness theorem  ;
%---------------------------------------------;
\subsection{A proof of the uniqueness theorem}
On the contrary, assume that
\[
\{\mathbf{v}_{\mathrm{free}}^{(1)}(\mathbf{x}),
p_{\mathrm{free}}^{(1)}(\mathbf{x}), 
\mathbf{v}_{\mathrm{por}}^{(1)}(\mathbf{x}),
p_{\mathrm{por}}^{(1)}(\mathbf{x})\}
\quad \mathrm{and} \quad 
\{\mathbf{v}_{\mathrm{free}}^{(2)}(\mathbf{x}), 
p_{\mathrm{free}}^{(2)}(\mathbf{x}),
\mathbf{v}_{\mathrm{por}}^{(2)}(\mathbf{x}), 
p_{\mathrm{por}}^{(2)}(\mathbf{x})\}
\]
are two solutions to the coupled
Stokes-Darcy-Brinkman equations
for the prescribed data. 
That is, $\{\mathbf{v}_{\mathrm{free}}^{(1)}(\mathbf{x}),
p_{\mathrm{free}}^{(1)}(\mathbf{x})\}$
  and $\{\mathbf{v}_{\mathrm{free}}^{(2)}(\mathbf{x}),
  p_{\mathrm{free}}^{(2)}(\mathbf{x})\}$ satisfy
  the Stokes equations in $\mathcal{K}_{\mathrm{free}}$, and 
  $\{\mathbf{v}_{\mathrm{por}}^{(1)}(\mathbf{x}), 
  p_{\mathrm{por}}^{(1)}(\mathbf{x})\}$ and 
  $\{\mathbf{v}_{\mathrm{por}}^{(2)}(\mathbf{x}), 
  p_{\mathrm{por}}^{(2)}(\mathbf{x})\}$ 
  satisfy the Darcy-Brinkman equations
  in $\mathcal{K}_{\mathrm{por}}$.
  Moreover, $\mathbf{v}_{\mathrm{free}}^{(1)}$,
  $\mathbf{v}_{\mathrm{free}}^{(2)}$,
  $\mathbf{v}_{\mathrm{por}}^{(1)}$ and
  $\mathbf{v}_{\mathrm{free}}^{(2)}$ satisfy
  \begin{align}
    \label{Eqn:divergence_free_for_free}
    &\mathrm{div}\left[\mathbf{v}_{\mathrm{free}}^{(1)}\right] = 0
    \quad \mathrm{and} \quad 
    \mathrm{div}\left[\mathbf{v}_{\mathrm{free}}^{(2)}\right] = 0
    \quad \mathrm{in} \; \mathcal{K}_{\mathrm{free}} \\
    \label{Eqn:divergence_free_for_por}
    &\mathrm{div}\left[\mathbf{v}_{\mathrm{por}}^{(1)}\right] = 0
    \quad \mathrm{and} \quad 
    \mathrm{div}\left[\mathbf{v}_{\mathrm{por}}^{(2)}\right] = 0
    \quad \mathrm{in} \; \mathcal{K}_{\mathrm{por}}
  \end{align}
  
  %----------------------------------;
  %  Kinematically admissible pairs  ;
  %----------------------------------;
  Since the pairs $\{\mathbf{v}_{\mathrm{free}}^{(1)}(\mathbf{x}),
  \mathbf{v}_{\mathrm{por}}^{(1)}(\mathbf{x})\}$ and 
  $\{\mathbf{v}_{\mathrm{free}}^{(2)}(\mathbf{x}),
  \mathbf{v}_{\mathrm{por}}^{(2)}(\mathbf{x})\}$ are 
  both kinematically admissible, the minimum power 
  theorem implies that:
  %-----------------------------------;
  %  Equation: Equality of P_coupled  ;
  %-----------------------------------;
  \begin{align}
    \label{Eqn:equality_of_P_coupled}
    \mathcal{P}_{\mathrm{coupled}}
    \left[\mathbf{v}_{\mathrm{free}}^{(1)}(\mathbf{x}), 
      \mathbf{v}_{\mathrm{por}}^{(1)}(\mathbf{x})\right]
    = \mathcal{P}_{\mathrm{coupled}}
    \left[\mathbf{v}_{\mathrm{free}}^{(2)}(\mathbf{x}),
      \mathbf{v}_{\mathrm{por}}^{(2)}(\mathbf{x})\right] 
  \end{align}
  Using the definition of
  $\mathcal{P}_{\mathrm{coupled}}$ given by equation
  \eqref{Eqn:Minimum_total_mechanical_power_Statement},
  the above equation can be expanded as follows:
  %-------------------------------------------------;
  %  Equation: Main statement with P_coupled terms  ;
  %-------------------------------------------------;
  \begin{align}
    \label{Eqn:P_coupled_Main}
    &\frac{1}{2}\left(\Phi_{\mathrm{free}}\left[
      \mathbf{v}_{\mathrm{free}}^{(1)}\right] 
    -\Phi_{\mathrm{free}}\left[\mathbf{v}_{\mathrm{free}}^{(2)}
      \right]\right)
    +\frac{1}{2}\left(\Phi_{\mathrm{por}}\left[
      \mathbf{v}_{\mathrm{por}}^{(1)}\right] 
    -\Phi_{\mathrm{por}}\left[\mathbf{v}_{\mathrm{por}}^{(2)}
      \right]\right) 
    \nonumber \\
    &+\int_{\Gamma_{\mathrm{int}}} \left(\Psi\left[
      \overset{*}{\mathbf{v}}_{\mathrm{free}}^{(1)},
      \overset{*}{\mathbf{v}}_{\mathrm{por}}^{(1)},
      v_n^{(1)}\right] 
    -\Psi\left[
      \overset{*}{\mathbf{v}}_{\mathrm{free}}^{(2)}, 
      \overset{*}{\mathbf{v}}_{\mathrm{por}}^{(2)},
      v_n^{(2)}\right] \right) \mathrm{d}\Gamma 
    \nonumber \\
    &=\int_{\mathcal{K}_{\mathrm{free}}} \gamma 
    \mathbf{b}_{\mathrm{free}} \cdot 
    \left(\mathbf{v}_{\mathrm{free}}^{(1)}-\mathbf{v}_{\mathrm{free}}^{(2)}\right) 
    \; \mathrm{d}\Omega 
    +\int_{\Gamma_{\mathrm{free}}^{t}} \mathbf{t}_{\mathrm{free}}^{\mathrm{p}} 
    \cdot \left(\mathbf{v}_{\mathrm{free}}^{(1)}-\mathbf{v}_{\mathrm{free}}^{(2)}\right) 
    \; \mathrm{d}\Omega \nonumber \\
    &+\int_{\mathcal{K}_{\mathrm{por}}} 
    \gamma \phi_{\mathrm{por}} \mathbf{b}_{\mathrm{por}} \cdot 
    \left(\mathbf{v}_{\mathrm{por}}^{(1)}-\mathbf{v}_{\mathrm{por}}^{(2)}\right) 
    \; \mathrm{d}\Omega 
    +\int_{\Gamma_{\mathrm{por}}^{t}} \mathbf{t}_{\mathrm{por}}^{\mathrm{p}} 
    \cdot \left(\mathbf{v}_{\mathrm{por}}^{(1)}-\mathbf{v}_{\mathrm{por}}^{(2)}\right) 
    \; \mathrm{d}\Omega
  \end{align}

  Noting the rate of internal dissipation in the Stokes
  model, it is easy to establish the following: 
  {\small
    \begin{align}
      \frac{1}{2}\left(\Phi_{\mathrm{free}}\left[
        \mathbf{v}_{\mathrm{free}}^{(1)}\right]
      -\Phi_{\mathrm{free}}\left[\mathbf{v}_{\mathrm{free}}^{(2)}
        \right]\right)
      = \frac{1}{2}\Phi_{\mathrm{free}}\left[\mathbf{v}_{\mathrm{free}}^{(1)}
        - \mathbf{v}_{\mathrm{free}}^{(2)}\right]
      + \int_{\mathcal{K}_{\mathrm{free}}} 2 \mu
      \mathbf{D}_{\mathrm{free}}^{(2)}
      \cdot \left(\mathbf{D}_{\mathrm{free}}^{(1)} -
      \mathbf{D}_{\mathrm{free}}^{(2)}\right) 
      \mathrm{d}\Omega 
    \end{align}
  }
  Using equation $\eqref{Eqn:divergence_free_for_por}_{2}$
  the above equation can be written as follows: 
  {\small
    \begin{align}
      \label{Eqn:Reduced_dissipation_free}
      \frac{1}{2}\left(\Phi_{\mathrm{free}}\left[
        \mathbf{v}_{\mathrm{free}}^{(1)}\right]
      -\Phi_{\mathrm{free}}\left[\mathbf{v}_{\mathrm{free}}^{(2)}
        \right]\right)
      = \frac{1}{2}\Phi_{\mathrm{free}}\left[\mathbf{v}_{\mathrm{free}}^{(1)}
        - \mathbf{v}_{\mathrm{free}}^{(2)}\right]
      + \int_{\mathcal{K}_{\mathrm{free}}} \mathbf{T}_{\mathrm{free}}^{(2)}
      \cdot \left(\mathbf{D}_{\mathrm{free}}^{(1)} -
      \mathbf{D}_{\mathrm{free}}^{(2)}\right) 
      \mathrm{d}\Omega 
  \end{align}}
  where
  \begin{align}
    \mathbf{T}_{\mathrm{free}}^{(2)} = -p_{\mathrm{free}}^{(2)}\mathbf{I}
    + 2 \mu \mathbf{D}_{\mathrm{free}}^{(2)}
  \end{align}

  On similar lines, one can establish the following relation:
  {\small
    \begin{align}
      \frac{1}{2}\left(\Phi_{\mathrm{por}}\left[
        \mathbf{v}_{\mathrm{por}}^{(1)}\right]
      -\Phi_{\mathrm{por}}\left[\mathbf{v}_{\mathrm{por}}^{(2)}
        \right]\right)
      = \frac{1}{2}\Phi_{\mathrm{por}}\left[\mathbf{v}_{\mathrm{por}}^{(1)}
        - \mathbf{v}_{\mathrm{por}}^{(2)}\right]
      &+ \int_{\mathcal{K}_{\mathrm{por}}} \mathbf{T}_{\mathrm{por}}^{(2)}
      \cdot \left(\mathbf{D}_{\mathrm{por}}^{(1)} -
      \mathbf{D}_{\mathrm{por}}^{(2)}\right) 
      \mathrm{d}\Omega \nonumber \\
      &+ \int_{\mathcal{K}_{\mathrm{por}}} \mu \mathbf{K}^{-1}
      \mathbf{v}_{\mathrm{por}}^{(2)} \cdot
      \left(\mathbf{v}_{\mathrm{por}}^{(1)}
      - \mathbf{v}_{\mathrm{por}}^{(2)}\right)
      \mathrm{d} \Omega 
  \end{align}}
  where
  \begin{align}
    \mathbf{T}_{\mathrm{por}}^{(2)} = -p_{\mathrm{por}}^{(2)}\mathbf{I}
    + 2 \mu \mathbf{D}_{\mathrm{por}}^{(2)}
  \end{align}
    
  We note the fields under the second
  solution satisfy the balance of linear
  momentum; that is: 
  \begin{alignat}{2}
    \label{Eqn:Interface_BoLM_free}
    &\mathrm{div}\left[\mathbf{T}_{\mathrm{free}}^{(2)}\right]
    + \gamma \mathbf{b}_{\mathrm{free}}
    = \mathbf{0}     
    &&\quad \mathrm{in} \; \mathcal{K}_{\mathrm{free}} \\
    \label{Eqn:Interface_BoLM_por}
    &\mathrm{div}\left[\mathbf{T}_{\mathrm{por}}^{(2)}\right]
    + \gamma \phi_{\mathrm{por}} \mathbf{b}_{\mathrm{por}}
    = \mu \mathbf{K}^{-1} \mathbf{v}_{\mathrm{por}}^{(2)}
    &&\quad \mathrm{in} \; \mathcal{K}_{\mathrm{por}}
  \end{alignat}
  and the prescribed tractions on
  the external boundary; that is:
  %--------------------------------------;
  %  Equation: Definitions of tractions  ;
  %--------------------------------------;
  \begin{align}
    \label{Eqn:Interface_traction_free}
    \mathbf{t}_{\mathrm{free}}^{(2)}
    := \mathbf{T}_{\mathrm{free}}^{(2)}
    \widehat{\mathbf{n}}_{\mathrm{free}}
    = \mathbf{t}_{\mathrm{free}}^{\mathrm{p}}
    \quad \mathrm{on} \; \Gamma_{\mathrm{free}}^{t} \\
        \label{Eqn:Interface_traction_por}
    \mathbf{t}_{\mathrm{por}}^{(2)}
    := \mathbf{T}_{\mathrm{por}}^{(2)}
    \widehat{\mathbf{n}}_{\mathrm{por}}
    = \mathbf{t}_{\mathrm{por}}^{\mathrm{p}}
    \quad \mathrm{on} \; \Gamma_{\mathrm{por}}^{t}
  \end{align}

  Using equations
  \eqref{Eqn:Reduced_dissipation_free}--\eqref{Eqn:Interface_BoLM_por}
  and the interface conditions
  \eqref{Eqn:normal_traction_continuity}--\eqref{Eqn:tangential_traction_porous}, 
  equation \eqref{Eqn:P_coupled_Main} reduces to the following:
  %----------------------------------------------;
  %  Equation: Substituting interface condition  ;
  %----------------------------------------------;
  \begin{align}
    &\frac{1}{2} \Phi_{\mathrm{free}}\left[\mathbf{v}^{(1)}_{\mathrm{free}} - 
      \mathbf{v}^{(2)}_{\mathrm{free}}\right] 
    +\frac{1}{2} \Phi_{\mathrm{por}}\left[\mathbf{v}^{(1)}_{\mathrm{por}} - 
      \mathbf{v}^{(2)}_{\mathrm{por}}\right] 
    +\int_{\Gamma_{\mathrm{int}}} \left(\Psi\left[
      \overset{*}{\mathbf{v}}_{\mathrm{free}}^{(1)},
      \overset{*}{\mathbf{v}}_{\mathrm{por}}^{(1)},
      v_{n}^{(1)}\right] 
    -\Psi\left[\overset{*}{\mathbf{v}}_{\mathrm{free}}^{(2)},
      \overset{*}{\mathbf{v}}_{\mathrm{por}}^{(2)},
      v_{n}^{(2)}\right]\right)
    \mathrm{d} \Gamma \nonumber \\
    &= \int_{\Gamma_{\mathrm{int}}} \left(\frac{\partial \Psi}{\partial
      \overset{*}{\mathbf{v}}_{\mathrm{free}}^{(2)}} 
    \cdot \left(\overset{*}{\mathbf{v}}_{\mathrm{free}}^{(1)}
    - \overset{*}{\mathbf{v}}_{\mathrm{free}}^{(2)} \right) 
    +\frac{\partial \Psi}{\partial
      \overset{*}{\mathbf{v}}_{\mathrm{por}}^{(2)}} 
    \cdot \left(\overset{*}{\mathbf{v}}_{\mathrm{por}}^{(1)}
    - \overset{*}{\mathbf{v}}_{\mathrm{por}}^{(2)} \right)
    +\frac{\partial \Psi}{\partial
     v_{n}^{(2)}} 
    \cdot \left(v_{n}^{(1)} - v_{n}^{(2)} \right)
    \right) \mathrm{d} \Gamma 
  \end{align}
  Noting the functional form of $\Psi$,
  the above equation reduces to the
  following: 
  %----------------------------------------------;
  %  Equation: Utilizing functional form of Psi  ;
  %----------------------------------------------;
  {\small
  \begin{align}
    \frac{1}{2} \Phi_{\mathrm{free}}\left[\mathbf{v}^{(1)}_{\mathrm{free}} - 
      \mathbf{v}^{(2)}_{\mathrm{free}}\right] 
    &+\frac{1}{2} \Phi_{\mathrm{por}}\left[\mathbf{v}^{(1)}_{\mathrm{por}} - 
      \mathbf{v}^{(2)}_{\mathrm{por}}\right] 
    +\int_{\Gamma_{\mathrm{int}}} \Psi\left[
      \overset{*}{\mathbf{v}}_{\mathrm{free}}^{(1)}
      - \overset{*}{\mathbf{v}}_{\mathrm{free}}^{(2)},
      \overset{*}{\mathbf{v}}_{\mathrm{por}}^{(1)}
      - \overset{*}{\mathbf{v}}_{\mathrm{por}}^{(2)}, 
      v_n^{(1)} - v_n^{(2)}\right] \mathrm{d} \Gamma
    = 0
  \end{align}}
  Using the fact that $\Phi_{\mathrm{free}}[\cdot]$,
  $\Phi_{\mathrm{por}}[\cdot]$ and $\Psi[\cdot]$ are
  individually norms (and hence individually
  non-negative), each term in the above equation
  is individually zero. This further implies that 
  \begin{subequations}
    \begin{alignat}{2}
      &\mathbf{v}_{\mathrm{free}}^{(1)}(\mathbf{x})
      =\mathbf{v}_{\mathrm{free}}^{(2)}(\mathbf{x}) 
      &&\quad \forall \mathbf{x} \in \mathcal{K}_{\mathrm{free}} \\
      &\mathbf{v}_{\mathrm{por}}^{(1)}(\mathbf{x}) =
      \mathbf{v}_{\mathrm{por}}^{(2)}(\mathbf{x})
      &&\quad \forall \mathbf{x} \in \mathcal{K}_{\mathrm{por}} \\
      &\overset{*}{\mathbf{v}}_{\mathrm{free}}^{(1)}(\mathbf{x})
      = \overset{*}{\mathbf{v}}_{\mathrm{free}}^{(2)}(\mathbf{x})
      &&\quad \forall \mathbf{x} \in \Gamma_{\mathrm{free}} \\
      &\overset{*}{\mathbf{v}}_{\mathrm{por}}^{(1)}(\mathbf{x})
      = \overset{*}{\mathbf{v}}_{\mathrm{por}}^{(2)}(\mathbf{x})
      &&\quad \forall \mathbf{x} \in \Gamma_{\mathrm{por}} \\
      &v_{n}^{(1)}(\mathbf{x}) = v_{n}^{(2)}(\mathbf{x})
      &&\quad \forall \mathbf{x} \in \Gamma_{\mathrm{int}} 
    \end{alignat}
  \end{subequations}
  The balance of linear momentum
  in $\mathcal{K}_{\mathrm{free}}$
  and $\mathcal{K}_{\mathrm{por}}$,
  respectively, implies that:
  \begin{subequations}
    \label{Eqn: Gradient_of_p}
    \begin{alignat}{2}
      \label{Eqn: Gradient_of_p_1}
      &\mathrm{grad}\left[p_{\mathrm{free}}^{(1)}(\mathbf{x})
        -p_{\mathrm{free}}^{(2)}(\mathbf{x})\right] = \mathbf{0} \qquad
      && \forall \mathbf{x} \in \mathcal{K}_{\mathrm{free}}\\
      \label{Eqn: Gradient_of_p_2}
      &\mathrm{grad}\left[p_{\mathrm{por}}^{(1)}(\mathbf{x})
        -p_{\mathrm{por}}^{(2)}(\mathbf{x})\right] = \mathbf{0} \qquad
      && \forall \mathbf{x} \in \mathcal{K}_{\mathrm{por}}
    \end{alignat}
  \end{subequations}
  which further implies that:
    \begin{align}
      p_{\mathrm{free}}^{(1)}(\mathbf{x})
      = p_{\mathrm{free}}^{(2)}(\mathbf{x})
      + C_1 
      \quad \forall \mathbf{x} \in \mathcal{K}_{\mathrm{free}}
      \quad \mathrm{and} \quad 
      p_{\mathrm{por}}^{(1)}(\mathbf{x})
      = p_{\mathrm{por}}^{(2)}(\mathbf{x})
      + C_2 
      \quad  \forall \mathbf{x} \in \mathcal{K}_{\mathrm{por}}
    \end{align}
  where $C_1$ and $C_2$ are arbitrary constants. 
  Using the interface condition given by equation
  \eqref{Eqn:normal_traction_continuity} and 
  noting that the velocity fields are continuous 
  fields, we conclude that $C_1 = C_2 = C$ and
  \begin{align}
    p^{(1)}_{\mathrm{free}}(\mathbf{x}) =
    p^{(2)}_{\mathrm{free}}(\mathbf{x}) + C
    \quad \mathrm{and} \quad 
    p^{(1)}_{\mathrm{por}}(\mathbf{x}) =
    p^{(2)}_{\mathrm{por}}(\mathbf{x}) + C
    \quad \forall \mathbf{x} \in \Gamma_{\mathrm{int}}
  \end{align}
  Physically, the constant $C$ fixes
  the datum for the pressure field.
  This completes the proof.

\end{document}